\pdfoutput=1 
\documentclass[traditabstract]{aa}  
\usepackage{graphicx}

\usepackage{txfonts}
\usepackage{rotating,tabularx}
\usepackage[margin=1.0in]{geometry} 
\usepackage{natbib,twoopt}
\bibpunct{(}{)}{;}{a}{}{,} 
\newcommandtwoopt{\citeads}[3][][]{\href{http://adsabs.harvard.edu/abs/#3}%
  {\citealp[#1][#2]{#3}}}
\newcommandtwoopt{\citepads}[3][][]{\href{http://adsabs.harvard.edu/abs/#3}%
  {\citep[#1][#2]{#3}}}
\newcommandtwoopt{\citetads}[3][][]{\href{http://adsabs.harvard.edu/abs/#3}%
  {\citet[#1][#2]{#3}}} 
\newcommandtwoopt{\citeyearads}[3][][]%
                 {\href{http://adsabs.harvard.edu/abs/#3}{\citeyear[#1][#2]{#3}}} 

\def\kms    {\ifmmode{{\rm \ts km\ts s}^{-1}}\else{\ts km\ts s$^{-1}$}\fi}
\def\lsol   {\ifmmode{{\rm L}_{\odot}}\else{L$_{\odot}$}\fi}
\def\msol   {\ifmmode{{\rm M}_{\odot}}\else{M$_{\odot}$}\fi}
\def\hi     {\ifmmode{{\rm H}{\rm \small I}}\else{H\ts {\scriptsize I}}\fi}
\def\hh   {\ifmmode{{\rm H}_2}\else{H$_2$}\fi}

\def\zsol   {\ifmmode{{\rm Z}_{\odot}}\else{Z$_{\odot}$}\fi}
\def\tex {\ifmmode{{T}_{\rm ex}}\else{$T_{\rm ex}$}\fi}
\def\tmb {\ifmmode{{T}_{\rm mb}}\else{$T_{\rm mb}$}\fi}
\begin{document} 

\title{Dense gas tracing the collisional past of Andromeda\thanks{Based on observations carried out with the IRAM 30m radio telescope. IRAM is supported by INSU/CNRS (France), MPG (Germany) and IGN (Spain).}}

\subtitle {An atypical inner region?}

\author{Anne-Laure Melchior\inst{1}, Fran\c coise Combes\inst{2}}

\institute{LERMA, Sorbonne Universit\'es, UPMC Univ. Paris 6,  Observatoire de Paris, PSL Research University, CNRS, Paris, France \\
  \email{A.L.Melchior@obspm.fr}
  \and
  Observatoire de Paris, LERMA, Coll\`ege de France, PSL, CNRS, Sorbonne Univ., UPMC, Paris, France\\
  \email{Francoise.Combes@obspm.fr}
  }

\date{Received April 3, 2015; accepted }

\abstract {The central kiloparsec region of the Andromeda galaxy is
  relatively gas poor, while the interstellar medium appears to be
  concentrated in a ring-like structure at about 10\,kpc radius. The
  central gas depletion has been attributed to a possible head-on
  collision 200\,Myr ago, supported by the existence of an offset
  inner ring of warm dust. We present new IRAM 30\,m radio telescope observations of the
  molecular gas in the central region, and the detection of CO and its
  isotopes $^{13}$CO$(2-1)$ and C$^{18}$O(2$-$1), together with the
  dense gas tracers, HCN(1$-$0) and HCO+(1$-$0).  A systematic study
  of the observed peak temperatures with non-LTE equilibrium
  simulations shows that the detected lines trace dense regions with
  n$_{H_2}$ in the range $2.5\times 10^4 - 5.6\times 10^5$\,cm$^{-3}$,
  while the gas is very clumpy with a beam filling factor of $0.5-2
  \times 10^{-2}$. This is compatible with the dust mass derived
  from the far-infrared emission, assuming a dust-to-gas mass ratio of
  0.01 with a typical clump size of 2\,pc. We also show that the
  gas is optically thin in all lines except for $^{12}$CO(1-0) and
  $^{12}$CO(2-1), CO lines are close to their thermal
  equilibrium condition at 17-20\,K, the molecular hydrogen density
  is larger than critical, and HCN and HCO+ lines have a
  subthermal excitation temperature of {9\,K} with a density
  smaller than critical.  The average $^{12}$CO/$^{13}$CO line
  ratio is high ($\sim$ 21), and {close} to the
  $^{12}$CO/C$^{18}$O ratio ($\sim 30$) that was measured in the north-western
  region and estimated in the south-east stacking. The fact that the
  optically thin $^{13}$CO and C$^{18}$O lines have comparable
  intensities means that the secondary element $^{13}$C is depleted
  with respect to the primary $^{12}$C, as is expected just after a
  recent star formation. This suggests that there has been a recent
  starburst in the central region, supporting the head-on collision
  scenario.  }
    
\keywords{galaxies: abundances; galaxies: individual: M31; galaxies: kinematics and dynamics; submillimeter: ISM; molecular data}

\maketitle
%
%

\section{Introduction}
While the outer parts of the Andromeda galaxy exhibit numerous relics
of past interactions \citep{2009Natur.461...66M}, the main disc
{exhibits an enhanced} star formation activity in a ring at 10\,kpc
\citep{2013ApJ...769...55F} and the central part is atypical. The M31 galaxy hosts
a very massive black hole with a mass of $0.7-1.4 \times 10^8 M_\odot$
\citep{2001A&A...371..409B,2005ApJ...631..280B} that is one of the
most under-luminous supermassive black holes \citep{2010ApJ...710..755G}, which is surrounded by very little gas
\citep{2013A&A...549A..27M}. Infrared data and optical ionised gas
display a 0.7\,kpc off-centred inner ring, and
\citet{2006Natur.443..832B} have proposed a frontal collision with M32
to explain this double ring structure. This collision would have occurred 200\,Myr
ago.  In this paper, we present observations of millimetre molecular
lines probing dense gas regions along the minor axis in the inner
ring.  In Sect. \ref{sec:obs}, we
present the data used in this paper. In Sect. \ref{sec:datared}, we
describe the data reduction. In Sect. \ref{sec:resu}, we describe
our results and which physical information we can extract. In Sect.
\ref{sec:disc}, we argue that our results support a 200\,Myr old
starburst triggered by the frontal collision with M32. For consistency with our previous analysis, we consider a distance of M31 of 780\,kpc and a pixel size 1\,arcsec $=$ 3.8\,pc.
\begin{figure*}
  \centering 
  \includegraphics[width=\textwidth]{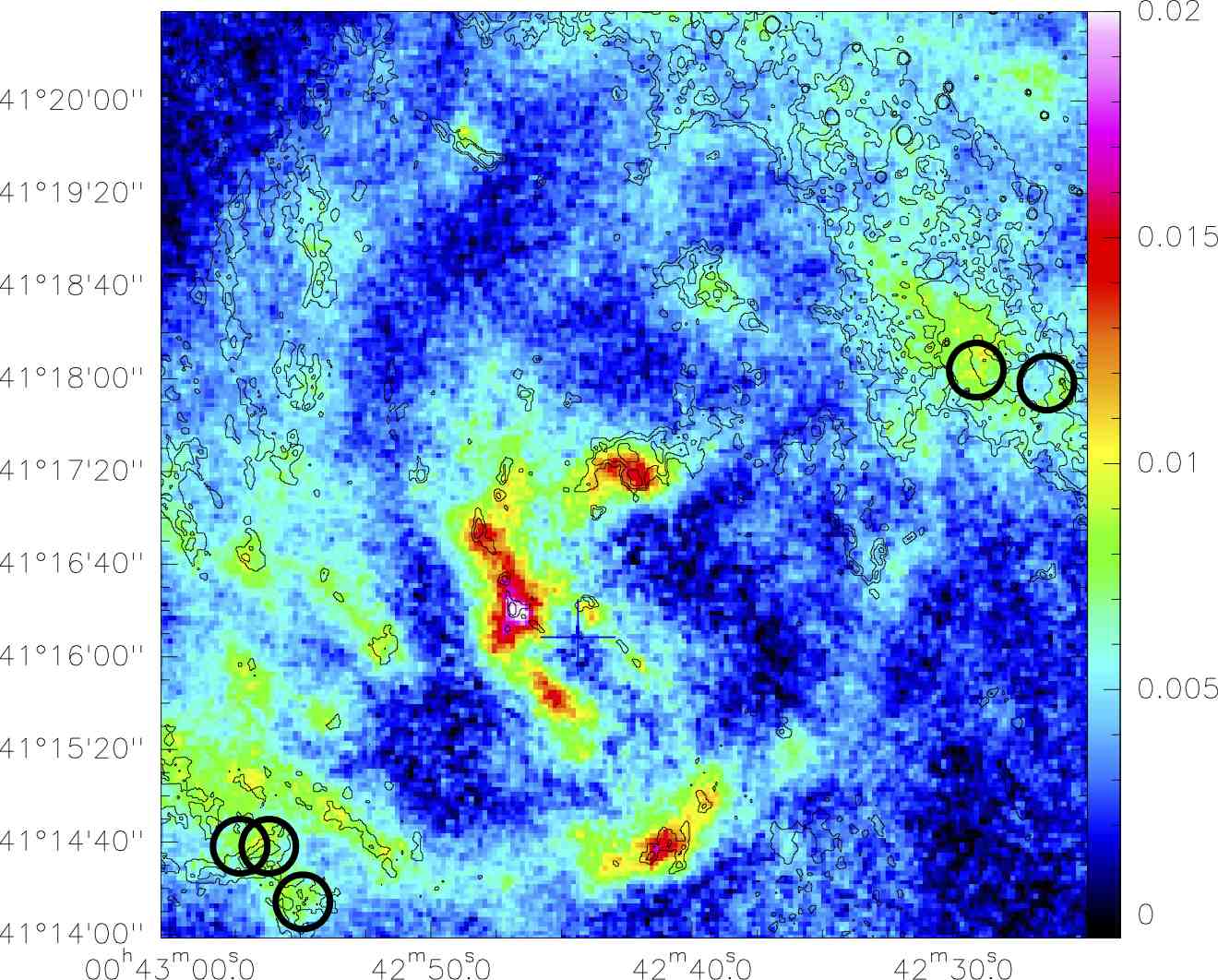}
  \caption{Positions observed in HCO+(1-0), HCN(1-0) and
    $^{13}$CO(2-1) (and C$^{18}$O) are indicated with circles
    superimposed on the PACS/Herschel 100$\mu$m map
    \citep{2014A&A...567A..71V}. Contour levels (0.05, 0.1, 0.2)
    correspond to A$_B$ extinction computed assuming all the dust is in
    front of the bulge as discussed in
    \citet{2000MNRAS.312L..29M}. Extinction is underestimated by an
    unknown factor due to the uncertainties in the geometrical
    configuration of the gas in this area.}
  \label{fig:ext}%
\end{figure*}


\section{The data}
\label{sec:obs}
\subsection{Millimetre observations}
\begin{table}
  \caption{M31 observed positions. }             
  \label{table:1}      
  \centering                          
  \begin{tabular}{llllll}        
    \hline\hline                 
    Pos & RA\hspace{1cm} DEC  & $\Delta \alpha$ & $\Delta \delta$  & T$_{int}$\\    
    & (J2000)\hspace{0.5cm}(J2000)  & (")& (") & (min) \\    
    \hline                        
    I     & 00 42 29.1 $+$41 18 03.6 & -172.1 & 115.3 & 284 \\      
    I-B   & 00 42 26.4 $+$41 17 58.0 & -202.5 & 109.7 & 173 \\
    2d-a & 00 42 56.2 $+$41 14 38.0 &  133.4 & -90.3 & 262 \\
    2d-b & 00 42 57.3 $+$41 14 38.0 &  145.8 & -90.3 & 204 \\
    2c-a & 00 42 54.9 $+$41 14 14.0 &  118.8 &-114.3 &  79 \\ 
    \hline                                   
  \end{tabular}
\tablefoot{We provide the J2000 coordinates, offsets and integration times of the five positions observed.}
\end{table} 
\begin{table*}
  \caption{Dust properties derived from the infrared data for the M31
    observed positions.}
  \label{tab:1b}      
  \centering                          
  \begin{tabular}{c|cccc|cccccc}        
    \hline\hline                 
& \multicolumn{4}{c|}{D14} & \multicolumn{6}{c}{V14} \\ \hline
    Pos &  U$_{SPIRE350}$ & U$_{M160}$ & T$^{cold\,dust}_{SPIRE350}$ & T$^{cold\,dust}_{M160}$  & T$^C_{dust}$ &T$^W_{dust}$ & M$_{dust}$ & M$_*$ & SFR & $\tau_v$\\    
    & && (K)  & (K)& (K) & (K) & ($10^3$\,M$_\odot$) & ($10^7$\,M$_\odot$) & ($M_\odot$/Myr) &\\    
\hline                        
 I    & 1.53$\pm$0.21 &2.1$\pm$0.5 &19.3 &20.3& 20.  & 42. & 2.90 & 3.2 & 3.3  & 0.52 \\      
 I-B  & 1.52$\pm$0.02 &1.3$\pm$0.3 &19.3 &18.7& 20.  & 57. & 2.86 & 2.5 & 2.8  & -- \\
 2d-a & 2.73$\pm$0.37 &2.6$\pm$0.5 &21.3 &21.1& 23.  & 34. & 0.98 & 3.2 & 13.4 & --\\
 2d-b & 1.78$\pm$0.15 &2.6$\pm$0.5 &19.8 &21.1& 23.  & 34. & 0.98 & 3.2 & 13.4 & --\\
 2c-a & 2.07$\pm$0.40 &3.1$\pm$0.4 &20.3 &21.8& 23.  & --  & 1.49 & 3.5 & 16.1 & 1.89\\ 
 \hline                                   
  \end{tabular}
\tablefoot{In the second and third columns, we provide the mean
  starlight heating rates provided by
  \citet[][D14]{2014ApJ...780..172D} for two different resolutions
  (24.$^{\prime\prime}$9 and 39\arcsec) at the pointing positions. In
  the fourth and fifth columns, we present cold dust temperatures
  derived from these mean starlight measurements. In the sixth through
  tenth columns, we provide cold and warm dust temperatures, dust and
  stellar masses, star formation rates and dust optical depths
  computed by \citet[][V14]{2014A&A...567A..71V} at these positions in
  36\arcsec pixels.}
\end{table*}

We observed five positions listed in Table \ref{table:1}, which also
gives the offsets and integration time. We computed the offsets with
respect to the position of the optical centre
((J2000) $00^h 42^m 44.371^s$\, $+$41$^\circ 16^\prime
08.34^{\prime\prime}$) provided by \citet{1992ApJ...390L...9C} and
used in previous papers
\citep{2000MNRAS.312L..29M,2011A&A...536A..52M,2013A&A...549A..27M}. The
observed positions lie in the inner 1\,kpc ring along the minor axis
(of the main disc inclined at 77\,$\deg$). The total integration time
for each position is provided in the last column. In Figure
\ref{fig:ext}, these positions are superimposed on the Herschel
100\,$\mu$m map provided by \citet{2014A&A...567A..71V} and A$_B$
extinction contours computed in \citet{2000MNRAS.312L..29M}. In the
100\,$\mu$m and A$_B$ maps, we can see the 0.7\,kpc off-centred inner
ring.
\begin{figure}
  \centering
  \includegraphics[width=7.95cm]{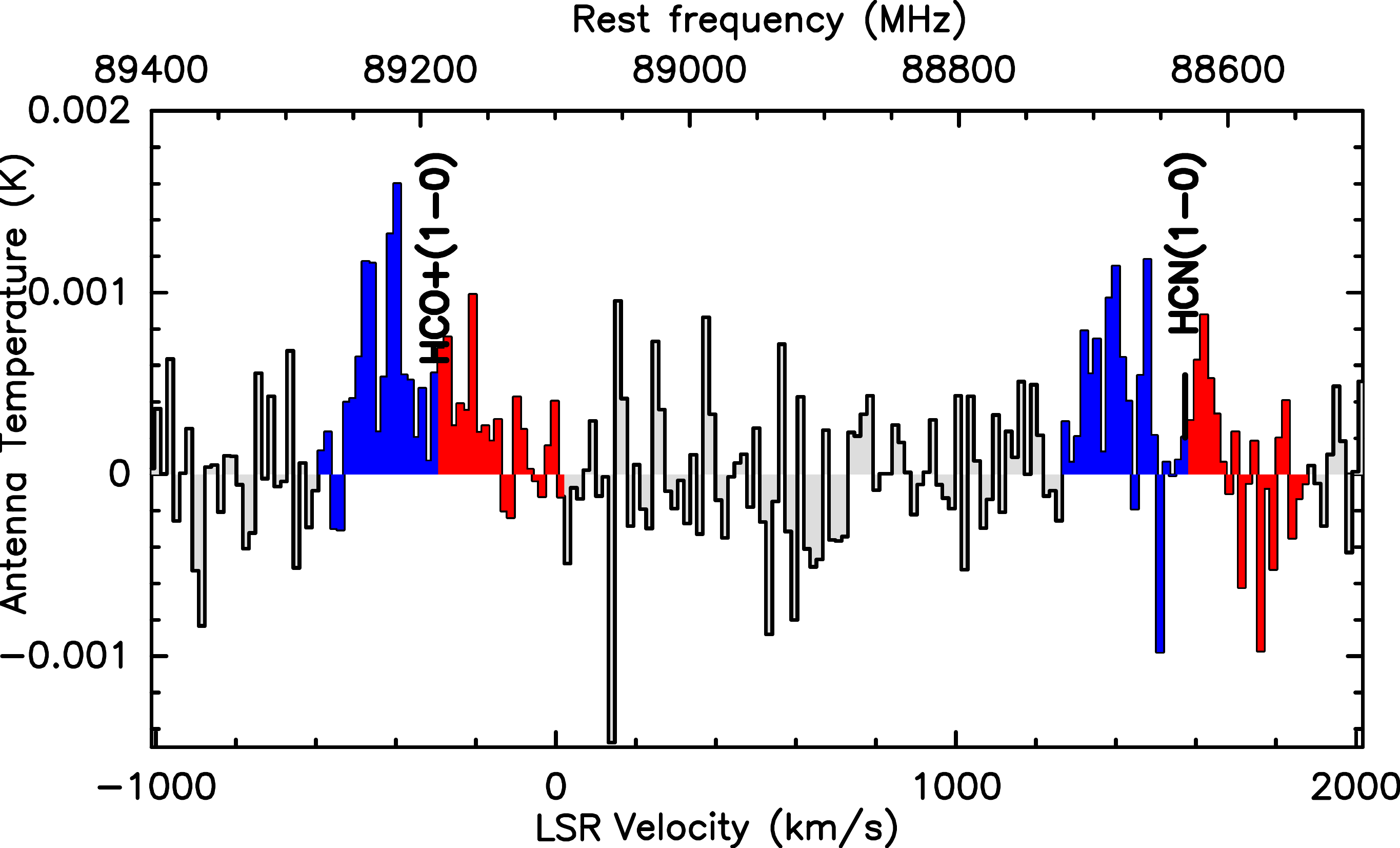}
  \caption{{Averaging} of the observations performed in the 88\,GHz
    bandwidth. The main astronomical lines present in the
    bandwidth are indicated (in black) in the rest-frame frequency of the
    Andromeda galaxy, whose systemic velocity is around
    -300\,km\,s$^{-1}$.  {For each line, the redshifted and  blueshifted parts of the expected velocity range are indicated.}  The noise level achieved is 0.2\,mK for a 
    frequency binning $\Delta \nu = 14$\,MHz. There are clear
    detections of HCO+(1-0) and HCN(1-0), { with possible components at both sides of the systemic velocity.}}
  \label{fig:stackpl2}%
\end{figure}
\begin{figure}
  \centering
  \includegraphics[width=7.95cm]{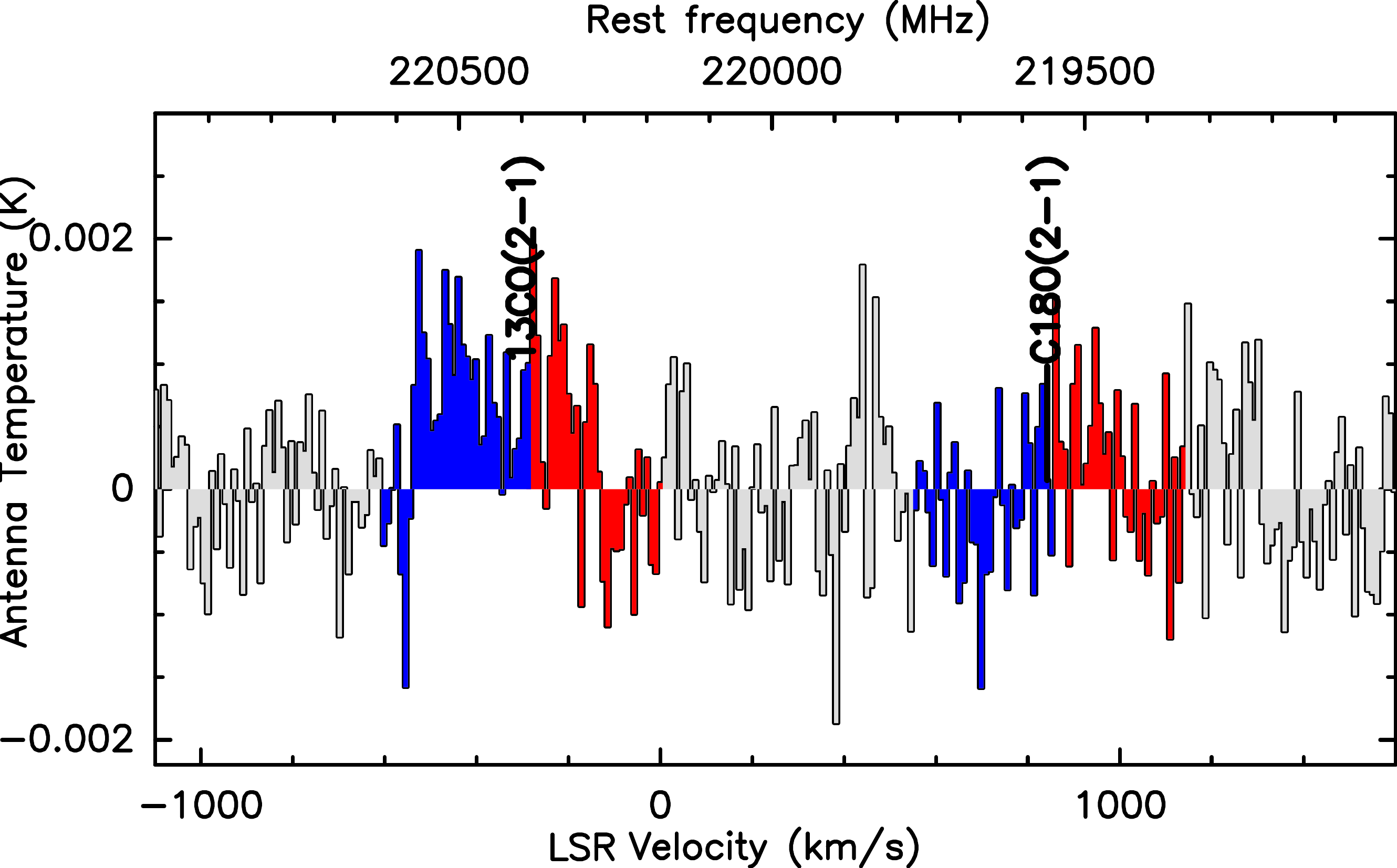}
  \caption{Averaging of the observations performed in the 216\,GHz
    bandwidth. The $^{13}$CO(2-1) and C$^{18}$O(2-1) lines present in
    this bandwidth are indicated in the rest-frame frequency of the
    Andromeda galaxy.  For each line, the redshifted and blueshifted
    parts of the expected velocity range are indicated. The noise
    level achieved is 3\,mK for a frequency binning $\Delta \nu =
    7$\,MHz ($\Delta v = 9.5$\,km\,s$^{-1}$). Apart from the
    multiple-component $^{13}$CO(2-1) signal, the C$^{18}$O(2-1)
    signal is weak in the receding part of the spectra.}
    \label{fig:stackpl}%
\end{figure}
\begin{figure}
  \centering
  \includegraphics[width=7.95cm]{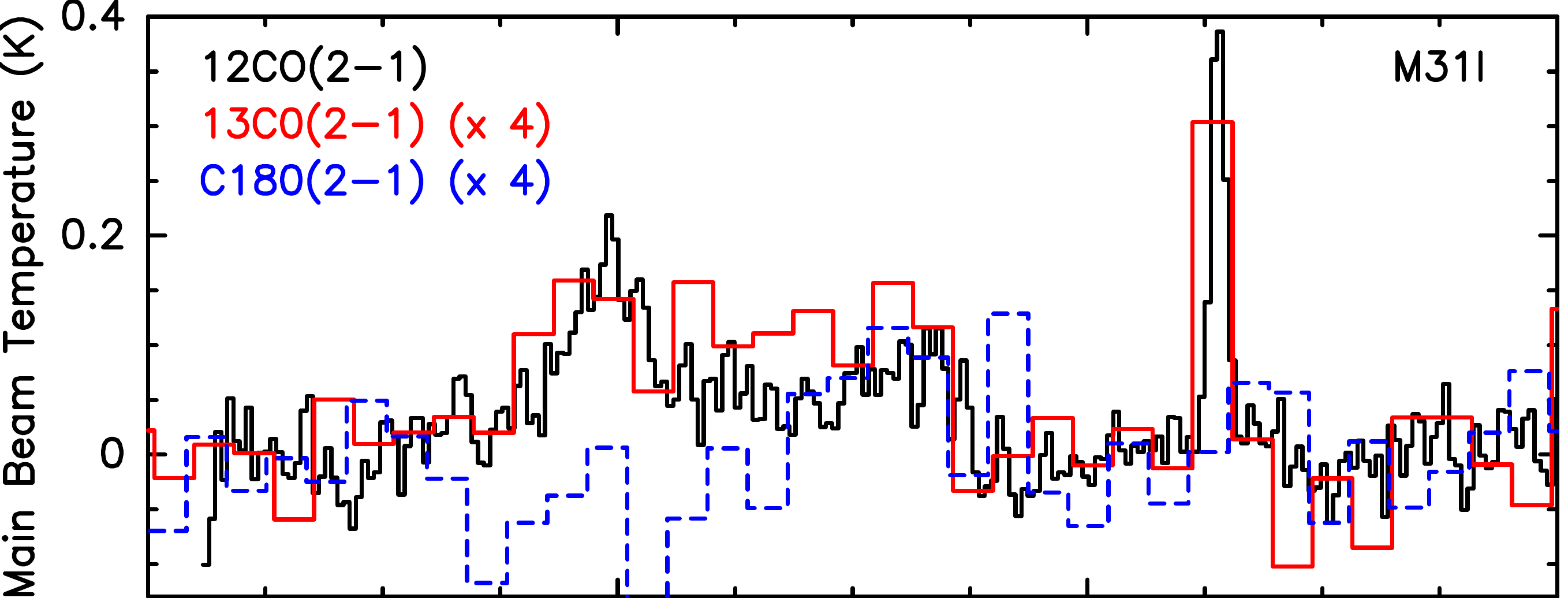}
  \includegraphics[width=7.95cm]{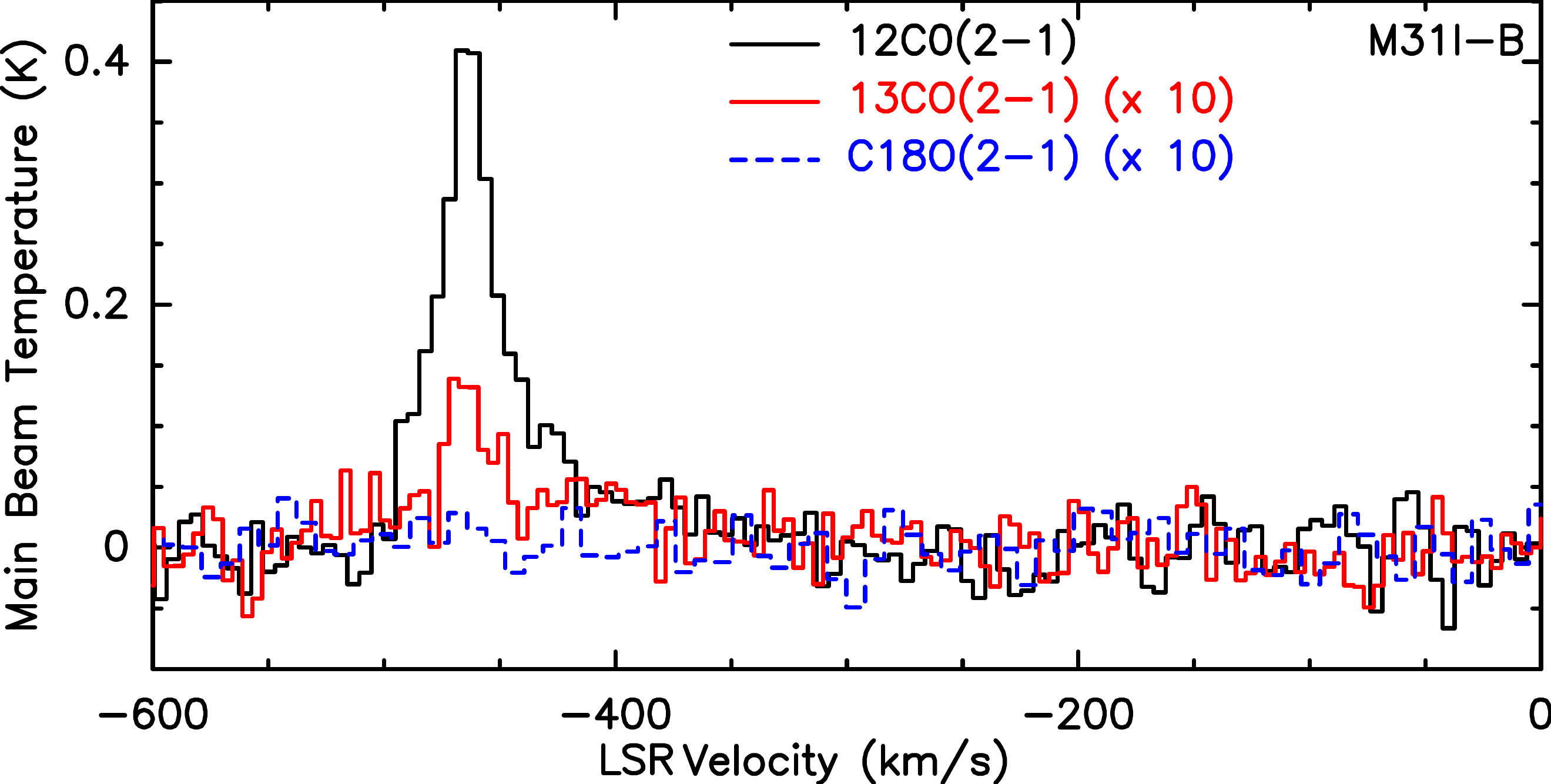}  
  \caption{Carbon monoxide spectra observed on the north-west side of
    the inner ring are superimposed in velocity. The main beam
    temperature is provided for $^{12}$CO, while the $^{13}$CO and
    C$^{18}$O spectra are multiplied by a factor of 10 and 4 in the bottom and top panels, respectively.}
  \label{fig:CObis}%
\end{figure}
\begin{figure}
  \centering
  \includegraphics[width=7.95cm]{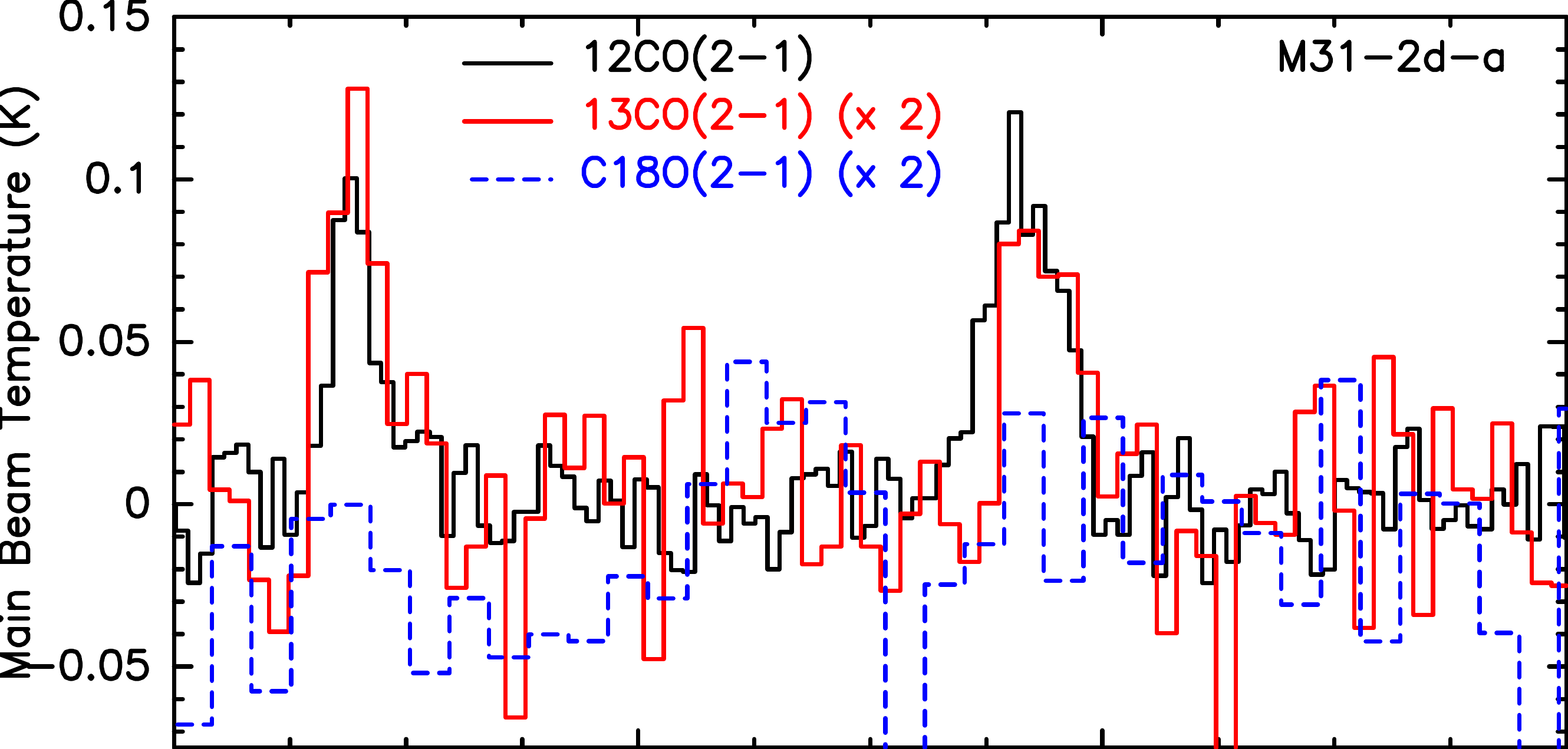}
  \includegraphics[width=7.95cm]{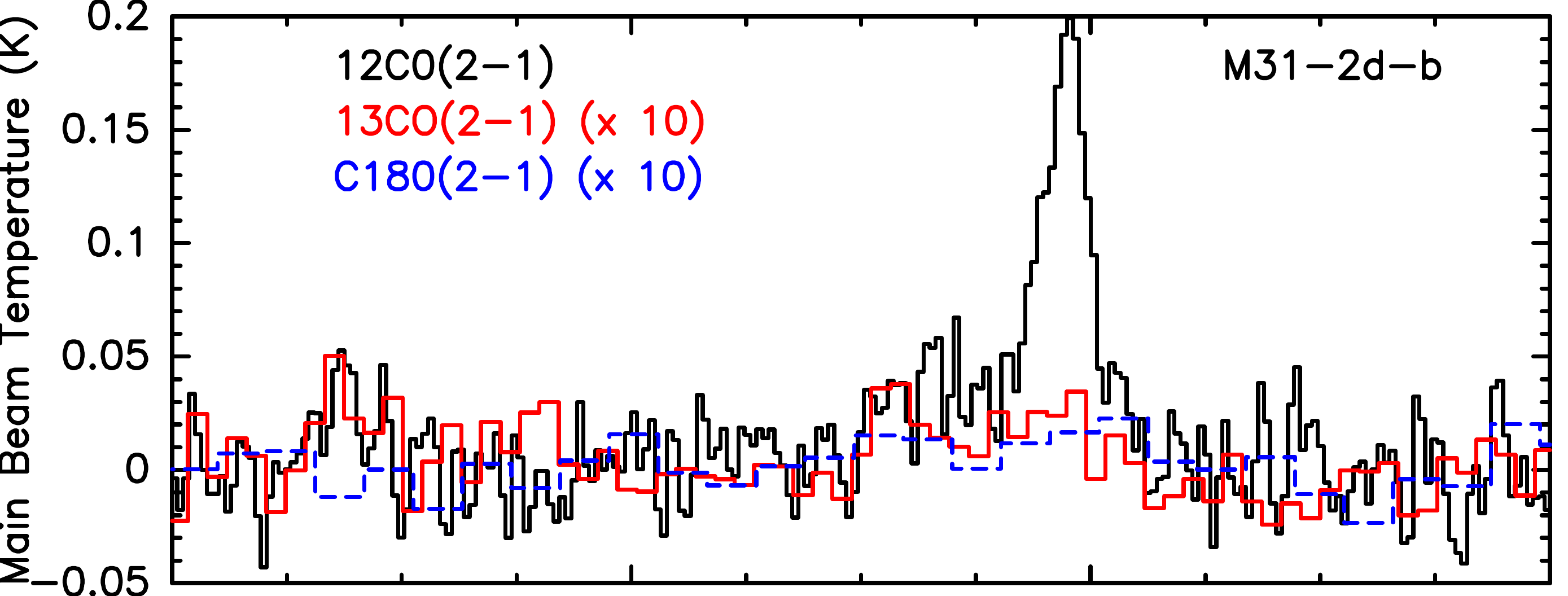}
  \includegraphics[width=7.95cm]{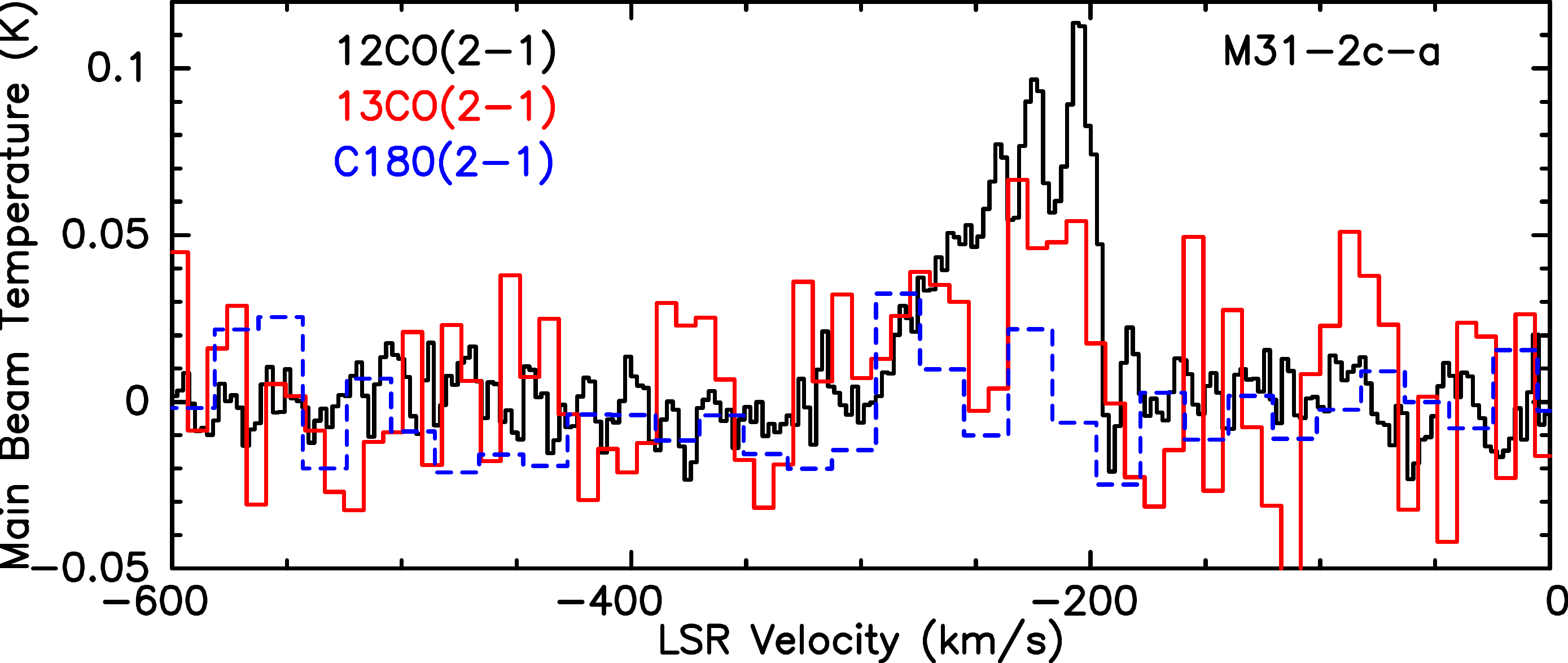}
  \caption{Carbon monoxide spectra observed on the south-east side of
    the inner ring are superimposed in velocity. The main beam
    temperature is provided for $^{12}$CO, while the $^{13}$CO and
    C$^{18}$O spectra are multiplied by a factor of 2, 10, and 1 for the top, middle, and bottom panels.}
  \label{fig:12CO(2-1)}%
\end{figure}

The observations have been performed with the IRAM 30\,m radio telescope
equipped with the EMIR receiver in the period 20$^{th}$-23$^{rd}$
December 2011. They were performed in dual polarisation and the E0/E2
band combination were tuned in order to cover the following
frequency bandwidths:
\begin{enumerate}
\item 87.4-91.4\,GHz and 83.8-87.8\,GHz (optimised for HCO+(1-0) and HCN(1-0))
\item 218.65-222.65\,GHz and 214.95-218.95\,GHz (optimised for
  $^{13}$CO(2-1) and C$^{18}$O(2-1)).
\end{enumerate}
Two backends were connected to two main receivers: FTS with a spectral
resolution of 0.19MHz and WILMA autocorrelator with a spectral
resolution of 2MHz. WILMA was non-operational in some observations,
and have mainly been used to check the quality of the observations. We
smoothed the FTS data to the optimised resolution of 2.1\,km\,s$^{-1}$
up to 34\,km\,s$^{-1}$.  Observations were performed in non-symmetric
wobbler mode to avoid sky subtraction from Andromeda's main disc. The
azimuthal radius of the wobbler was set to 240\,arcsec or set
manually, relying on the extinction map of the area
\citep{2000MNRAS.312L..29M}. As discussed by
\citet{1996RaSc...31.1053G}, large wobbler throws reduce the
sensitivity to point sources. However, this is restored to a large
extent with an active near-focus correction. It is also possible that
some emission is subtracted by the OFF signal. In principle, this
would occur at a velocity different from the signal velocity, even
though this cannot be excluded given the complex velocity pattern. One
can also argue, on the basis of the detections discussed in the next
section, that the emitted signal is weak, so the probability of
chopping a strong source is small. As shown in the figures discussed
below, there is no obvious sign of emission in the OFF measurements,
so this is in the worst case a second order effect. In addition, these
effects are expected to affect $^{13}$CO(2-1) and C$^{18}$O(2-1)
similarly, as well as HCO+(1-0) and HCN(1-0). They could nevertheless
affect the normalisation with respect to $^{12}$CO(2-1) and
$^{12}$CO(1-0) obtained independently.

We retrieved $^{12}$CO(2-1) and $^{12}$CO(1-0) data from previous IRAM
observations described in
\citet{2011A&A...536A..52M,2013A&A...549A..27M} (and in
prep.). $^{12}$CO(1-0) detection is only available for M31I.

\subsection{Infrared data}
The five positions observed at IRAM 30\,m were observed by Herschel/SPIRE instrument. The 1-kpc ring hosts cold temperature gas, while inside the ring the gas is hotter and not
detected at large wavelengths. We extracted different parameters relevant for our
study from the work of \citet{2014A&A...567A..71V} and \citet{2014ApJ...780..172D}. These parameters are summarised in Table \ref{tab:1b}.

\citet{2014ApJ...780..172D} adjusted a physical dust model on Herschel and Spitzer observations to estimate the mean intensities of starlight heating the dust.  These authors computed these mean intensities for the SPIRE 350/Herschel and MIPS 160/Spitzer resolutions of $24.^{\prime\prime}9$ and $39^{\prime\prime}$, respectively. \citet{2014A&A...567A..71V} account for two more photometric channels. The mean starlight recovered for both resolutions are provided in the second and third columns of Table \ref{tab:1b}. These differences have an amplitude comparable to the $30\%$ discrepancy between the inferred heating rate and the value predicted by a simple model of bulge starlight discussed by \citet{2014ApJ...780..172D}. This provides an idea of the real uncertainty of these estimates, which smooth the substructures, while the corresponding resolutions (24.\arcsec 9 and 39\arcsec) are comparable and larger than those of our millimeter observations (11\arcsec, 21\arcsec, and 28\arcsec; see Table \ref{table:2}). We then provide the cold dust temperatures computed according to the \citet{2007ApJ...657..810D} model as $T^{cold}_{dust}=T^{1/6}$. 
In the studied area, \citet{2014ApJ...780..172D} find that the
starlight heating intensity is compatible with the stars in
the bulge, as found by \citet{2012MNRAS.426..892G}. Relying on Herschel data, as well as GALEX, SDSS, WISE, and Spitzer, \citet{2014A&A...567A..71V} model the spectral energy distribution to extract physical quantities. They estimate cold dust temperatures, which are similar to those derived from \citet{2014ApJ...780..172D}, as shown in Table \ref{tab:1b}. The differences could be accounted for by the different datasets used as well as different modelling, \citet{2014A&A...567A..71V} directly fit the spectral energy distribution, while \citet{2014ApJ...780..172D} relies on a physical model. \citet{2014A&A...567A..71V} have also  provided a warm dust temperature (labelled "birth
cloud"  in their paper) larger than in the surrounding regions. This suggests that there are substructures linked to recent star formation, corresponding to (weak) star-forming activity
(3-15M$_\odot/$Myr). We can expect the kinetic temperatures to be equal at most to the dust temperature, depending on the molecular hydrogen density. In the following, we consider a kinetic temperature of 20\,K, which corresponds to the cold gas temperature. 

\section{Data reduction}
\label{sec:datared}
\begin{figure}
  \centering
  \includegraphics[width=7.8cm]{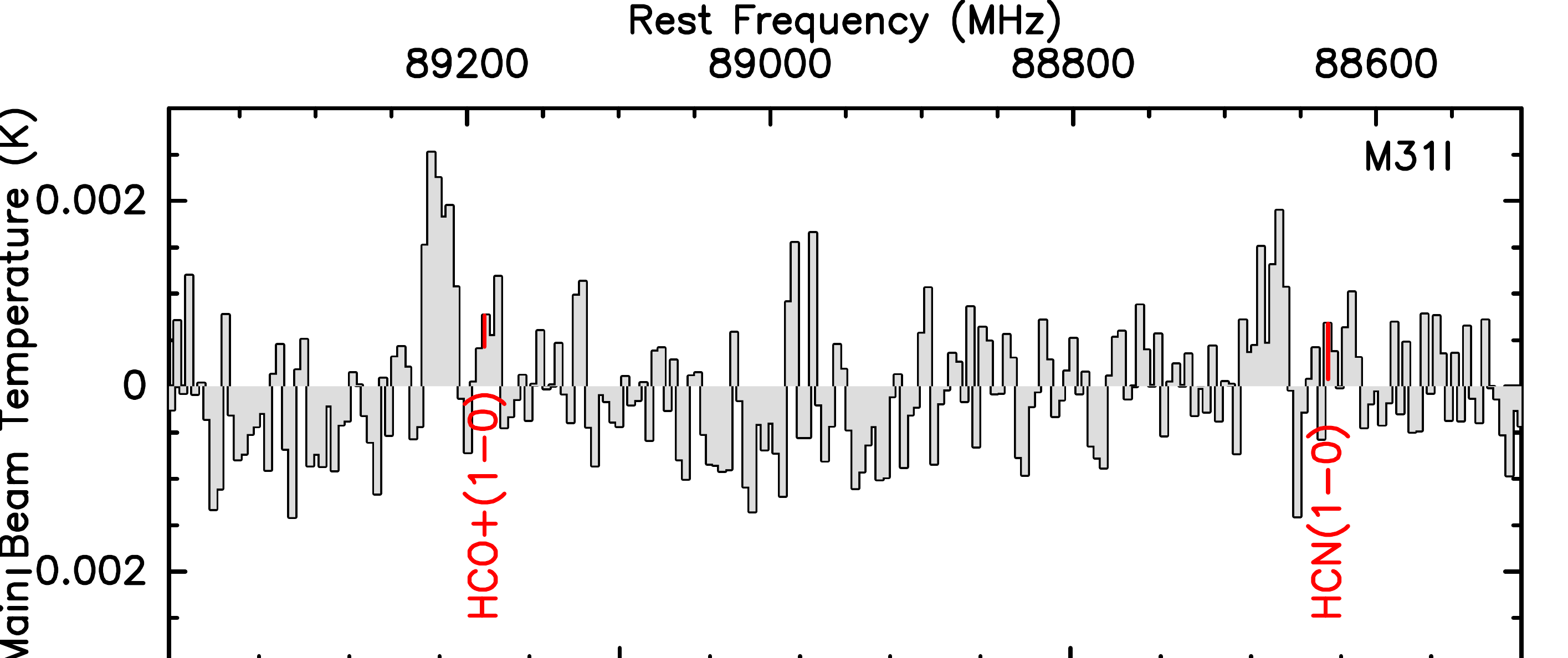}
  \includegraphics[width=7.8cm]{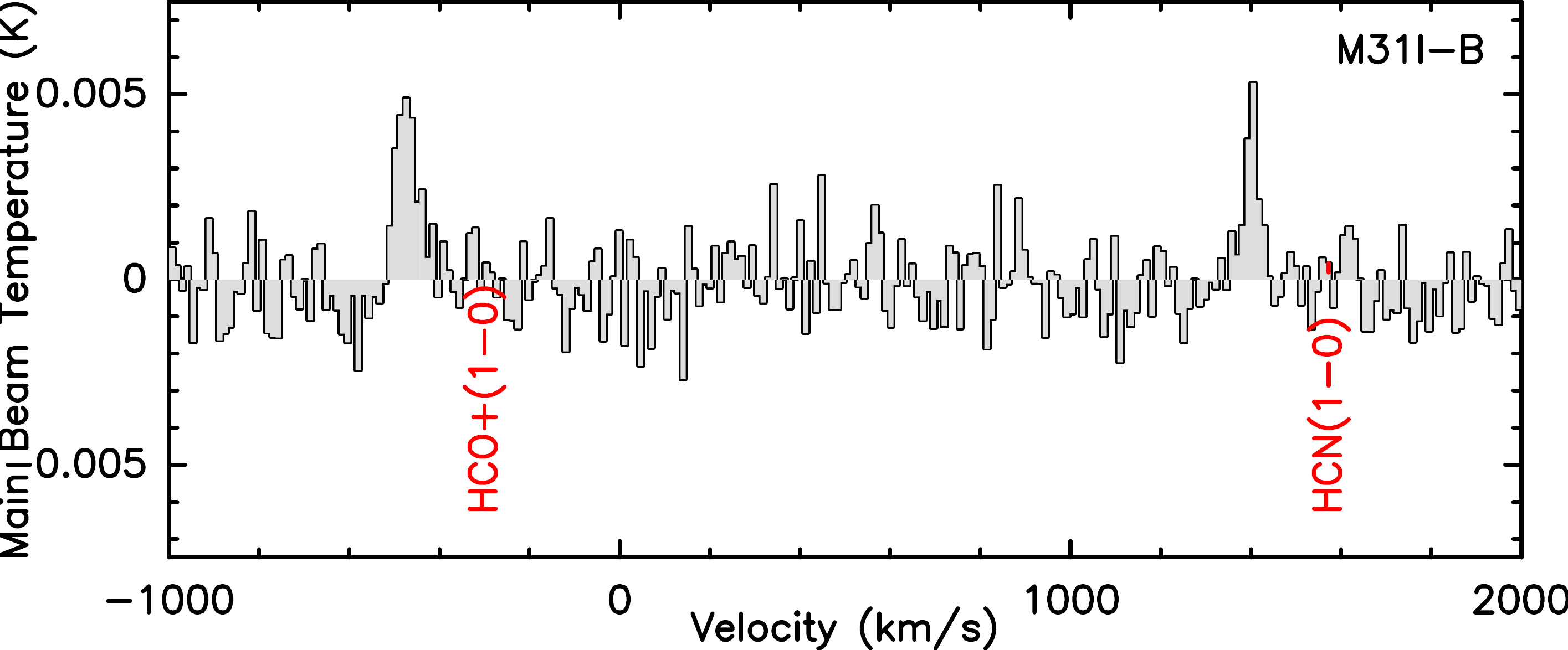}
  \caption{Signal detected in HCO+(1-0) and HCN(1-0) { at the position of M31I (top) and M31I-B (bottom)}. The main-beam temperatures are displayed as a function of the rest-frame frequency of the Andromeda galaxy (corresponding to -300\,km\.s$^{-1}$).}
  \label{fig:HCOpbis}%
\end{figure}
\begin{figure*}
    \centering
   \includegraphics[width=\textwidth]{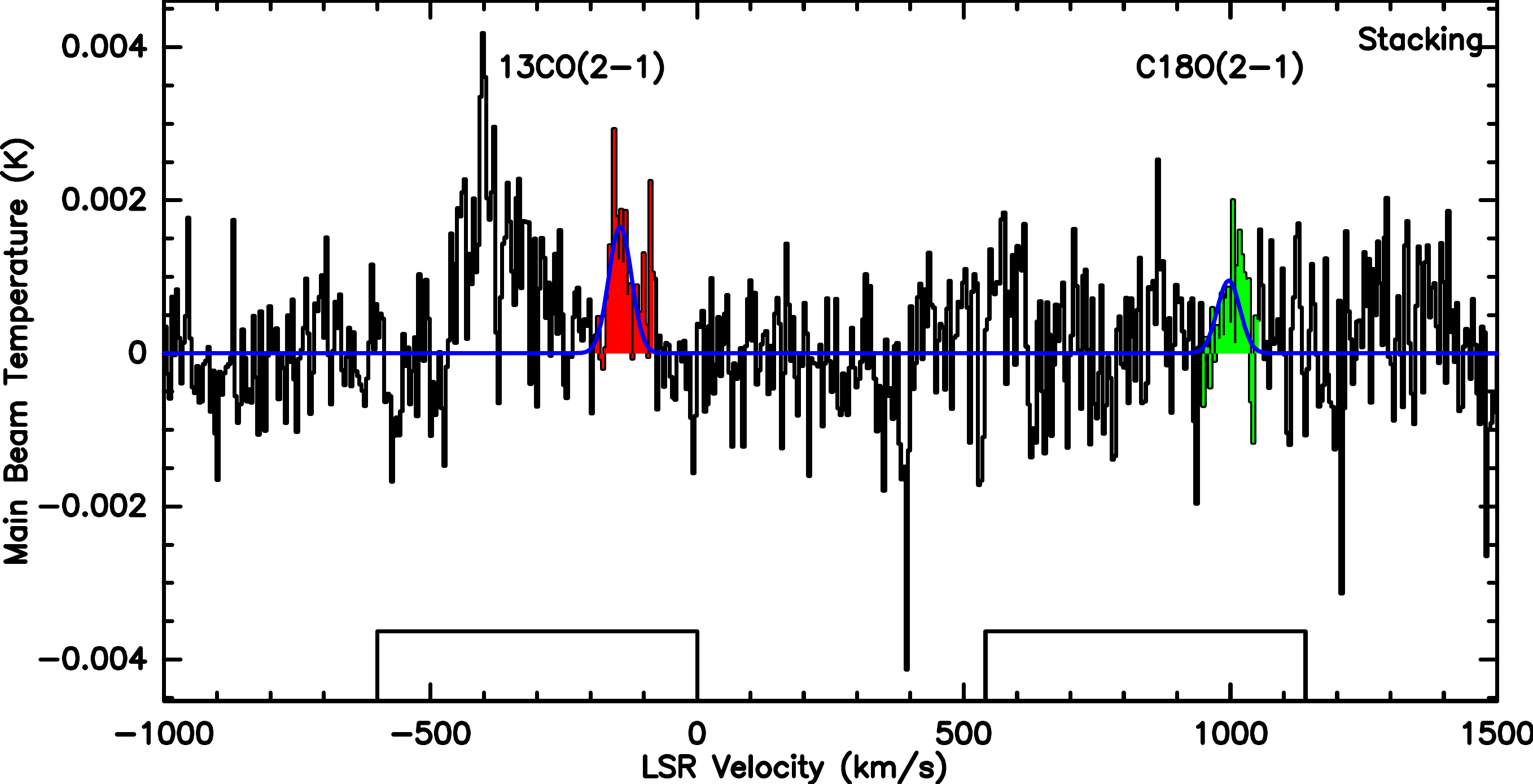}
   \caption{Stacking {centred at $-144.3$\,km\,s$^{-1}$ of each $^{13}$CO(2-1) velocity component with $v>-300$\,km\,s$^{-1}$. The $^{13}$CO component is displayed in red (at $-144$\,km\,s$^{-1}$)}. There is a component (displayed in green at $-1285$\,km\,s$^{-1}$) in C$^{18}$O which appears at the same velocity. The lines adjustment are displayed in blue. The velocity ranges corresponding to the two lines are displayed as rectangles. { Standard weights ($\propto T_{int}\,\delta v / {T_{sys}}^2$) for each observation have been used for the averaging.}}
 \label{fig:mystack13co}
\end{figure*}
We used the CLASS\footnote{Continuum and Line Analysis Single-dish Software,
  http://www.iram.fr/IRAMFR/GILDAS} package for the data reduction. We inspected individual spectra. Various interference lines are present far from the expected
astronomical signals. These interference lines affect the determination of the
baselines, and they have been carefully removed on each spectra. The M31I
position has suffered an interference next to the strongest peak of
the $^{13}$CO(1-0) line. This interference only affects the data acquired on
20$^{th}$ December 2011 (18/30). We thus set to zero the corresponding
channels on data taken with the FTS backend with the best spectral
resolution (0.26\,km\,s$^{-1}$).

The observations averaged in the two main bandwidths are presented
in the Figs. \ref{fig:stackpl2} and \ref{fig:stackpl}. We used default weighting with exposure time and system temperature. These
spectra reveal multicomponent signals corresponding to the Andromeda
velocity range (given its systemic velocity of
-300\,km\,s$^{-1}$). 

Characteristics of the main astronomical molecular lines
\citep{2005A&A...432..369S} present in the observed bandwidths are
provided in Table \ref{table:2} as follows: half power beam width (HPBW),
beam efficiency, rest-frame frequency, and velocity bin
corresponding to 14\,MHz. We also provide the following characteristics of these
molecules: energy level, Einstein coefficient A$_{uk}$,
collisional de-excitation rate C$_{ul}$, and critical density n$_c$
computed at 20\,K. If we consider a collisional temperature of 50\,K,
the critical densities of HCN and HCO+ increase by 56$\%$
and 21$\%$, respectively, but by 7.4$\%$ for CO. As indicated in Figs. \ref{fig:stackpl2} and \ref{fig:stackpl}, there are obvious detections of
HCO+(1-0), HCN(1-0), and $^{13}$CO(2-1) in the summed spectra and a
weak signal in C$^{18}$O(2-1). There is no { obvious} contamination from the
Milky Way, which could be possible as M31 is 20\,$\deg$ below the
Galactic plane. The spectra obtained in the lower outer bands are
presented in Fig. \ref{fig:stackpllo} but do not exhibit any
signal. They are used to derived upper limits.  For each position, the
detections are adjusted with a Gaussian function. Up to three
different CO lines are detected at different central velocity at each
position. As discussed in \citet{2011A&A...536A..52M}, this is due to
several intervening molecular components along the line of sight.

All of the measurements are summarised in Table \ref{table:3}. The
columns on the right side provide the CO line measurements.  We
provide, for each CO line, the central LSR velocity, the FWHM velocity
dispersion $\Delta$v, the peak (main-beam) temperature T$_{peak}$, and
the integral I$_{CO}$ of the line.
In the last two columns, the $^{12}$CO/$^{13}$CO(2-1) and $^{12}$CO/C$^{18}$O(2-1) line
ratios and lower limits are shown. We also provide (in the bottom of this table) our $^{12}$CO(1-0)
measurements described in \citet{2011A&A...536A..52M} for M31-I with a beam
resolution of 21\,arcsec, similar to the resolution achieved  for HCN(1-0) and HCO+(1-0), { and the N$_{H_2}$ column density derived as discussed below but directly computed from $^{12}$CO(1-0). These column density estimates differ by up to a factor of 3 with respect to $^{12}$CO(2-1)-based estimates, due to the difference in resolution and corresponding dilution effects.  }

The columns on the left side of Table \ref{table:3} list HCN(1-0) and HCO+(1-0)
measurements and 3$\sigma$ upper limits (for
$\Delta$v$=$32\,km\,s$^{-1}$) for some of the lines presented in Table
\ref{table:2}. The central table provides the remaining upper limits,
as well as column densities (and upper limits) of the species with
some detections, namely $^{13}$CO, C$^{18}$O, HCN, and HCO+,
assuming local thermal equilibrium (LTE) and optically thin
conditions. The column densities have been corrected for the beam filling factor, as further discussed in the next section. Given the velocity dispersion of the observed HCN lines ($>$25\,km\,s$^{-1}$), the HCN hyperfine line structure ($\sim$ 10\,km\,s$^{-1}$) will not be detected. { We also provide the  N$_{H_2}$ column densities of molecular hydrogen relying on X$_{CO}$ and CO(2-1)/(1-0) line ratio values.  We assume a Galactic ratio
X$_{CO}=N_{H_2}/I^\prime_{CO} = 2. \times
10^{20}$cm$^{-2}$(K\,km\,s$^{-1}$)$^{-1}$ following
\citet{1988A&A...207....1S} and \citet{2001ApJ...547..792D}, which is
compatible with the findings of \citet{2012ApJ...756...40S} for
Andromeda and \citet{2003A&A...403..561M} for local galaxies. As further discussed in Sect. \ref{ssect:prop}, we take $I^\prime_{CO}=I_{CO}/\eta_{bf}$. We assume a CO(2-1)/(1-0) line ratio of 0.8, as measured by
\citet{2011A&A...536A..52M} in a clump located within 10\,\arcsec of
M31I (and named M31G), to compute $N_{H_2}$ column densities.}
\begin{table*}
  \caption{Properties of the main lines in the two observed bands.}
  \label{table:2}      
  \centering                          
  \begin{tabular}{c c c c c c c c c}        
    \hline\hline                 
    Transition & HPBW & $\eta_{mb}$ & $\nu_0$ & $\delta$ v (1\,MHz) & E$_{up}$ & A$_{ul}$& C$_{ul}$&  n$_c$ (20\,K)\\    
    &  (arcsec)& & (GHz)& (km\,s$^{-1}$) & (K) & s$^{-1}$& (cm$^3$\,s$^{-1}$)& (cm$^{-3}$)  \\    
    \hline                        
    SiO\,$J=2 \rightarrow 1$                                    & 28. & 81 & 86.847  & 3.5 & 6.25  & 2.93\,10$^{-5}$ & 1\,1$^{-10}$    & 2.93\,10$^{5}$\\
    C$_2$H\,$N=1 \rightarrow 0$                                 & 28. & 81 & 87.329  & 3.4 & 4.19  & 0.26\,10$^{-6}$ & 5.19\,10$^{-12}$ & 5.01\,10$^{4}$\\ 
    ($J=3/2\rightarrow 1/2$,$F=1-1$)                            &     &    &         &      &       &                 &                  & \\ 
    HNCO(4$_{04}$ $\rightarrow$ 3$_{03}$)                       & 28. & 81 & 87.925  & 3.4 & 10.55 & 9.02\,10$^{-6}$ & 9.2\,10$^{-12}$  & 9.80\,10$^{5}$\\   
    HCN\,$J=1 \rightarrow 0$                                    & 28. & 81 & 88.632  & 3.4 & 4.25  & 2.41\,10$^{-5}$ & 1.92\,10$^{-11}$ & 1.26\,10$^6$\\
    HCO+\,$J=1 \rightarrow 0$                                   & 28. & 81 & 89.189  & 3.4 & 4.28  & 4.25\,10$^{-5}$ & 2.3\,10$^{-10}$  & 1.85\,10$^5$ \\
    HNC\,$J=1 \rightarrow 0$                                    & 27. & 81 & 90.664  & 3.3 & 4.35  & 2.69\,10$^{-5}$ & 8.95\,10$^{-11}$ & 3.01\,10$^5$ \\
    HC$_3$N\,$J=10 \rightarrow 9$                               & 27. & 81 & 90.979  & 3.3 & 24.01 & 5.81\,10$^{-5}$ & 1.1\,10$^{-10}$  & 5.28\,10$^{5}$\\
    \hline                                   
    SiO\,$J=5 \rightarrow 4$                                    & 11. & 60 & 217.105 & 1.4 & 31.26 & 5.20\,10$^{-4}$ & 1.1\,10$^{-10}$  & 4.73\,10$^{6}$\\
    H$_2$CO(3$_{03}$ $\rightarrow$ 2$_{02}$)                    & 11. & 60 & 218.222 & 1.4 & 21.0  & 2.82\,10$^{-4}$ & 1.1\,10$^{-10}$  & 2.56\,10$^{6}$\\
    H$_2$CO(3$_{22}$ $\rightarrow$ 2$_{21}$)                    & 11. & 60 & 218.476 & 1.4 & 68.1  & 1.57\,10$^{-4}$ & 5.3\,10$^{-11}$  & 2.96\,10$^{6}$ \\
    H$_2$CO(3$_{21}$ $\rightarrow$ 2$_{20}$)                    & 11. & 60 & 218.760 & 1.4 & 68.1  & 1.58\,10$^{-4}$ & 4.3\,10$^{-11}$  & 3.67\,10$^{6}$\\ 
    C$^{18}$O\,$J=2 \rightarrow 1$                              & 11. & 60 & 219.560 & 1.4 & 15.81 & 6.01\,10$^{-7}$ & 6.44\,10$^{-11}$ & 9.33\,10$^3$ \\
    $^{13}$CO\,$J=2 \rightarrow 1$                              & 11. & 60 & 220.399 & 1.4 & 15.87 & 6.04\,10$^{-7}$ & 6.44\,10$^{-11}$ & 9.38\,10$^3$ \\
    $^{12}$CO\,$J=2 \rightarrow 1$                              & 11. & 60 & 230.538 & 1.3 & 16.60 & 6.91\,10$^{-7}$ & 6.44\,10$^{-11}$ & 1.07\,10$^4$ \\
    $^{12}$CO\,$J=1 \rightarrow 0$                              & 21. & 71 & 115.271 & 2.6 & 5.53  & 7.20\,10$^{-8}$ & 3.25\,10$^{-11}$ & 2.22\,10$^3$ \\
    \hline
  \end{tabular}
\tablefoot{ The
    beam sizes (HPBW) and the main beam efficiencies ($\eta_{mb}$) are
    presented with the rest frequencies $\nu_0$ of the
    brightest lines in each band and the velocity bin corresponding to
    1\,MHz. The energy levels E$_{up}$ and Einstein coefficients
    A$_{ul}$ are based on the Leiden Atomic and Molecular Database
    \citep{2005A&A...432..369S}; the collisional
    de-excitation rate coefficients C$_{ul}$ are computed at a kinetic temperature of 20\,K. The
    critical density is computed as n$_c=A_{ul}/C_{ul}$.
}
\end{table*}


\begin{table*}
  \caption{Results. }  
  \label{table:3}      
  \vspace{-0.5cm}
  \rotatebox{-90}{
    \noindent\begin{minipage}[l]{\textheight}
    \begin{flushleft}  
      \begin{tabular}{c|cc|cc|cc|cccccc}       
        \hline\hline                 
        Pos. & \multicolumn{2}{c}{V$_0$ (km/s)} & \multicolumn{2}{c}{$\Delta v$ (km/s)} & \multicolumn{2}{c}{T$_{Peak}$ (mK)} & \multicolumn{6}{c}{I$_{line}$ (K\,km\,s$^{-1}$)}  \\    
      \hline   
        & HCO+& HCN& HCO+& HCN& HCO+& HCN& HCO+& HCN& HNC & HC3N &HOC+ & HNCO \\ 
        & (1-0)& (1-0)& (1-0)&(1-0)& (1-0)& (1-0)& (1-0)& (1-0)& (1-0) & (10-9) & (1-0) & (4-3) \\ \hline
        I(2)    & { -402$\pm$4} & { -425$\pm$10} & { 54$\pm$8} & { 57$\pm$18} & { 3.0} & { 1.7} & { 0.17$\pm$0.03} & { 0.11$\pm$0.03} & { $<$0.12}& { $<$0.12}& { $<$0.10}& { $<$0.10}\\
        I-B(1)  &  -473$\pm$4& { -471$\pm$3}&  { 54$\pm$8 } & 32$\pm$8 & { 5.8} & { 5.8} &{  0.34$\pm$0.04}  & { 0.20$\pm$0.04} & { $<$0.07} &{ $<$0.07} &{ $<$0.07} &{ $<$0.08}\\
        2d-a &  &  &  &  & &  & { $<$0.11} & { $<$0.13}  & { $<$0.10} & { $<$0.12} & { $<$0.13} & { $<$0.13} \\
        2d-b &  &  &  &  & &  & { $<$0.13} & { $<$0.13}  & { $<$0.09} & { $<$0.10} & { $<$0.11} & { $<$0.07}\\
        2c-a &  &  &  &  & &  & { $<$0.20} & { $<$0.18}  & { $<$0.21} & { $<$0.22} & { $<$0.20} & { $<$0.18}\\
        \hline                                   
       \end{tabular}  
    \end{flushleft}  
    \end{minipage}
  }
  \hspace{0.1cm}
  \rotatebox{-90}{
    \noindent\begin{minipage}[l]{\textheight}
    \begin{flushleft}  
      \begin{tabular}{c|cccccc|cccc|c}       
        \hline\hline                 
        Pos. &  \multicolumn{6}{c}{I$_{Line}$} & \multicolumn{4}{c|}{Molecular column densities} & \multicolumn{1}{c}{N$_{H_2}$}  \\    
        & \multicolumn{6}{c}{(K\,km\,s$^{-1}$)} & \multicolumn{4}{c|}{(hypothesis: LTE, optically thin, $\eta_{bf}$,  20K) (cm$^{-2}$)} & \multicolumn{1}{c}{(cm$^{-2}$)} \\ \hline   
     & SiO& CCH & SiO&H$_2$CO& H$_2$CO & H$_2$CO&                                         \multicolumn{1}{c}{N$_{^{13}CO}$ } &  \multicolumn{1}{c}{N$_{C^{18}O}$  }& \multicolumn{1}{c}{N$_{HCN}$ }& \multicolumn{1}{c|}{N$_{HCO+}$ } &\\ 
    & (2-1)& (1-0) & (5-4)&(3$_{03}$-2$_{02}$) & (3$_{22}$-2$_{21}$)    & (3$_{21}$-2$_{20}$)  &  ($10^{16}$) & ($10^{16}$)  &  ($10^{13}$) &   ($10^{13}$) & (10$^{22}$)  \\ \hline                                                              
I (1)   & { $<$0.06} & { $<$0.09} & { $<$0.12} & $<$0.13       & { $<$0.16}  & { $<$0.09} &  { 0.29$\pm$0.04} & $<$0.10             & { $<$0.77}          & { $<$0.44}         & 3.0$\pm$0.2  \\         
I (2)   & { $<$0.06} & { $<$0.09} & { $<$0.12} & $<$0.13       & { $<$0.16}  & { $<$0.09} &  { 1.64$\pm$0.32} & { $<$1.16}	       & { 2.73$\pm$0.75}    & { 2.41$\pm$0.43}   & 24.3$\pm$1.3       \\
I (3)   & { $<$0.06} & { $<$0.09} & { $<$0.12} & $<$0.13       & { $<$0.16}  & { $<$0.09} &  { 3.11$\pm$0.60} & { 2.10$\pm$ 0.60}& { $<$4.24}          & { $<$2.42}         & 28.8 $\pm$4.8\\
I-B (1) & { $<$0.08} & { $<$0.09} & { $<$0.10} & { $<$0.09} & { $<$0.10}  & { $<$0.09} &  { 1.02$\pm$0.09} & { $<$0.22}       & { 2.04$\pm$0.41}    & { 1.98$\pm$0.23}   & 14.9$\pm$0.2 \\   
I-B (2) & { $<$0.08} & { $<$0.09} & { $<$0.10} & { $<$0.09} & { $<$0.10}  & { $<$0.09} &  { 4.40$\pm$0.78} & { $<$7.99}       & { $<$8.85}         & { $<$5.06}         & 26.3$\pm$2.3 \\ 
2d-a(1) & $<$0.08& { $<$0.09}   & { $<$0.08} & { $<$0.08} & { $<$0.10} & { $<$0.08}  &  { 1.16$\pm$0.23} & { $<$1.00}	           & { $<$4.72}          & { $<$2.29}         & 8.2 $\pm$0.6  \\
2d-a(2) & $<$0.08& { $<$0.09}   & { $<$0.08} & { $<$0.08} & { $<$0.10} & { $<$0.08}  &  { 1.40$\pm$0.29} & { $<$1.25}            & { $<$4.49}          & { $<$2.17}         & 14.4$\pm$1.1\\
2d-b(1) & $<$0.08& { $<$0.061}  & { $<$0.11} & { $<$0.09} & { $<$0.09} & { $<$0.09}  &  { 0.44$\pm$0.14} & { $<$0.26}	           & { $<$2.68}          & { $<$1.53}         & 11.6$\pm$0.6     \\
2d-b(2) & $<$0.08& { $<$0.061}  & { $<$0.11} & { $<$0.09} & { $<$0.09} & { $<$0.09}  &  { 2.28$\pm$0.70} & { $<$1.58}	           & { $<$10.72}         & { $<$6.13}         & 17.7$\pm$2.5     \\
2d-b(3) & $<$0.08& { $<$0.061}  & { $<$0.11} & { $<$0.09} & { $<$0.09} & { $<$0.09}  &  { 2.03$\pm$0.60} & { $<$1.83}            & { $<$12.41}	       & { $<$7.09}         & 7.8 $\pm$1.9     \\   
2c-a(1) & $<$0.08&$<$0.24          &$<$0.19        & $<$0.05       & $<$0.14       & $<$0.08        &  { 5.57$\pm$1.39} & { $<$3.79}	           & { $<$11.89}         & { $<$6.80}         & 22.4$\pm$1.8\\
2c-a(2) & $<$0.08&$<$0.24          &$<$0.19        & $<$0.05       & $<$0.14       & $<$0.08        &  { $<$1.28}       & {$ <$1.57}	           & { $<$13.44}	       & { $<$7.68}         & 4.0 $\pm$1.0 \\   
2c-a(3) & $<$0.08&$<$0.24          &$<$0.19        & $<$0.05       & $<$0.14       & $<$0.08        &  $<$0.71             & { $<$0.94}            & { $<$7.40}	       & { $<$4.23}         & 4.9 $\pm$0.6\\ 
\hline                                   
	\end{tabular}
    \end{flushleft}  
    \end{minipage}  
    }
  \hspace{0.1cm}
  \rotatebox{-90}{
    \noindent\begin{minipage}[l]{\textheight}
    \begin{flushleft}  
      \begin{tabular}{c|ccc|ccc|ccc|ccc|cc}       
\hline\hline                 
Pos. &  \multicolumn{3}{c}{V$_0$ (km/s)} & \multicolumn{3}{c}{$\Delta v$ (km/s)} & \multicolumn{3}{c}{T$_{Peak}$ (mK)} & \multicolumn{3}{c}{I$_{Line}$ (K\,km\,s$^{-1}$)}& I$_{^{12}CO}$/I$_{^{13}CO}$ &I$_{^{12}CO}$/I$_{C^{18}O}$   \\    
\hline   
 (2-1)  & $^{12}$CO      & $^{13}$CO       & C$^{18}$O& $^{12}$CO   & $^{13}$CO          & C$^{18}$O& $^{12}$CO & $^{13}$CO & C$^{18}$O&$^{12}$CO    & $^{13}$CO          & C$^{18}$O   &     & \\ \hline
I(1)    & -144.2$\pm$0.2 & { -144.2$*$} &          &7.5$\pm$0.5  & { 6.3$\pm$1.5}  &          & 422       & 20.0      &          &3.4$\pm$0.2  & { 0.16$\pm$0.02}& $<$0.056             &{ 22.1$\pm$2.9}& { $>$61}\\
I(2)    & -402.$\pm$3    &      -402$*$    &          & 62$\pm$12   &       62$*$             &          & 146       & 4.8       &          &9.6$\pm$0.5  & { 0.31$\pm$ 0.06} & $<$0.22          &   { 31.0$\pm$6.2}            & { $>$44.}\\
I(3)    & -291.$\pm$9    &  -291$*$        & {-291$*$}& 69$\pm$18   & 69$*$              & 69$*$    &  77       &  4.2      &   2.9    & 6$\pm$1     & { 0.31$\pm$0.06} &{ 0.21$\pm$0.06}&    { 19.3$\pm$4.9}        & { 28.6$\pm$9.5} \\
I-B(1)  &-463.0$\pm$0.7  & -463$*$         &          & 38$\pm$2    & 38$*$              &          & 356       & { 11.5}&          &14.3$\pm$0.6 & { 0.47$\pm$0.04}& $<$0.10     &  { 30.4$\pm$2.9}   & $>$143.  \\
I-B(2)  &-397.$\pm$7     & -397$*$         &          & 66.$\pm$14  & 66$*$              &          &  49.2     & { 4.2} &          &3.5$\pm$0.7  & { 0.28$\pm$0.05}&$<$0.51      &  { 12.5$\pm$3.4}   & $>$6.9 \\
2d-a(1) &-523.0$\pm$0.9  & -522$\pm$3      &          & 21.$\pm$3  & { 25$\pm$7}     &          &  99.8     & { 5.7} &          &2.2$\pm$0.2  & 0.15$\pm$0.03      & $<$0.13      &  15.$\pm$3    & $>$16.9 \\
2d-a(2) &-234.$\pm$1     & { -225$\pm$4}&          & 37.$\pm$3   &{ 40$\pm$11}     &          & 105       & { 5.2} &          &4.1$\pm$0.3  & 0.19$\pm$0.04      & $<$0.17      & 22.$\pm$5     & $>$24.1\\
2d-b(1) & -211.5$\pm$0.7 & -211$*$        &          & 30$\pm$2    & 30$*$             &          & 176       &  3.2      &          &5.5$\pm$0.3  & 0.10$\pm$0.03      & $<$0.06      & 55.0$\pm$17   & $>$91.7\\
2d-b(2) & -269.$\pm$3    & -269$*$        &          & 46$\pm$6    & 46$*$             &          &  44       &  2.7      &          &2.1$\pm$0.3  & 0.13$\pm$0.04      & $<$0.09      & 16.$\pm$5     & $>$23.3\\
2d-b(3) & -526.$\pm$2    & -526$*$        &          & 20$\pm$8    & 20$*$             &          &  38       &  4.8      &          &0.8$\pm$0.2  & 0.10$\pm$0.03      & $<$0.09     &  8.$\pm$3     & $>$8.9 \\
2c-a(1) & -243$\pm$3     &{ -241$\pm$13}&          & 57$\pm$4    & { 97$\pm$28}    &          &  61       &  4.2      &          &3.7$\pm$0.3  & { 0.44$\pm$0.11}& $<$0.30     &{ 8.4$\pm$2.2} & $>$12.3\\
2c-a(2) & -223.9$\pm$0.6 &                 &          & 10$\pm$2    &                    &          &  54       &           &          &0.58$\pm$0.15& $<$0.09            & $<$0.11     & { $>$5.8}        & { $>$4.8}\\
2c-a(3) & -204.7$\pm$0.4 &                 &          & 12.6$\pm$0.9&                    &          &  98       &           &          &1.31$\pm$0.15& $<$0.09            & $<$0.12      & $>$14.6       & { $>$9.4}\\ \hline                                   
 (1-0)    & $^{12}$CO      && & $^{12}$CO    &&  & $^{12}$CO &      &          & $^{12}$CO    & &  & \multicolumn{2}{l}{N$_{H_2}$ (10$^{22}$cm$^{-2}$)} \\ \hline
I(1)      & -144.2$\pm$0.2 && &  9.3$\pm$0.4 &&  & 322       &      &          & 3.2$\pm$0.1  & &  & 2.24$\pm$0.07& \\
I(2)      & -404.2$\pm$0.9 && &  49$\pm$3    &&  & 140       &      &          & 7.3$\pm$0.3  & &  & 14.8$\pm$0.6 & \\
I(3)      & -278$\pm$3     && &  77$\pm$14   &&  & 54        &      &          & 2.7$\pm$0.2  & &  & 10.4$\pm$0.8 & \\ \hline                                   
     \end{tabular}     
    \end{flushleft}  
    \end{minipage}
  }
  \vspace{-1cm} 
  \tablefoot{{\em Right:} CO(2-1) line fit results:
    I$_{CO}$ upper limits are provided at $3\sigma$ { for the
      $^{13}$CO (or $^{12}$CO) velocity dispersion}; {\em Middle:}
    $3\sigma$ upper limits on molecular lines observed in the LO
    (Lower Outer) band (with $\sigma=32$km/s); column densities are
    corrected for beam-filling factors; { hydrogen} column density {
      are derived from $^{12}$CO measurements} as discussed in
    Sect. \ref{sec:datared} with a Galactic X$_{CO}$ factor. {\em
      Left:} HCO+(1-0) and HCN(1-0) Line fit results; HNC(1-0),
    HNCO(4-3), HOC+(1-0) and HC3N(10-9) upper limits provided at
    $3\Delta v$ (assuming { $\sigma=54$km/s}).}
\end{table*}

\section{Results}
\label{sec:resu}
In this section, we describe our results. In Sect. \ref{ssec:desc}, we describe the detections of five different molecular emission lines in the two observed positions along the minor axis. In Sect \ref{ssec:obse}, we compute the line ratios relying on previous detections and compare them with previous results achieved in other galaxies, in particular, Galactic giant clouds and in M31's disc. In Sect. \ref{ssect:prop}, we discuss the basic properties we can derive from the detections, including excitation temperatures, optical depths, beam filling factors, and column densities, and the uncertainties and limitations of this type of analysis. In Sect. \ref{ssec:lte}, we discuss the parameters estimated under LTE conditions  and compare them with Galactic values. In Sect. \ref{ssec:radex}, we run RADEX \citep{2007A&A...468..627V} simulations to relax the LTE conditions and test the validity of our previous assumptions. While CO emission lines correspond to gas in thermal equilibrium, gas traced by HCN and HCO+ emission lines is under subthermal excitation conditions.

\subsection{Description of the detections}
\label{ssec:desc}
The CO lines detected in the north-west (south-east) region are
shown in Fig. \ref{fig:CObis} (Fig. \ref{fig:12CO(2-1)}). The
HCN(1-0) and HCO+(1-0) lines detected on the north-west side are shown
in Fig. \ref{fig:HCOpbis}.
\begin{table*}
  \caption{ Line measurements for the stacking detection shown in Fig. \ref{fig:mystack13co}, based on the red (v$>-300$\,km/s) velocities of both regions (excluding the M31-I (3) detection and M31-2c-a(1) tentative $^{13}$CO(2-1) detection).  }  
  \label{table:stack}      
  \rotatebox{0}{
    \noindent\begin{minipage}[l]{\textheight}
    \begin{flushleft}  
      \begin{tabular}{c|c|cc|cc|cc|cc}       
        \hline\hline                 
        Pos. & V$_{Stack}$&\multicolumn{2}{c|}{$\langle$ V$_0$ $\rangle$ (km\,s$^{-1}$)} & \multicolumn{2}{c|}{$\Delta v$ (km\,s$^{-1}$)} & \multicolumn{2}{c|}{T$_{Peak}$ (mK)} & \multicolumn{2}{c}{I$_{line}$ (K\,km\,s$^{-1}$)} \\    
      \hline   
        & & $^{13}$CO & C$^{18}$O & $^{13}$CO & C$^{18}$O & $^{13}$CO & C$^{18}$O & $^{13}$CO & C$^{18}$O\\ 
        & & (2-1)& (2-1)& (2-1)&(2-1)& (2-1)& (2-1)& (2-1)& (2-1)\\ \hline
        {  Stack (red)}    & -144.2 & -212 & -212         & 50* & 50*     & 1.65   &0.95 &  0.088$\pm$0.015      & 0.051$\pm$0.015\\       
            &  &  &       & {\em 31} &    &   & &{\em 0.15$\pm$0.03}     & \\       
        {  SE stack (red)}    & -226. & -218. & -218.         & 39* & 39*     & 2.26   &1.28 &  0.094$\pm$0.015      & 0.053$\pm$0.015\\       
             &  &  &       & {\em 39} &    &   & & {\em 0.14$\pm$0.04}    & \\       
       \hline                                   
        \hline\hline                 
        Pos. & V$_{Stack}$&\multicolumn{2}{c|}{$\langle$ V$_0$ $\rangle$ (km\,s$^{-1}$)} & \multicolumn{2}{c|}{$\Delta v$ (km\,s$^{-1}$)} & \multicolumn{2}{c|}{T$_{Peak}$ (mK)} & \multicolumn{2}{c}{I$_{line}$ (K\,km\,s$^{-1}$)} \\    
      \hline   
        & & HCO+ & HCN & HCO+ & HCN & HCO+ & HCN & HCO+ & HCN\\ 
        & & (1-0)& (1-0)& (1-0)&(1-0)& (1-0)& (1-0)& (1-0)& (1-0)\\ \hline
        { SE Stack (blue)}    & -523 & $-$         &  { -524.5} & 89$*$     & { 89$\pm$25}   &$-$ &  { 1.17} & { $<0.075$ }     &  { 0.11$\pm$0.03}\\              
        { SE Stack (red) }  & -234 & { -231} &  { -231} &  { 78$\pm$29} &  { 78$*$} & 0.74 & 0.85 & { 0.061$\pm$0.026} &  { 0.070$\pm$0.024}\\        
        \hline                                   
      \end{tabular}
    \end{flushleft}  
    \end{minipage}
  }
      \tablefoot{We present the central velocities, velocity dispersions, peak temperatures, and integrated lines. $\langle V_0 \rangle$ is the averaged velocity of the spectra used for each stacking. Similar stacking based on the $^{12}$CO(2-1) velocities is performed for HCN and HCO+ spectra of the south-east (SE) positions. As shown in Fig. \ref{fig:appstack}, redshifted (blueshifted) velocities are considered separately. }
\end{table*}
\begin{table*}
  \caption{Observed line ratios.}             
  \label{table:4}      
  \centering                          
  \begin{tabular}{c c c c c c c c}        
    \hline\hline                 
    Position& V$_0$ (km/s) & HCO+/HCN & HNC/HCN & HOC+/HCO+ & $^{12}$CO/HCO+ & $^{12}$CO/HCN & $^{12}$CO/$^{13}$CO\\    
& & (1-0) & (1-0)& (1-0) & (1-0) & (1-0) & (2-1) \\    \hline                        
    { I(2)}   & $\sim -402$ & 1.5$\pm$0.5 & $<$1.09 & $<$0.59 & 43.$\pm$8 & 62.$\pm$18 & 24.$\pm$5\\
    { I-B(1)} & $\sim -470$ & 1.7$\pm$0.4 & $<$0.35 & $<$0.21 & {\em 53.$\pm$ 7} & {\em 89.$\pm$ 18}  & 30.$\pm$3
    \\    
     \hline       
    \\ \hline\hline                            
     Position& V$_0$ (km/s)  &  HCO+/HCN & $^{12}$CO/C$^{18}$O & $^{13}$CO /C$^{18}$O& & & $^{12}$CO/$^{13}$CO \\
     & & (1-0) & (2-1) & (2-1) &  & & (2-1) \\
    \hline                        
    { I(3)}        &   $\sim -291$ & &  28.6$\pm$9.5 & 1.48$\pm$0.51   &  & & 19.4$\pm$5.0 \\  \hline        
   {  Stack (red)} &           &    &   {\em 30$\pm$12}         & 1.73$\pm$0.59 & & & {\em 17.4$\pm$4.1} \\  
    { SE stack (red)} &             &   0.87$\pm$0.49  & {\em 33$\pm$14}       & 1.77$\pm$0.58 &  & &  {\em{18.7$\pm$5.1}}\\  
    \hline                                   
  \end{tabular}
  \tablefoot{These ratio are computed with the velocity-integrated
intensities measured within similar beams for the same component. Numbers in italics correspond to tentative estimates: $^{12}$CO(1-0) are derived from $^{12}$CO(2-1) with a 0.8 line ratio, stacked $^{12}$CO(2-1) and $^{13}$CO(2-1) are computed as the sum of the detected lines; and $^{13}$CO(2-1) and C$^{18}$O(2-1) are measured on stacked spectra.}
\end{table*}
As shown in Table \ref{table:3}, we detected up to five different molecular lines (two
to four per position and per velocity component). For
some positions (M31I, M31I-B, M31-2d-a), there are clearly several
well-identified peaks with different velocities, while for other
positions the signal is very wide and best fitted with several
components. The most striking cases are M31-2c-a, and the low-velocity
component of M31-2d-b. The 4$\sigma$ $^{13}$CO(2-1) signal measured for M31-2c-a is affected by a small integration time and the detection integrates several $^{12}$CO$(2-1)$ components. Hence, this signal is not used for subsequent stacking. Apart from the $^{12}$CO and some $^{13}$CO lines,
the detections are weak with peak temperatures at the mK level. We detected HCO+(1-0) (HCN(1-0))  at the 6$\sigma$ and 9$\sigma$ (4$\sigma$ and 5$\sigma$) levels, while most CO(2-1) detections are above 4$\sigma$ with a strong one at 11$\sigma$; but three weak $^{13}$CO detections in the south-east region are kept at the 3.3$\sigma$ level.
  Knowing the low intensities of these lines, we can infer that it is extremely challenging to perform multi-transition observations of these molecules.  

 While $^{12}$CO and $^{13}$CO are detected on
both regions, C$^{18}$O, HCN(1-0) and HCO+(1-0) are only detected on
the north-west side. If the dust-to-gas mass ratio is similar on both
sides, according to \citet{2014A&A...567A..71V} we expect a gas mass a
factor of 3 larger on this side (see Table \ref{tab:1b}). Similarly,
when one averages the observed column densities of molecular hydrogen
provided in Table \ref{table:3} along the two lines of sight, the
north-west positions have a column density a factor of 2 higher than
the south-east positions. Also, the radiation field (U) is larger on the south-east side than in the north-west side: this may be related to larger CO column densities and to the fact that dense gas (C$^{18}$O, HCN, HCO+) is not { directly} detected in the south-east positions.

The C$^{18}$O detection corresponds to the component at the systemic velocity, while
the HCN and HCO+ detections are in the blueshifted components.   However, as shown in Fig. \ref{fig:mystack13co}, stacking of the $v>-300$\,km\,s$^{-1}$ $^{13}$CO detections  (excluding the M31-I(3) detection) also reveals a  redshifted component in C$^{18}$O with an amplitude comparable with $^{13}$CO. The characteristics of this stacking detection is presented in Table \ref{table:stack}. We thus have at least two different C$^{18}$O components in two different velocity ranges.

 Figure \ref{fig:appstack} denotes the stacking of v$<-300$\,km/s and v$>-300$\,km/s of HCN(1-0) and HCO+(1-0) spectra. The detected signals are summarised in Table \ref{table:stack}. The south-east spectra have been shifted to the $^{12}$CO velocities and averaged with standard weights ($\propto T_{int}\,\delta v / {T_{sys}}^2$). For HCN, there is 3.7$\sigma$ detection in the blue stacking while there is a possible 2.6$\sigma$ detection in the red stacking. In contrast to the direct observations, the stacked HCN signal is stronger than the HCO+ signal, with a HCO+/HCN line ratio of $0.87\pm 0.49$ in the red stacking and $<0.68$ in the blue stacking. 

The difficulty of the interpretation of these dense gas detections is
due to a complicated configuration. We integrate the signal over
a relatively large beam (42-106\,pc in projection), and the
velocity dispersions are significantly
larger than those ($\sim$2\,km/s) detected in M31's main disc
(Schruba et al., in prep) with $\Delta v > 30$\,km\,s$^{-1}$ for most of our detections. This is because of the strong velocity gradient present in this region and to the superposition of several clouds.   This also points towards inner elongated orbits observed with a large radial velocity range (in contrast to the outer orbits observed in the main disc where small velocity dispersions ($\sim$\,2\,km\,s$^{-1}$) are expected and observed). M31-I(1) has a smaller velocity dispersion, which is still four times larger than disc clouds. The other components of M31-I are very large and could correspond to an underlying (possibly continuous) component with a 150\,km\,s$^{-1}$ width with non-circular orbits.  This could be a relic of a spiral arm or dissolved bar.

\begin{figure*}
    \centering
   \includegraphics[width=\textwidth]{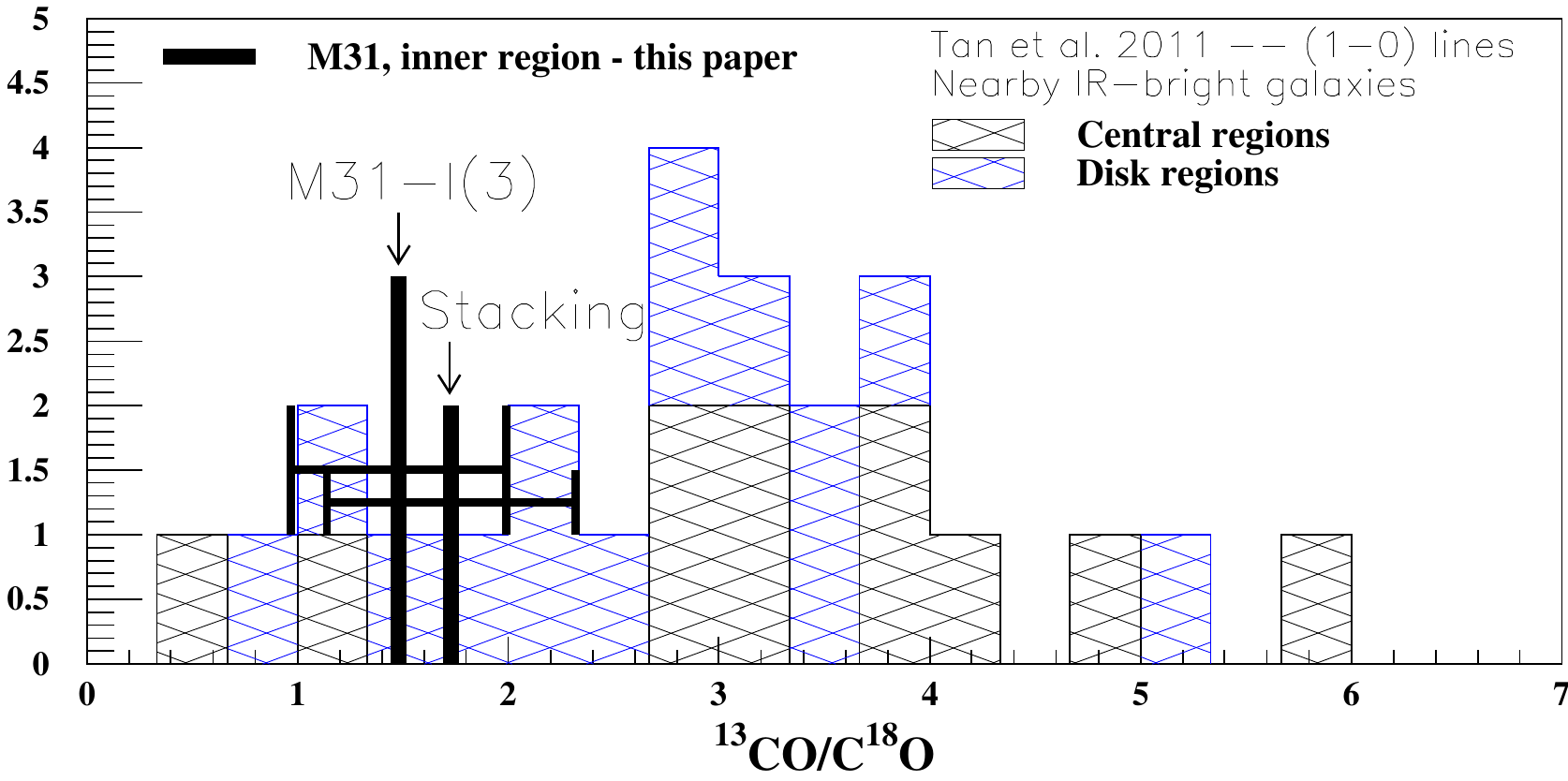}
   \caption{Comparison of our $^{13}$CO/C$^{18}$O(2-1) line ratio measurements (M31-I(3) and stacking) corresponding to different clouds (i.e. with different velocities) with
     previous measurements in nearby galaxies from
     \citet{2011RAA....11..787T}. The hatched histogram displayed the
     original $^{13}$CO/C$^{18}$O(1-0) line ratio. }
 \label{fig:co1318}
\end{figure*}
\begin{figure*}
    \centering
   \includegraphics[width=\textwidth]{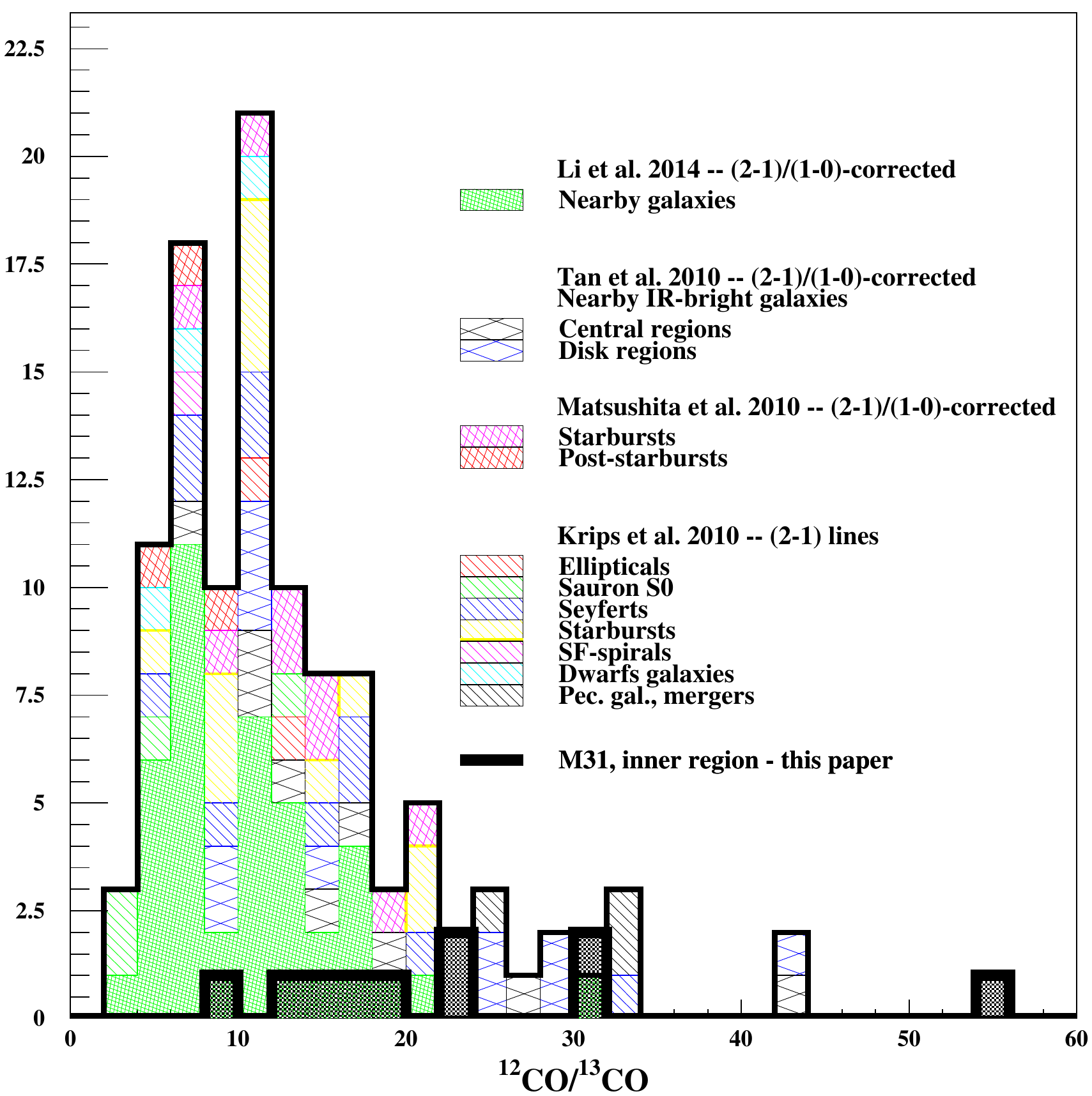}
   \caption{Comparison of our $^{12}$CO/$^{13}$CO line ratio with
     previous measurements in nearby galaxies. The hatched histogram
     displayed the $^{12}$CO/$^{13}$CO line ratios with available
     morphological types. The \citet{2011ApJ...728L..10L},
     \citet{2010PASJ...62..409M} and \citet{2011RAA....11..787T}
     samples measured the (1-0) transition and have been corrected by
     a factor 1.33 as discussed in the text. The
     \citet{2010MNRAS.407.2261K} sample and our measurements were
     directly measured the (2-1) transition. The black full line histogram
     gathers all the four samples corrected to (2-1). The black thick line
     histogram in the bottom corresponds to the measurements performed
     in this paper. The average $^{12}$CO/$^{13}$CO detected in this paper tends to be a bit larger than the main distribution from other areas and other galaxies (but well below the standard isotopic ratio of $53$ \citep{1994ARA&A..32..191W}). The largest value above $50$ might corresponds to an optically thin cloud, but it is affected by a large error bar (30$\%$, see Table \ref{table:3}).}
 \label{fig:corat}
\end{figure*}
\begin{figure*}
    \centering
   \includegraphics[width=\textwidth]{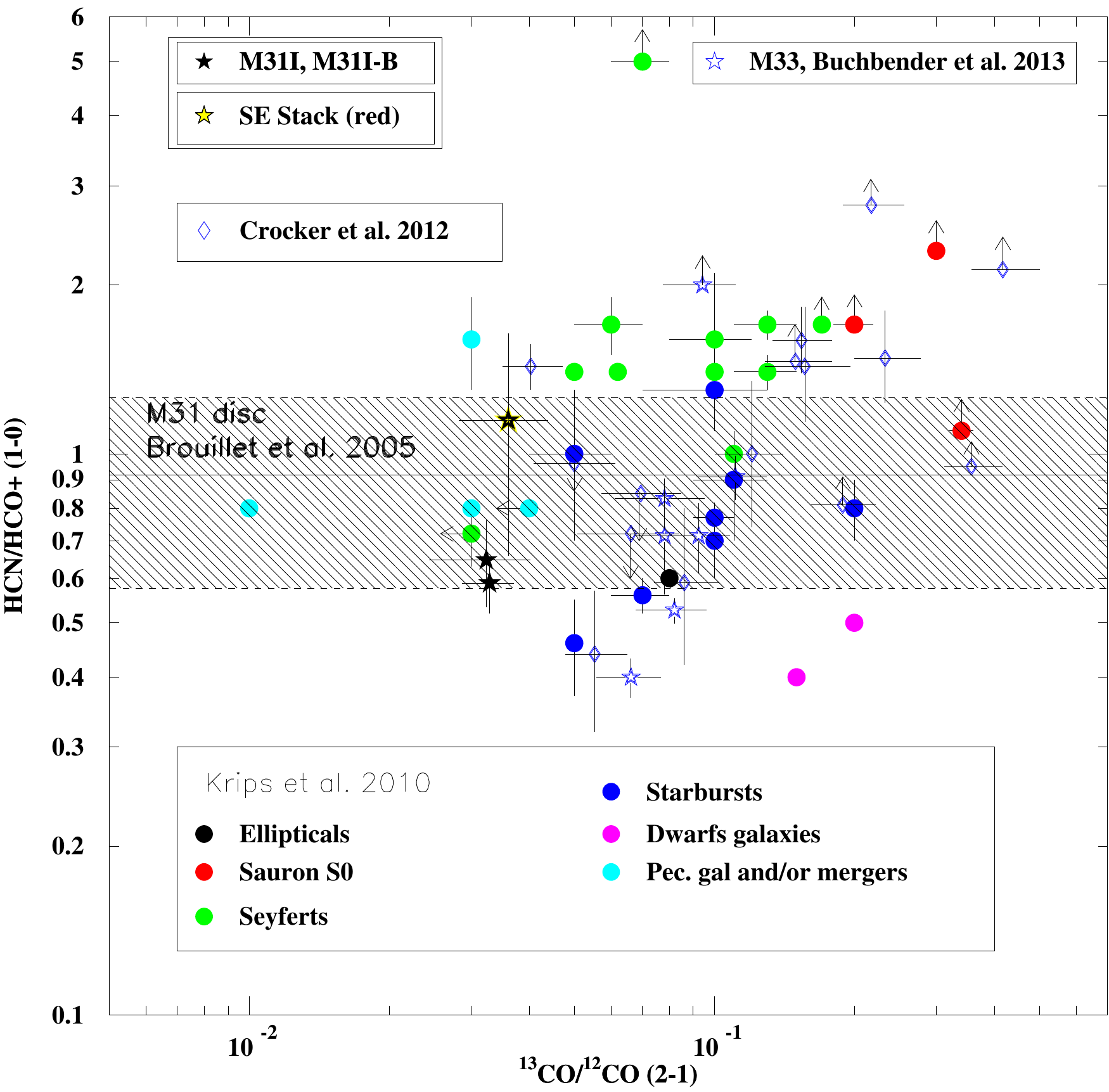}
   \caption{Comparison with other catalogues (1). HCN/HCO+(1-0) line
     ratio versus $^{13}$CO/$^{12}$CO(2-1) line ratio. The two
     north-west detections presented in this paper (M31I and M31I-B)
     together with the red south-east stacking measurement are
     superimposed on previous measurements, namely: a compilation of
     various types of galaxies from \citet{2010MNRAS.407.2261K},
     early-type galaxies from \citet{2012MNRAS.421.1298C} and M33's
     giant molecular clouds from \citet{2013A&A...549A..17B}. The
     hatched area corresponds to the mean HCN/HCO+(1-0) line ratio
     obtained by \citet{2005A&A...429..153B} in M31's
     disc. Surprisingly, our detections, compatible with the M31
       disc measurements of \citet{2005A&A...429..153B}, lie in "the
      peculiar/merger region".}
 \label{fig:lrat_dense2b}
\end{figure*}

\subsection{Observed line ratios}
\label{ssec:obse}
{ In this section, we analyse the observed line ratios, which are directly derived from the observations.}

When one computes line ratios based on velocity-integrated intensities
as shown in Table \ref{table:4}, { dilution effects due to the unknown beam filling factors should be removed at least at first order.  Empirical comparisons with similar targets can be performed.  
 
Figure \ref{fig:co1318} shows} that our measured $^{13}$CO/C$^{18}$O(2-1) line ratios lie below the (1-0) measurements of
\citet{2014MNRAS.443.2264L} in infrared Galactic dark clouds (average
4.4$\pm$1.3; see their Fig. 2), but corresponds to the lower end of
the \citet{2011RAA....11..787T} starburst sample distribution based on
$^{13}$CO/C$^{18}$O(1-0) line ratio. Our measured line ratios are also significantly lower than the M51's spiral arms gas detected by \citet{2010ApJ...719.1588S} (their average $^{13}$CO/C$^{18}$O line ratio: 3.5). 

It is remarkable that the C$^{18}$O detection is at a level comparable
to the $^{13}$CO detection, in contrast to what is observed in the
Galaxy \citep[e.g.][]{2015ApJS..216...18N,2014A&A...564A..68S}, where
the C$^{18}$O is usually 10 to 150 times weaker than $^{13}$
CO. $^{13}$CO/C$^{18}$O line ratios close to one are expected if
$^{13}$CO is optically thick or if $^{13}$C is depleted with respect
to $^{12}$C, such as after a starburst.

In Fig. \ref{fig:corat}, we compare
our $^{12}$CO/$^{13}$CO(2-1) line ratios with values measured
previously in other galaxies. We consider the following catalogues:
\begin{itemize}
\item the (2-1) line ratio compilations for different morphological types from \citet{2010MNRAS.407.2261K} (We provide the different morphological types under the form of hatched histogram);
\item the (1-0) line ratio measured by \citet{2015RAA....15..785L} in nearby galaxies;
\item the (1-0) line ratio measured by \citet{2010PASJ...62..409M} in starburst and post-starburst galaxies; and
\item the (1-0) line ratio measured by \citet{2011RAA....11..787T} in nearby infrared-bright galaxies and starbursts.
\end{itemize}
The (1-0) line ratios have been corrected with the average
$^{12}$CO/$^{13}$CO(1-0)/(2-1) ratio (1.33) from
\citet{2010MNRAS.407.2261K}. These authors found that the $^{13}$CO/$^{12}$CO(2-1) ratios of three SAURON galaxies are somewhat higher than those in galaxies of different Hubble types. The hatched histogram of Fig. \ref{fig:corat} shows the
different morphological types when available, and indicates that there is no trend with
morphology.  As discussed by
\citet{2015RAA....15..785L}, the isotopologue ratio does not depend on
the morphological type of the galaxy. Our
measurements (with an average { $^{12}$CO/$^{13}$CO$=20$}), shown with a thick line hatched histogram, lie below the standard isotopic ratio ($53$) \citep{1994ARA&A..32..191W} but fall in the same range as the other measurements. This is consistent with the fact that $^{12}$CO is optically thick. As a result of very different cloud geometries, it is difficult to compare our $^{13}$CO/$^{12}$CO(2-1) line ratio measurements in M31 with studies performed in Galactic regions. For instance, Orion regions observed by \citet{2015ApJS..216...18N} or Planck cold dust clumps \citep{2012ApJ...756...76W} exhibits relatively low $^{12}$CO/$^{13}$CO line ratio ($\sim 2-3$) because the dense clouds fill the beam and $^{12}$CO is very optically thick.

Figure \ref{fig:lrat_dense2b} shows the HCN/HCO+(1-0) and $^{13}$CO/$^{12}$CO(2-1) ratios we measured in the components
approaching us in the north-western positions and the south-east stacking detection, which is redshifted. We compare these line ratios with previous observations from the literature.
By superposing our measurements on the wide range of galaxy types observed by \citet{2010MNRAS.407.2261K}, we find
that our detections seem to lie in the peculiar/merger region, rather than in the star-forming (M33) and starburst part of the diagram. The HCN/HCO+ line ratio is compatible with the average ratio computed for M31's disc by \citet{2005A&A...429..153B}. Again, the Galactic clouds of \citet{2015ApJS..216...18N} lie on the right hand side of this figure and are difficult to compare with M31 measurements. Our measurements are not compatible with Seyfert galaxies, which makes sense because there is no sign of active galactic nuclei (AGN) in M31 \citep{2011ApJ...728L..10L}. Interestingly, on the one hand, the HCN/HCO+ line ratio detected here is compatible with the value (1.5) measured by \citet{2013MNRAS.431...27L} in infrared dark clouds. The red stacking ratio has a larger value but it is affected by a large error bar. On the other hand, HCN/HNC(1-0) line ratio is systematically larger than that measured in dark clouds by these authors.  The HCN(1-0) line detections enable us to estimate lower limits on the
HCN/HNC(1-0) line ratio towards the positions in the north-western area (M31I and M31I-B), which are  0.9 and 2.9, respectively.  Figure
\ref{fig:lrat_dense2} compares the HCO+/CO(1-0) and HCN/CO(1-0) line ratios detected on the north-western side of the inner ring with other
positions in M31's main disc \citep{2005A&A...429..153B} and with other galaxies. Our detections { are compatible with the} \citet{2005A&A...429..153B} measurements performed in M31's disc, but lie in the { lower end of the} overall distribution of \citet{2010MNRAS.407.2261K}.  Figure \ref{fig:coli} shows the detections and the upper limits of our observed CO line ratios.

Our detections differ from measurements performed in M33's giant molecular clouds \citep{2013A&A...549A..17B}.  They  are compatible with line ratios available for clouds in M31's disc studied by \citet{2005A&A...429..153B}, but there is no information on the CO isotopic ratio in the disc. The main trends regard the $^{13}$CO/C$^{18}$O line ratio, which is depleted, while the HCN/HCO+(1-0) line ratio is very typical of what is observed in other local galaxies.\citet{2013MNRAS.431...27L} have studied infrared dark clouds and argue that the abundance ratio of HCN and HCO+ may be unaffected by the environment. Even though the 100$\mu$m and A$_B$ maps shown in Figure \ref{fig:ext} do not exhibit strong features, we argue in Sect. \ref{ssect:prop} that our detections correspond to dense cores. These ratios depend non-linearly on physical parameters, such as optical depth, excitation temperature, density, and  some lines are probably optically thick and not necessarily in thermal equilibrium. This is explored further in  Sects. \ref{ssect:prop}, \ref{ssec:lte}, and \ref{ssec:radex}.

\subsection{Estimate of the beam filling factors}
\label{ssect:prop}
The temperatures we have detected for $^{12}$CO(2-1) are well below
the cosmological background, even though the molecular lines are detected in emission.
This is due to the low surface filling factor of the gas detected in the beam. In this section, we try to derive estimates of the beam filling factors affecting our observations. These estimates affect all absolute parameters (temperature, density, abundance, column density, etc.) that could be determined later, and thus the possible interpretation of the state of the detected gas.

As the average $^{12}$CO/$^{13}$CO line ratio (20) is well below the standard Galactic $^{12}$C/$^{13}$C  abundance ratio (53) of \citet{1994ARA&A..32..191W} and the C$^{18}$O line intensity comparable to the $^{13}$CO intensity, we  consider that the $^{12}$CO line is 
optically thick. As the gas  is detected in emission, we can then derive a minimal excitation temperature
of about 3\,K from the intrinsic brightness temperature $T_B$, with the
usual formula
\begin{eqnarray}
T_{B}=(f(T_{ex})-f(T_{bg})) \times (1 - e^{-\tau})
\label{eq:tpeak}
\end{eqnarray}
where
\begin{eqnarray}
f(T)=\frac{\frac{h\nu}{k}}{exp(\frac{h\nu}{kT})-1}
\end{eqnarray}
Relying on the modelling of infrared measurements,
\citet{2014A&A...567A..71V} and \citet{2014ApJ...780..172D}
estimated a cold dust temperature of the order of 20\,K (see Table \ref{tab:1b}). Since we  measured the line ratio CO(2-1)/CO(1-0)$\sim$0.8 \citep{2011A&A...536A..52M},  which is relatively high,  we assume that the line is thermalised { and that we can define a single excitation temperature.  This makes sense for cold gas in the central part of a large galaxy. 
 Most importantly, one has
to take into account the beam filling factor $\eta_{bf}$, which
is linked to the intrinsic brightness temperature as follows:
\begin{eqnarray}
T_{B}=T_{peak}/\eta_{bf} 
\label{eq:tpeak2}
\end{eqnarray}
where $T_{peak}$ is the peak temperature measured for our
detections, i.e. the observed diluted main beam temperature. Given our assumptions about the excitation temperature and the $^{12}$CO optical depth, we
can derive $\eta_{bf}$ from our $^{12}$CO(2-1) detections as
\begin{eqnarray}
\eta_{bf}=T_{peak}/\left[f(T_{ex})-f(T_{bg})\right]
\label{eq:etabfT}
\end{eqnarray}
\begin{table*}
\caption{Estimate the beam filling factor $\eta_{bf}(T_{ex})$ for
  different excitation temperatures, assuming that the $^{12}$CO(2-1) line is optically thick.}
\begin{tabular}{c|ccccc|cccc|cc}
\hline\hline
Positions & $\eta_{bf}(25\,K)$  &     $ {\eta_{bf}(20\,K)}$  &              $\eta_{bf}(15\,K)$  &    $\eta_{bf}(10\,K)$  &     $\eta_{bf}(5\,K)$  & $\tau_{^{13}CO}$ & $\tau_{C^{18}O}$ & $\tau_{HCO+}$ & $\tau_{HCN}$ & $\tau^{MR}_{^{12}CO}$ & $\tau^{GC}_{^{12}CO}$ \\
\hline
I(1)    & 2.1$\times 10^{-2}$ & 	${2.9\times 10^{-2}}$ & 		 4.2$\times 10^{-2}$ & 	8.0$\times 10^{-2}$ & 	3.6$\times 10^{-1}$ & { 0.049} &      & & &{ 2.57} & { 0.97}\\
I(2)    & 7.4$\times 10^{-3}$ & 	${9.9\times 10^{-3}}$ & 		 1.5$\times 10^{-2}$ & 	2.8$\times 10^{-2}$ & 	1.2$\times 10^{-1}$ & { 0.033} &      & { 0.021} & { 0.012}& { 1.77} & { 0.67}\\
I(3)    & 3.9$\times 10^{-3}$ & 	${5.2\times 10^{-3}}$ & 		 7.7$\times 10^{-3}$ & 	1.5$\times 10^{-2}$ & 	6.6$\times 10^{-2}$ & { 0.056} & { 0.038}& & & { 2.97} & { 1.12}\\
I-B(1)  & { 1.8$\times 10^{-2}$} & 	${2.4\times 10^{-2}}$ & 		 3.6$\times 10^{-2}$ & 	6.7$\times 10^{-2}$ & 	3.0$\times 10^{-1}$ & { 0.033} &      & { 0.016} & { 0.016} & { 1.74} & { 0.66}\\
I-B(2)  & { 2.5$\times 10^{-3}$} & 	${3.3\times 10^{-3}}$ & 		 4.9$\times 10^{-3}$ & 	9.3$\times 10^{-3}$ & 	4.2$\times 10^{-2}$ & { 0.089} &      & & & { 4.73} & { 1.78}\\
2d-a(1) & 5.1$\times 10^{-3}$ & 	${6.7\times 10^{-3}}$ & 		 1.0$\times 10^{-2}$ & 	1.9$\times 10^{-2}$ & 	8.5$\times 10^{-2}$ & { 0.059} &      & & & 3.12 & 1.18\\
2d-a(2) & 5.3$\times 10^{-3}$ & 	${7.1\times 10^{-3}}$ & 		 1.1$\times 10^{-2}$ & 	2.0$\times 10^{-2}$ & 	9.0$\times 10^{-2}$ & 0.051       &      & & & { 2.69} & { 1.02}\\
2d-b(1) & 8.9$\times 10^{-3}$ & 	${1.2\times 10^{-2}}$ & 		 1.8$\times 10^{-2}$ & 	3.3$\times 10^{-2}$ & 	1.5$\times 10^{-1}$ & 0.018       &      & & & 0.97 & 0.37 \\
2d-b(2) & 2.2$\times 10^{-3}$ & 	${3.0\times 10^{-3}}$ & 		 4.4$\times 10^{-3}$ & 	8.3$\times 10^{-3}$ & 	3.8$\times 10^{-2}$ & 0.063       &      & & & { 3.36} & 1.27\\
2d-b(3) & 1.9$\times 10^{-3}$ & 	${2.6\times 10^{-3}}$ & 		 3.8$\times 10^{-3}$ & 	7.2$\times 10^{-3}$ & 	3.2$\times 10^{-2}$ & 0.135       &      & & & { 7.16} & 2.70\\
2c-a(1) & 3.1$\times 10^{-3}$ & 	${4.1\times 10^{-3}}$ & 		 6.1$\times 10^{-3}$ & 	1.2$\times 10^{-2}$ & 	5.2$\times 10^{-2}$ & { 0.071} &      & & & { 3.78} & { 1.43}\\
2c-a(2) & 2.7$\times 10^{-3}$ & 	${3.7\times 10^{-3}}$ & 		 5.4$\times 10^{-3}$ & 	1.0$\times 10^{-2}$ & 	4.6$\times 10^{-2}$ &             &      & & & &\\
2c-a(3) & 5.0$\times 10^{-3}$ & 	${6.6\times 10^{-3}}$ & 		 9.8$\times 10^{-3}$ & 	1.9$\times 10^{-2}$ & 	8.4$\times 10^{-2}$ &             &      & & & &\\
\hline
\end{tabular}
\tablefoot{The
  next four columns provide the optical depths for $^{13}$CO(2-1),
  C$^{18}$O(2-1), HCO+(1-0), and HCN(1-0) assuming an excitation
  temperature of 20\,K and a beam filling factor equal to that
  derived from $^{12}$CO, i.e. $\eta_{bf}(20\,K)$.  The next two columns provide the $^{12}$CO(2-1) column densities computed for two standard $^{12}CO/^{13}CO$ line ratios for the molecular ring (MR) and Galactic centre (GC) from \citet{1994ARA&A..32..191W}.}
\label{tab:etabf}
\end{table*}
If the molecular lines are close to thermal equilibrium around 20\,K, the beam filling factor is
below one percent. This order of magnitude is consistent with several different
qualitative arguments, as follows: (1) The beam size is 44\,pc for $^{12}$CO(2-1), while
typical interstellar filaments have a size of 0.1\,pc
\citep{2014prpl.conf...27A}. { (2)} We are detecting gas complexes, which are
probably extended and patchy according to the optical image. { (3)} The
large velocity dispersions also suggest that we are integrating gas
along a large depth. { (4)} Last, if one considers a standard gas-to-dust relation
N$_H=0.94 \times 10^{21} \times A_V$ cm$^{-2}$
\citep{1978ApJ...224..132B}, we estimate A$_V=2.6$ and $1.4$ for M31-I(2) and M31-IB(1) (with no beam-filling factor corrections), while lower limits on
extinction estimated on the optical images are of the order of
A$_B=0.2$ \citep{2011A&A...536A..52M}.  This order of magnitude of the
beam filling factor is also compatible with values previously discussed in
the literature, for example \citet{1988LNP...315..405C} estimated a volume
filling factor smaller than $10^{-3}$ in M82, while
\citet{2015MNRAS.448.2469B} estimate a filling factor of approximately 4$\%$ in M31 and M51.

In the following, we consider that $^{12}$CO(2-1) is optically
thick to determine the beam filling
factors at thermal equilibrium, as provided in Table \ref{tab:etabf}. The values we estimate for $\eta_{bf}$ are consistent with \citet{2011ApJ...736..149G}, who discuss that giant molecular clouds consist of very low volume-filling factor ($\sim 5 \times 10^{-3}$) high-density (n(H$_2$)$\sim 3 \times 10^{4}$\,cm$^{-3}$) clumps, as well as typical starburst galaxies (with $\eta_{bf}<0.01$).

For a given excitation temperature, we can derive the optical depth corresponding to each optically thin line:
\begin{eqnarray}
\tau_{mol}=-ln\left( 1 - \frac{T_{peak}}{\eta_{bf}} \frac{1}{f(T_{ex})-f(T_{bg})}\right)
\end{eqnarray}
We can then compute the column density associated to each molecular line assuming an excitation temperature $T_{ex}$:
\begin{equation}
N_{mol} = \frac{8\pi k}{h c^3}\frac{Q(T)}{g}e^{E_{up}/T_{ex}} \frac{\tau_{mol}}{1-e^{-{\tau_{mol}}}}\frac{\nu^2 I_{CO}}{\eta_{bf}}
\label{eq:nmol}
\end{equation}
where $g$ is the level degeneracy $g=5$ for CO(2-1) molecules and $g=3$ for HCN(1-0) and HCO+(1-0), $Q(T)$ is the rotational partition function and $E_{up}$ is the energy level provided in Table \ref{table:2}.
 We also consider the beam filling factors defined in Table \ref{tab:etabf}. The factor $\frac{\tau_{mol}}{1-e^{-{\tau_{mol}}}}$ enables to account for non-optically thin conditions \citep[e.g.][]{2015PASP..127..266M}. However, in practice, we have to assume an unknown abundance { ratio to determine $\tau_{mol}$ if it is optically thick}, which limits the use of corrected column density in this analysis.

The optical depths thus computed and provided in Table \ref{tab:etabf} show that our hypotheses are consistent: $^{13}$CO is optically thin.
In Sect. \ref{ssec:lte}, we discuss abundances estimated  assuming LTE conditions. In Sect. \ref{ssec:radex}, we explore  how to derive molecular hydrogen densities for lower excitation temperatures with RADEX simulations \citep{2007A&A...468..627V}. 

\begin{table*}
  \caption{ Abundance ratios.  }               \label{table:3b}      
  \centering                          
  \begin{tabular}{c | c c | c c | c c |c c }        
    \hline\hline                 
              & \multicolumn{4}{c}{Measured values (LTE)}       & \multicolumn{4}{|c}{Galactic values} \\ \hline
             & \multicolumn{2}{c}{I(3)} &     \multicolumn{2}{|c}{Red Stacking} &  \multicolumn{4}{|c}{}\\ \hline
    Transition&  X$^{Obs}$        & $^{13}$CO      & X$^{Obs}$        & $^{13}$CO      & \multicolumn{2}{|c}{X$^{GISM}$} &  \multicolumn{2}{c}{$^{13}$CO  }  \\  
   (2-1)      &                   & $/$Mol   &   &         $/$Mol      & \multicolumn{2}{|c}{}  & \multicolumn{2}{c}{/Mol  } \\    
         &    &         &  && (MR)    & (GC)  &    (MR)    & (GC)    \\\hline    
$^{13}$CO  &10.8$\pm$2.7$\times 10^{-8}$       & 1    & {\em 8.0$\pm$1.5$\times 10^{-8}$}  & 1 & 5.3$\times 10^{-7}$        & 1.4$\times 10^{-6}$ &    1 & 1\\
C$^{18}$O  & 7.3$\pm$2.4$\times 10^{-8}$       & 1.48$\pm$0.51  & {\em 4.6$\pm$1.4$\times 10^{-8}$}  & 1.73$\pm$0.58 & 8.6$\times 10^{-8}$        & 1.1$\times 10^{-7}$ &  6.2& 12.7\\
\hline \hline  
 
              & \multicolumn{4}{c}{Measured values (LTE)}       & \multicolumn{4}{|c}{Galactic values} \\ \hline
             & \multicolumn{2}{c}{I(2)} &     \multicolumn{2}{|c}{SE red Stacking} &  \multicolumn{4}{|c}{}\\ \hline
   Transition&  X$^{Obs}$        & Mol     & X$^{Obs}$        & Mol     & X$^{GISM}$ &  \multicolumn{1}{c}{Mol} &  &\\  
   (1-0)      &                   & $/$HCO+ &   &         $/$HCO+      & & \multicolumn{1}{c}{$/$HCO+}  & \multicolumn{2}{c}{ } \\  \hline  
HCN        & 1.8$\pm 0.3 \, 10^{-10}$ & \multicolumn{1}{c|}{1.11$\pm$0.22 }   & {\em 2.6$\pm$0.5$\times 10^{-10}$} & \multicolumn{1}{c|}{1.96$\pm$0.64}  &1.9$\times 10^{-9}$        & \multicolumn{1}{c}{4.0}  & \multicolumn{2}{c}{ }   \\
HCO+       & 1.6$\pm 0.2 \, 10^{-10}$ & \multicolumn{1}{c|}{1} & {\em 1.3$\pm$0.3$\times 10^{-10}$} & 1& 4.7$\times 10^{-10}$       & \multicolumn{1}{c}{1} & \multicolumn{2}{c}{ }    \\
    \hline                                   
             & \multicolumn{2}{c}{I-B(1)} &     \multicolumn{6}{|c}{} \\ \hline
   Transition&  X$^{Obs}$        & Mol     &     \multicolumn{6}{|c}{ } \\  
   (1-0)      &                   & $/$HCO+ &   \multicolumn{6}{|c}{ }   \\  \hline  
HCN        & {\em 1.3$\pm 0.2 \, 10^{-10}$} & \multicolumn{1}{c|}{1.01$\pm$0.14 }   & \multicolumn{6}{|c}{ }   \\
HCO+       & {\em 1.3$\pm 0.1 \, 10^{-10}$} & \multicolumn{1}{c|}{1} & \multicolumn{6}{|c}{ }   \\
    \hline                                   
  \end{tabular}
  \tablefoot{For each molecule, we provide the expected abundances, assuming LTE and optically thin conditions, to compare them with Galactic values.   We provide abundances and abundance ratios for two positions M31-I(3) and M31-I(2) (columns 2-3)  and the stacking detections (columns 4-5), and then compare  these values with those expected for the Galactic interstellar medium, at 4\,kpc (MR) and at the Galactic centre (GC), derived from \citet{1995ApJ...441..222B} and \citet{1994ARA&A..32..191W} (columns 6-9). When $^{12}$CO(1-0) is not measured directly, the numbers are in italics.}
\end{table*}
\begin{figure*}
\centering
\includegraphics[width=\textwidth]{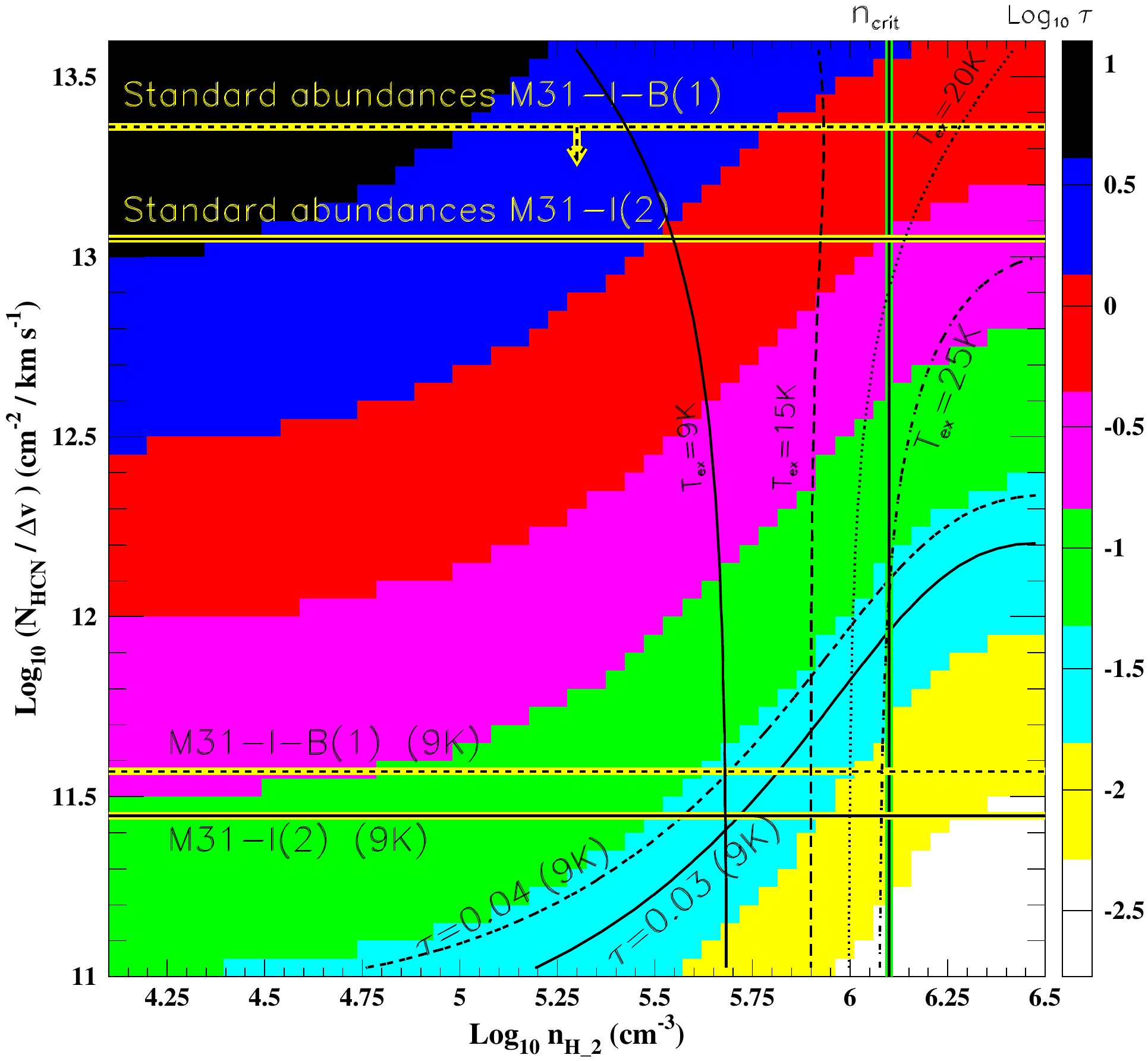}
\caption{Main physical parameters corresponding to the HCN gas observed in M31-I(2) and M31-I-B(1). The optical depths are shown as function of the HCN column density (per km\,s$^{-1}$) and the molecular hydrogen density $n_{H_2}$. The corresponding contours for these two clumps are shown with a dashed line for M31-I-B(1) and with a full line for M31-I(2). The vertical (green) line indicates the critical density computed for a collisional temperature of 20\,K (see Table \ref{table:2}). The lower horizontal full (dashed) line shows the column density measured for M31-I(2) (M31-I-B(1)) at 9\,K. The upper horizontal lines correspond to the column densities computed with standard abundances \citep{1995ApJ...441..222B} and the corresponding molecular hydrogen column densities. For M31-I-B(1), we derive an upper limit as the molecular hydrogen column density is only available for a smaller beam (see Table \ref{table:3}) and we expect some dilution of the signal.   The contour levels correspond to excitation temperatures of 9\,K (full line), 15\,K (dashed line), 20\,K (dotted line), and 25\,K (dash-dotted line). }
\label{fig:hcn}
\end{figure*}
\begin{figure*}
\centering
\includegraphics[width=7.8cm]{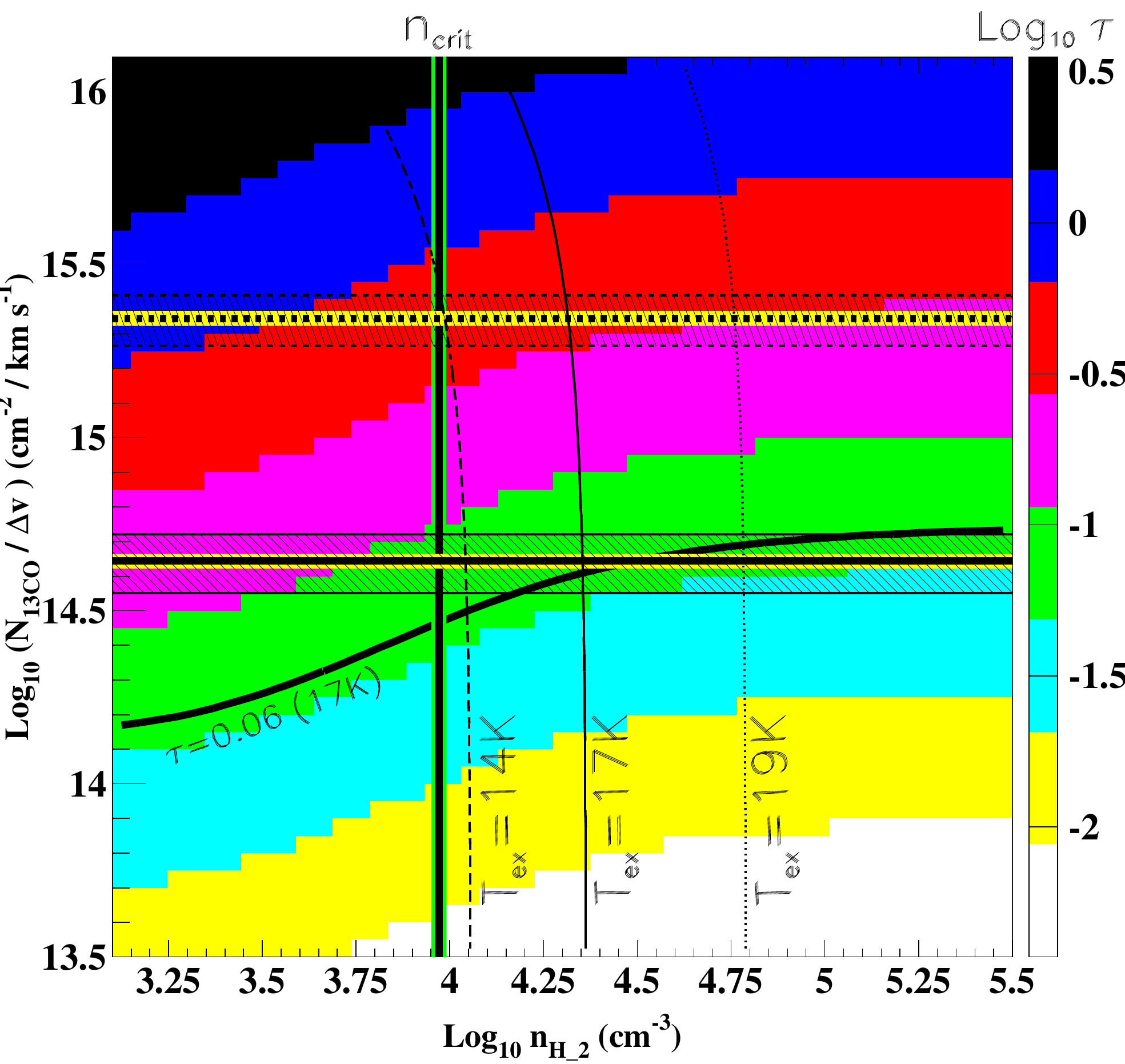}
\includegraphics[width=7.8cm]{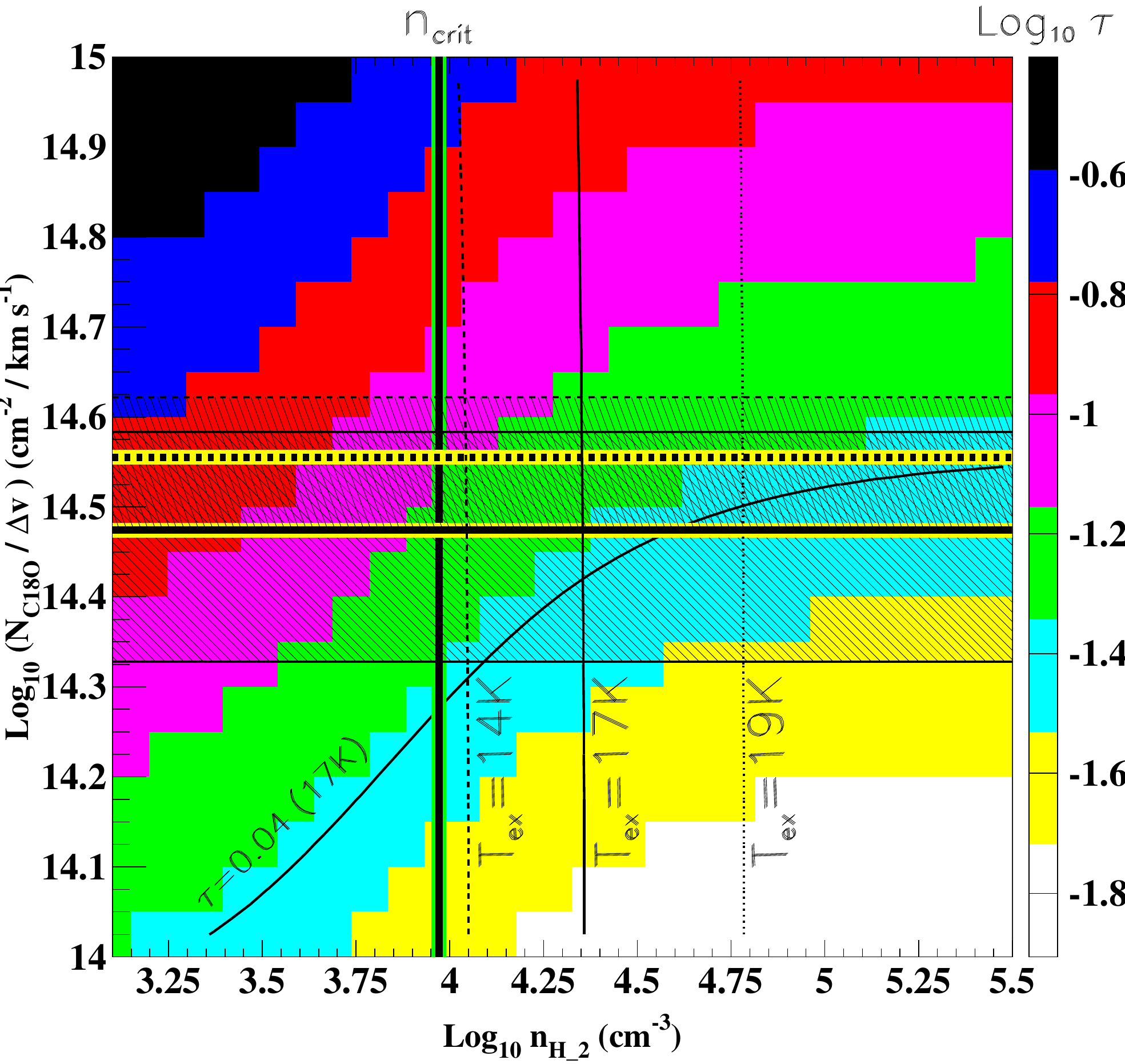}
\caption{Main physical parameters corresponding to the $^{13}$CO(2-1) gas (left panel)  
and the C$^{18}$O(2-1) gas (right panel) detected in M31-I(3). The optical depths are shown as
 function of the $^{13}$CO(2-1) column density (per km\,s$^{-1}$) and the molecular hydrogen
  density $n_{H_2}$. The vertical green-black lines indicate the critical densities computed 
  for a collisional temperature of 20\,K (see Table \ref{table:2}). The horizontal thick, full
   yellow-black lines indicate the $^{13}$CO(2-1)  and the C$^{18}$O(2-1) averaged column
    densities (per unit velocity), while the horizontal thin black lines indicate the
     corresponding 1\,$\sigma$ standard deviation; the hatched area corresponds to the M31-I(3)
      detection. The  horizontal thick, dashed yellow-black lines (thin black lines) correspond
       to the column densities (1\,$\sigma$ standard deviation) computed with standard abundances
        \citep{1995ApJ...441..222B} and the corresponding molecular hydrogen column densities.
         The full line (dashed and dotted) contour levels correspond to an excitation temperature
          of 17\,K (14\,K and 19\,K). The optical depths,  corresponding to 17\,K, are shown as
           a thick line contour level.  }
\label{fig:13co}
\end{figure*}

\subsection{Local thermal equilibrium conditions}
\label{ssec:lte}
We consider here that the different detected lines correspond to some gas in LTE with an excitation temperature of 20\,K. 
This enables us to derive physical parameters from the observations with an analytical formula and to give an initial idea of the physical configurations of the observed complexes. 

\subsubsection{Optical depths and column densities}
With these assumptions, we derived optical depths for $^{13}$CO(2-1), C$^{18}$O(2-1), HCO+(1-0) and HCN(1-0) with the filling factor $\eta_{bf}(20\,K)$, computed in Sect. \ref{ssect:prop}. These lines are optically thin, as shown in Table \ref{tab:etabf}. 

If we consider standard $^{12}$CO/$^{13}$CO line ratios from \citet{1994ARA&A..32..191W} of $53$ for the molecular ring ($20$ for the Galactic centre) and the $^{13}$CO optical depths provided in Table \ref{tab:etabf}, we estimate an average optical depth for $^{12}$CO of $3.2$ ($1.2$). The $^{12}$CO$(2-1)$ transition is indeed optically thick.
These different optical depths discussed above are provided for each line of sight in Table \ref{tab:etabf}.

In the middle panel of Table \ref{table:3}, we provide column
densities of our main detections ($^{13}$CO, C$^{18}$O, HCN, HCO+),
some upper limits { for} optically thin gas ($\tau << 1$)  in LTE, { and}  the column density of molecular hydrogen derived empirically from the $^{12}$CO line intensity with a X$_{CO}$ factor as discussed in Sect. \ref{sec:datared}. All column densities account for the beam filling factor $\eta_{bf}$.

\subsubsection{Abundance ratios} 
In Table \ref{table:3b}, we tentatively compute the abundances (i.e. the ratio of the column density of the molecule with respect to the column density of molecular hydrogen) for { the direct and stacking detections.} In order to calculate the ratio of lines measured with similar HPBW resolutions, we use $^{12}$CO(2-1) to derive the molecular hydrogen column density for the $^{13}$CO(2-1) and C$^{18}$0(2-1)  measurements and $^{12}$CO(1-0) (available for M31-I) for HCN(1-0) and HCO+(1-0) abundances. { We thus study M31I(3) and M31I(2). We also include the red stacking detections for which we adopt an {\em ad-hoc} estimate of N$_{H_2}$ based on the averaging of the measured I$_{^{12}CO}$ and I$_{^{13}CO}$ measurements performed with the spectra used for the stacking: a scaling factor is derived from these averaged measurements and applied to the stacked $^{13}$CO(2-1) measurement to estimate $ {^{12}CO}$(2-1). It is then scaled to $^{12}$CO(1-0) with the 0.8 factor, as discussed in Sect. \ref{sec:datared}. For the sake of discussion, we also provide (in italics) M31I-B(1) and the stacking measurements for which we have HCN(1-0) and HCO+(1-0) measurements to no direct $^{12}$CO(1-0) measurements. As shown in Table \ref{table:3b}, the molecular hydrogen column density exhibits variations up to a factor 3 owing to different resolutions (11$^{\prime\prime}$ and 21$^{\prime\prime}$).}

 We use the column densities presented in Table \ref{table:3} and described in Sect. \ref{sec:datared}.   We also provide for comparison the abundance values for the Galaxy computed in the molecular ring (MR) at 4\,kpc from the Galactic centre and in the Galactic centre (GC) by \citet{1995ApJ...441..222B} and \citet{1994ARA&A..32..191W}. 

The abundances are affected by the uncertainties in the molecular hydrogen column densities computed empirically with the X$_{CO}$ factor and by the uncertainties in the excitation temperatures. For the CO transitions, we can expect that variations of the excitation temperatures are a secondary effect, and we can thus rely on abundance ratios for optically thin  transitions. 

The detected $^{13}$CO/C$^{18}$O abundance ratios are very different from the Galactic values. There is a discrepancy of a factor $4$ with the molecular ring (MR), and a factor $9$ with respect to the Galactic centre (GC) where gas has been processed regularly. Obviously, the gas is not in the same state as these obvious regions in the Galaxy. Considering the C$^{18}$O abundances in M31-I(3), it is relatively close to the { molecular region} value, while the discrepancy affects the $^{13}$CO. For the stacking, the C$^{18}$O abundance { is a bit smaller and intermediate between GC and MR values, even though it is compatible with the M31I(3) abundance within 1\,$\sigma$. While} the C$^{18}$O abundance lies in the same range as the Galactic value, the $^{13}$CO abundance is deficient. 
 
The HCN and HCO+ abundances differ from the Galactic values, and the
abundance ratio. These discrepancies could be due to different
excitation temperatures.  However, our different measurements are
globally compatible within 3$\sigma$. { \citet{2013MNRAS.431...27L}
  measured an HCN/HCO+(1-0) abundance ratio of $1.2\pm0.4$ in infrared
  dark clouds (IRDC) and discuss that this ratio may be unaffected by
  the environment. In parallel, they measured an average HNC/HCN
  abundance ratio of 1.47$\pm$0.50 in IRDC, while the $3\sigma$ upper
  limit we derived on the HNC/HCN abundance ratio for M31I-B(1) and
  M31I(2) is 1.01 and 1.14. As these authors discuss, we can argue
  that our detections are different from the IRDC. Our upper limits
  allow us} to put some constraints on the kinetic temperature
\citep[e.g.][]{1992A&A...256..595S,1998ApJ...503..717H}. This
corresponds to kinetic temperatures in the range 38-100\,K and
14-50\,K, and is compatible with the results of
\citet{2014A&A...567A..71V} based on the analysis of infrared data.
Similarly, we get HCO+/HOC+ abundance ratios larger than { 1.7 and
  4.9}, but this is not restrictive enough as previous detections are
in the range 50-12500 \citep{2004ApJ...616..966S}.

In the next section, we explore with non-LTE simulations how different excitation temperatures could affect the different parameters and help to get a coherent set of parameters.

\subsection{RADEX simulations}
\label{ssec:radex}
In order to check the consistency of our parameters, we ran some { 
simulations with the radiative transfer code} RADEX  \citep{2007A&A...468..627V} for several
molecules. The idea is to relax the LTE conditions { to scan a range 
of excitation temperatures and of molecular hydrogen densities. Knowing the column densities of each emission line, we can estimate the excitation temperature and the density. As indicated in the figures discussed below, the standard abundances correspond to optically thick conditions and can be excluded.}

In Fig. \ref{fig:hcn}, we show the results of the simulations run
for HCN and superimpose the various parameters discussed in the
previous section, namely, the optical depth and the molecule column
densities per unit velocity derived from our observations and from
standard abundances. These results support the view that this gas is
subthermally excited at 9\,K with a density smaller than the critical
density, which is provided in Table \ref{table:2}. 
This critical density plotted here (and in Figs. \ref{fig:rdxhc} and \ref{fig:13co}) does not match $T_{ex}=20K,$ confirming the arguments of \citet{2015PASP..127..299S} that this quantity is an oversimplified and incorrect interpretation of the excitation density of a transition. 
The results of the constraints of RADEX simulations with our observations are also consistent with the HCN(1-0) transition being
optically thin. We detected two dense clumps with a molecular
hydrogen density $n_{H_2}=4.8\times 10^5$\,cm$^{-3}$ on the north-west
side of the inner ring.  Our abundances are more than a factor of 10
lower than the column densities derived from the standard abundances
estimated by \citet{1995ApJ...441..222B} in the molecular ring (MR),
and the discrepancy would be even larger if one considers the Galactic
centre conditions.

We performed similar computations for HCO$+$(1-0), as shown in Figure \ref{fig:rdxhc}. We also find subthermal conditions with a molecular
hydrogen density smaller than the critical density and an excitation
temperature of { 9\,K}.  The abundances are also smaller { (factor 5-6)}
than the column densities derived from the standard MR abundances
\citep{1995ApJ...441..222B}. The gas associated with the HCO+(1-0)
detection has a density of { $1.0\times 10^5$\,cm$^{-3}$}.

We applied the same procedure to M31-I(3) where $^{13}$CO and
C$^{18}$O have been detected, as shown in Fig.
\ref{fig:13co}. On the one hand, the best excitation temperature is
consistent with 17\,K but we cannot exclude
20\,K given the error bars. On the other hand, the simulations support a molecular hydrogen
density larger than the critical density. These lines are close to LTE
conditions and trace clumps with a density larger than { 2.24$\times
10^4$\,cm$^{-3}$.} The $^{13}$CO measured abundance is smaller than the
standard abundances by a factor { 5}, while the C$^{18}$O measured
abundance is compatible within the error bars of the standard
abundances estimated in the molecular ring.

The abundances we have derived are systematically smaller than the
standard abundances \citep{1995ApJ...441..222B} for the molecular
ring, but for C$^{18}$O. This is consistent with our previous
discussion that $^{12}$CO is optically thick and $^{13}$CO depleted,
with LTE and optically thin conditions for $^{13}$CO and
C$^{18}$O. HCN and HCO+ abundances are consistent with subthermal
excitation conditions with excitation temperatures of { 9\,K}
and are smaller than the standard abundances. As shown in Figs \ref{fig:lrat_dense2} and \ref{fig:lrat_dense2b},  the HCO+/HCN line ratio measured in the clouds studied here is similar to that measured  in M31's
disc by \citet{2005A&A...429..153B}.

\section{Discussion}
\label{sec:disc}
In this section, we discuss our dense gas detection, how it can improve our knowledge of this inner region, and whether the properties of this gas are consistent with our understanding of previous observations. In Sect. \ref{ssect:geomconf}, we discuss the geometrical configuration of the gas: the information derived from dust extinction affecting point sources, and the properties of various gas components observed at different wavelengths. In Sect. \ref{ssect:heat}, we discuss our knowledge of the heating sources and the properties of the interstellar medium in this region. In Sect. \ref{ssect:kine}, we discuss the kinematical information obtained with our millimeter observations. 
\begin{figure*}
  \centering
  \includegraphics[width=0.49\textwidth]{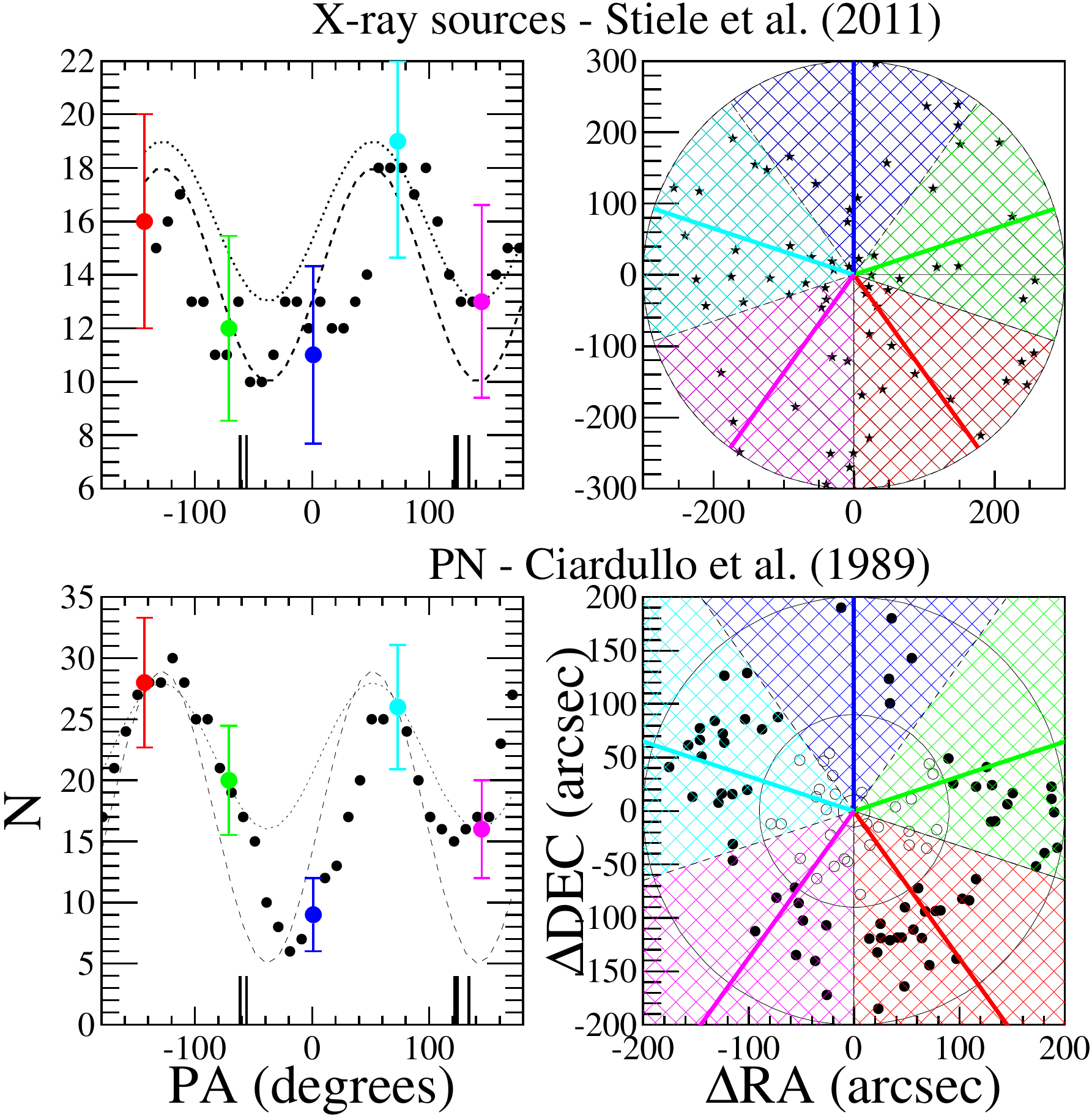}  
  \includegraphics[width=0.49\textwidth]{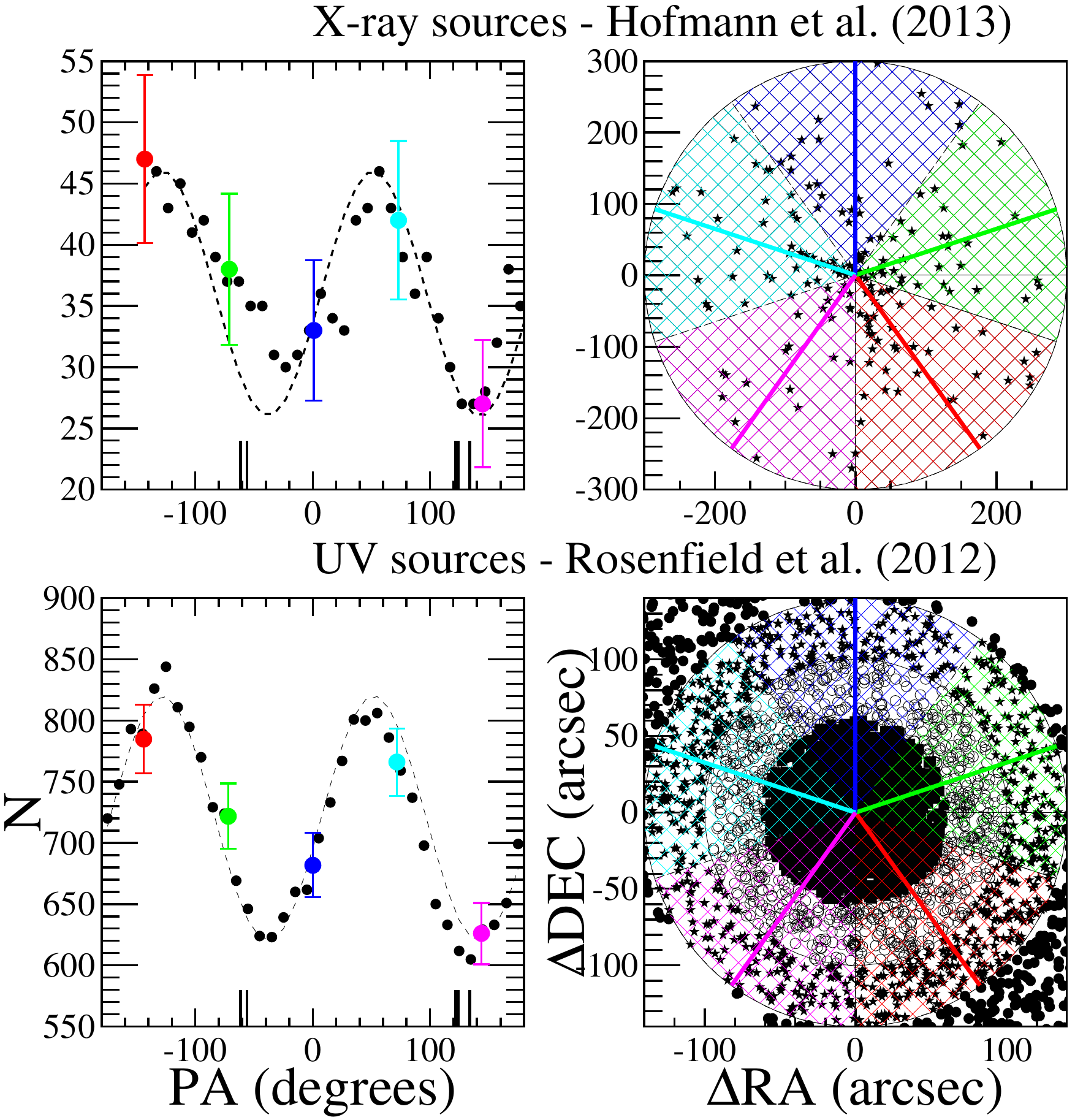}  
 \caption{Azimuthal (1$^{st}$ and 3$^{rd}$ columns) and spatial (2$^{nd}$ and 4$^{th}$) distribution of X-ray sources (top) and planetary nebulae and UV sources (bottom). The top left panels show the deep Chandra-detected X-ray sources from \citet{2011A&A...534A..55S}. The top right panels show the X-ray sources from \citet{2013A&A...555A..65H}. The bottom left panels indicate the planetary nebulae from \citet{1989ApJ...339...53C} (see also \citet{2013A&A...549A..27M}). The bottom right panels show the HST UV sources from \citet{2012ApJ...755..131R}. The dashed line exhibits the best-fit profile from \citet{2002ApJ...578..114K} corresponding to a disc with a position angle PA$=52$\,$\deg$. In the right panels, the two minima are roughly at the same level, while in the left panels the first is deeper as indicated qualitatively with the dotted line. The ticks in the 1$^{st}$ and 3$^{rd}$ figure columns correspond to the position angles of our detections. }
  \label{fig:azimuth}%
\end{figure*}

\subsection{Geometrical configuration of the gas}
\label{ssect:geomconf}
M31 has been targeted by different surveys.  In Sect. \ref{sssect:point}, we study the distribution of point sources from different catalogues in search of possible signs of extinction that trace the gas. In Sect. \ref{ssect:gascomp}, we summarise the arguments we can derive from the diffuse components.

\subsubsection{Point sources and extinction}
\label{sssect:point}
In Fig.  \ref{fig:azimuth}, we considered four catalogues: X-ray
sources from \citet{2011A&A...534A..55S}, deeper X-ray sources from
\citet{2013A&A...555A..65H}, planetary nebulae from
\citet{1989ApJ...339...53C}, and UV sources from
\citet{2012ApJ...755..131R}.  The UV sources studied by
\citet{2012ApJ...755..131R} trace the old stellar population, and they
have the same radial distribution as low-mass X-ray binaries, which
correspond to the majority of the X-ray sources in
\citet{2013A&A...555A..65H}. The stellar kinematics has been studied
by \citet{2010A&A...509A..61S} with long slits: bulge stars exhibit a
rotation with a maximum velocity of 100\,km\,s$^{-1}$ along the major
axis (considered at PA$=48$\,$\deg$) and a zero (systemic) velocity
along the minor axis). These velocities clearly do not correspond to
the gas component we detected.

For each catalogue, we plotted the azimuthal distribution of point sources within a radius of 300, 200, and 140$\arcsec$, depending on the area covered. These populations all belong to the same ellipsoid of PA$\simeq 52$\,$\deg$ as first discussed by \citet{2002ApJ...578..114K}. The maxima of the sinusoids correspond to the major axis, while the minima (where our dense gas has been detected) correspond to the minor axis. 

The number of sources is significantly larger for \citet{2013A&A...555A..65H} than for \citet{2011A&A...534A..55S}; the former work is more sensitive as the data are stacked over a longer period.  The work of  \citet{2012ApJ...755..131R}  based on HST data contains an unprecedented number of sources thanks to deep integration and a gain in spatial resolution. We can observe that in these deep observations the two minima of the sinusoid are about at the same level, while the previous observations exhibit an asymmetry. This asymmetry is not really significant when compared to the Poisson error bars shown in the Fig. \ref{fig:azimuth}. 
While in radio we do not expect much extinction, the UV sources in \citet{2012ApJ...755..131R}  should be very sensitive to extinction. However, we do not detect any strong asymmetry: if we compute the ratio of the numbers of sources from the near and far sides, we find 0.9676 (0.9662) if we consider 52 deg (77 deg). These ratios are significant as they have a Poisson noise of 0.012. If this asymmetry is due to extinction, it affects 4$\%$ of the studied area. This order of magnitude is consistent with the gas beam filling factors computed in Sect. \ref{ssect:prop}. However, we do not have the sensitivity to derive whether this patchy dust component belongs to the large scale disc (PA=$77$\,$\deg$) or the more face-on component detected in the stellar bulge population (PA$=52$\,$\deg$). We can exclude the possibility that young stellar clusters from the disc component could create this inhomogeneity, as no HII region is detected in this area \citep{2011AJ....142..139A} beside the cluster close to the centre \citep{2012ApJ...745..121L}. The (young) stellar UV cluster \citep{2012ApJS..199...37K} closest to our detection has an age of 1\,Gyr and is at the offset ($-111''$,$152''$) from the centre and at a closest separation of 71\,arcsec (271\,pc in projection) to our north-western dense gas complex. 
\citet{2015MNRAS.451.4126D} detected an intermediate age (300\,Myr to 1\,Gyr old) stellar population at solar metallicity associated with the bulge, which could contribute up to 50$\%$ of the V intensity. However, they only studied the radial dependency, while we expect a smooth spatial distribution as the stars have performed numerous galactic rotations since their birth. 

Assuming a uniform distribution in the stellar ellipsoid, one can estimate the ellipticity as done by \citet{2002ApJ...578..114K}. For the \citet{2012ApJ...755..131R} \citep{2013A&A...555A..65H} sample studied here within 140$\arcsec$ (300$\arcsec$), we estimate $\epsilon=0.24$ ($\epsilon=0.44$). This is consistent with the measurements of the radial variation of the ellipticity and position angle performed on J-2MASS photometry by \citet[see Figure A.1][]{2013A&A...549A..27M}. This supports our previous finding that there is very little extinction.

While these populations trace the bulge, there is an unexpected X-ray
source alignment in the \citet[figure 14, ][]{2013A&A...555A..65H}
catalogue very close to the centre. Similar to that observed in the
outskirts of M31 \citep[e.g.][]{2009Natur.461...66M}, these X-ray
sources could be part of a tidal stream whose stellar population could
be hidden by the crowding at other wavelengths. The facts that it
corresponds to an old stellar population and that it is aligned are
strong arguments to argue that they are far from the centre and seen
in projection. However, it is not aligned with our positions nor any
diffuse components discussed below.

\subsubsection{Gas components}
\label{ssect:gascomp}
While the stellar population mainly belongs to the bulge, the diffuse components, apart from the UV to near-IR stellar continuum, are gaseous.

\paragraph{Radio component}
\label{sssect:radio}
 The analysis of the radio emission \citep{2014A&A...571A..61G}
 reveals complementary information. There is a significant continuum
 flux at 3.6 cm and 6.2 cm, and a weak polarised component
 ($\sim$10$\%$). There is a relatively flat spectral index
 ($\nu^{\beta=-0.4}$), while the contribution from the black hole is
 very weak and localised \citep{2011ApJ...728L..10L}. This flat
 spectral index cannot be accounted for by an inverted synchrotron
 emission due to a present AGN, but we cannot exclude relic emission
 from past AGN activity \citep[e.g.][]{2015MNRAS.451...93I}.  The flat
 spectral index could be due to the combination of synchrotron and
 free-free emissions. In the \citet{2006Natur.443..832B} scenario, the
 inner ring forms and expands to its actual position in
 $\sim$150\,Myr. The corresponding past starburst activities
 (150-200\,Myr old) could account for the continuum emission stronger
 close to the centre.  The polarisation map of
 \citet{2014A&A...571A..61G} is illustrated as contours in
 Fig. \ref{fig:xray}. The polarisation is blurred by the Faraday
 rotation of the electrons. On the north side, no polarisation is
 detected, while it is detected on the south side. This polarisation
 pattern corresponds to a good approximation to the simple modelling
 proposed in \citet[see Fig. 16 and Tab.  7,][]{2011A&A...536A..52M}
 of the inner ring with a PA of -35\,$\deg$ and an inclination of
 40\,$\deg$, together with the two discs components strongly
 depolarising the north-western side.  In addition in the south side,
 we clearly see some inhomogeneities in the polarised emission
 detected: there is a gap that corresponds to the
 H$_{\alpha}$($+[NII]$) emission. This confirms the parameters of the
 inner ring's location on the near side, i.e. in the south-east
 direction.  This supports some previous tentative interpretations
 \citep{1988AJ.....95..438C} that we are observing structures more
 face-on than the 77\,$\deg$ main disc with the near side on the north
 side and a more edge-on structure on the south side.

Supporting the post-starburst scenario, the central field exhibits a
remmant tidal swirl observed in radio at 3.6\,cm and 6.2\,cm
\citep{2014A&A...571A..61G} and in ionised gas
\citep{1988AJ.....95..438C}. These radio emissions could trace relic
activities of stellar populations younger than 200\,Myr
\citep{1992ARA&A..30..575C} i.e. following the impact and expansion of
the inner ring. This remnant tidal swirl could be dominated by shocks,
e.g. from supernovae as [NII] is larger than H$_\alpha$
\citep{1972ApJ...177...31R}, while the inner ring is possibly stronger
in H$_\alpha$ tracing a 200\,Myr old starburst
\citep{2006Natur.443..832B}.

\paragraph{Ionised gas}
\label{sssect:ionised}
While we know that the observed extinction is weak and the overall
amount of gas is small, \citet{2000MNRAS.312L..29M} has estimated a
lower limit on A$_B$ assuming all the gas is in front of the bulge.
Even in locations where the extinction detected is small
(i.e. probably underestimated by the previous assumption), it
corresponds well with features detected in the infrared (e.g. at
100\,$\mu$m as indicated in Figure \ref{fig:ext}). As discussed by
\citet{2013ApJ...769...55F}, the central gas is hotter than the inner
ring. The 100\,$\mu$m map exhibits hot dust and cold dust, while the
500\,$\mu$m map mainly shows cold dust in the inner ring (close to our
observed positions). This central hot component corresponds to the
ionised gas \citep{1988AJ.....95..438C}. The ionsed gas is very
inhomegeneous and its distribution does not exhibit any clear sign of
extinction. This is compatible with the fact that the gas is very
clumpy with a filling factor smaller than 1\,$\%$and that the
near/far-side asymmetry, possibly due to extinction, is at the level
of 4\,$\%$.

\paragraph{Molecular gas}
\label{ssect:molgas}
Our dense gas detections discussed here exhibit a surprising velocity pattern. This is the wavelength range where the best velocity resolution is achieved (few km/s), and we observe in two positions along the minor axis where the systemic velocity is expected atvelocities between -530 and -140 km\,s$^{-1}$. This kinematical information demonstrates the presence of several molecular gas components along these lines of sight. In contrast to HI measurements \citep{2009ApJ...695..937B,2009ApJ...705.1395C}, we do not have {\em a priori} the sensitivity to detect molecular gas in the warp or the external parts of the galaxy.

\subsection{Heating sources}
\label{ssect:heat}
We investigate here the possible heating sources, which could explain why the signal is stronger on one side and what heats the gas in general.
Following \citet{2008MNRAS.388...56B}, we know that the diffuse X-ray gas
 detected in this area within 150-200\arcsec\, from the centre is
weak. The detected { 0.5-1.2keV} flux lies in the range 1.6 - 9$\times
10^{-18}$erg\,s$^{-1}$cm$^{-2}$arcsec$^{-2}$, considering a molecular cloud of 1\,pc with a distance of M31 of 780\,kpc, we can
estimate a rough X-ray incident radiation intensity of 6-34$\times
10^{-7}$ \,erg\,s$^{-1}$cm$^{-2}$. This is at least five orders of
magnitude weaker than the radiation field considered by
\citet{2007A&A...461..793M} for XDR modelling.  There is only one
ROSAT (Second ROSAT PSPC catalogue) source within 12\arcsec (22\arcsec)
from M31I (M31I-B) with a 14\,arcsec positional error, with a
flux of 10$^{-12}$\,erg\,s$^{-1}$cm$^{-2}$arcsec$^{-2}$. 
\begin{figure*}
  \centering
  \includegraphics[width=7.9cm]{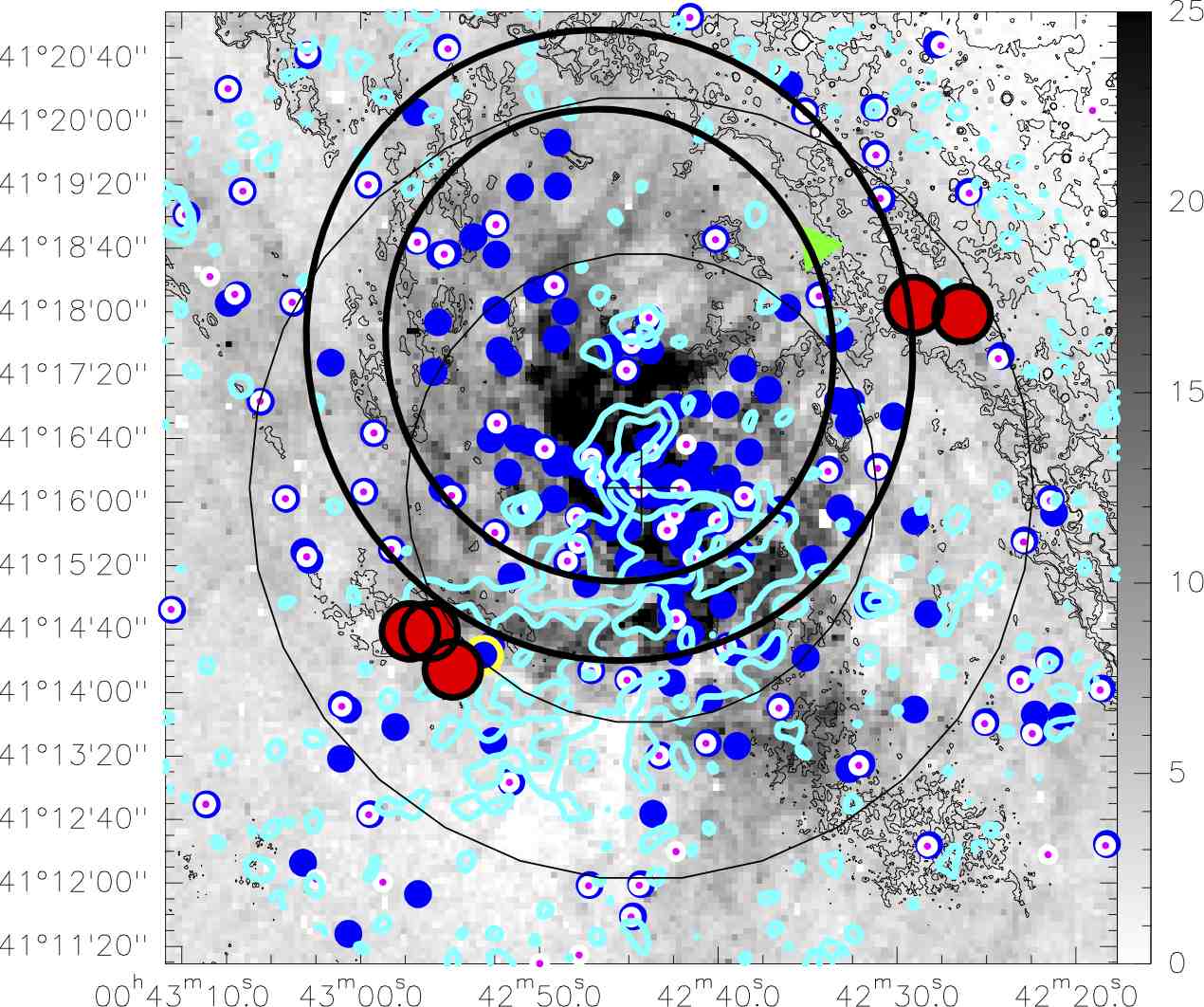}
  \includegraphics[width=6.9cm]{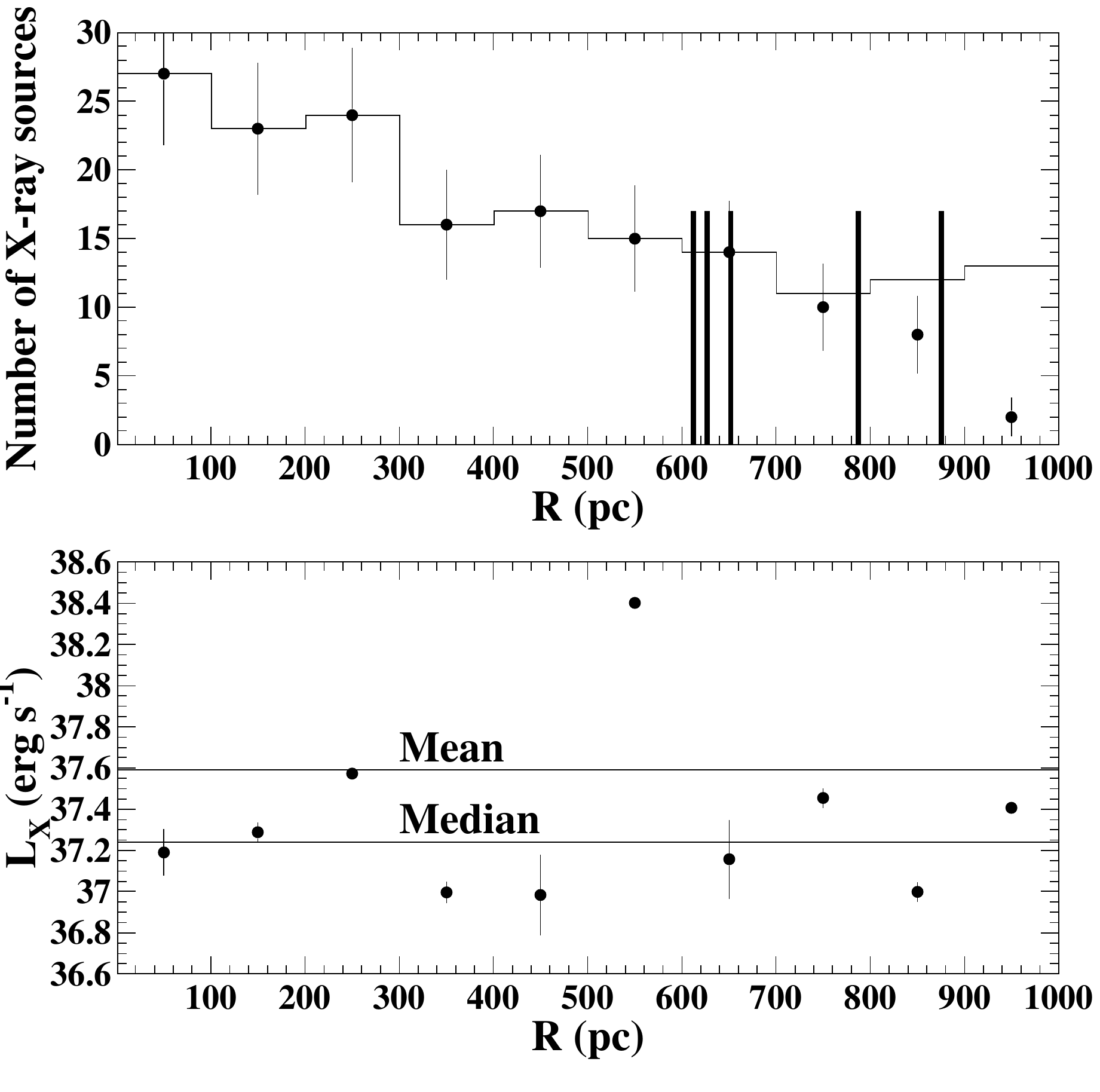}  
  \caption{Known heating sources. On the left side, we superimpose on
    the H$\alpha$+$[$NII$]$ map from \citet{1988AJ.....95..438C}: the
    X-ray sources detected by \citet{2013A&A...555A..65H} with blue
    dots; the 1 Gyr young stellar cluster detected by
    \citet{2012ApJS..199...37K} with the green triangle; and the X-ray
    sources detected by \citet{2011A&A...534A..55S} and
    \citet{2014ApJ...780...83B} with white and purple dots. The blue
    dots with a yellow circle correspond to the first (and more
    central) ultra-luminous X-ray transient (ULX) detected in M31
    \citep[e.g.][]{2012A&A...538A..49K}. The black contours correspond
    to the apparent A$_B$ extinction map discussed in
    \citet{2000MNRAS.312L..29M}. The light blue contours correspond to
    the 6\,cm polarised intensity detected by
    \citet{2014A&A...571A..61G}. The large red circles correspond to
    our detections. Light black circles provide typical projected
    radial distance from the centre (x pc and y pc). Thick ellipses
    show the approximate position of the inner ring. On the right
    side, we indicate, in the top panel, the number of X-ray sources
    \citep{2013A&A...555A..65H} detected at each radius; in the bottom
    panel, the X-ray luminosity at each radius is shown as a function
    of the radius. The mean and median values of the X-ray luminosity
    (averaged in 100pc radial bin) is also provided.}
  \label{fig:xray}%
\end{figure*}

There is an ultra-luminous X-ray source \citep{2012A&A...538A..49K} located 22\arcsec\, from M31-2c-a, which has exhibited an outburst that lasted three months. ULX are correlated with star formation \citep{2009MNRAS.395L..71M}. As in the Cartwheel, this ULX is detected next to the inner ring. The presence of ULX in the ring supports the starburst scenario, as several ULX are detected in ring galaxies suspected of frontal collision \citep{2012ApJ...747..150P}. ULX are expected in low-metallicity environments \citep{2013ApJ...769...92P}, which could explain why only one has been detected. \citet{2012ApJ...756...32B} measured an outburst flux of 50$\times 10^{37}$erg\,s$^{-1}$, which could contribute at the level of $5\times 10 ^{-3}$\,erg\,s$^{-1}$\,cm$^{-2}$.

 In this inner region, the contribution of the central black hole could be significant. However, \citet{2011ApJ...728L..10L} measured  a
burst of activity in January 2006 with Chandra X-Ray Observatory at a level of
4.3$\times10^{37}$\,erg\,s$^{-1}$ and a quiescent level of 4.8$\times
10^{36}$ erg\,s$^{-1}$. At the distance of M31I and M31I-B (supposed
in the sky plane), this would contribute to the X-ray incident
radiation intensity at a level of 6.9 and 5.5$\times
10^{-8}$erg\,s$^{-1}$\,cm${-2}$ (6.2 and 5.0$\times
10^{-7}$erg\,s$^{-1}$\,cm$^{-2}$) in the quiescent state (during
the outburst). This is four to five orders of magnitude weaker than the possible contribution of the ULX discussed above.

The X-ray sources shown in Fig. \ref{fig:xray} are strongly concentrated towards the centre. The catalogue \citet{2013A&A...555A..65H} based on Chandra HRC is more sensitive than other catalogues \citep{2011A&A...534A..55S,2014ApJ...780...83B}. The surface density varies as (R/1\,kpc)$^{-2.5}$. The contribution of the central part is thus more important than the black hole activity. 

We can thus exclude that the excitation mechanism of the gas is due to
X-ray radiation from the gas accreted by the black hole.  The $[OIII]$ line emissions detected by \citet{2010A&A...509A..61S} and M. Sarzi (private communication) can be accounted for by planetary nebulae and no LINER activity is necessary.

The heating is mainly due to the old stellar population as discussed by \citet{2014ApJ...780..172D} and \citet{2014A&A...567A..71V}. Transient X-ray sources, such as ULX, could contribute inhomogeneously to the heating.  The heating rate U, based on \citet{1983A&A...128..212M} and computed by
\citet{2014ApJ...780..172D}, are provided in Table \ref{table:1}. The area around M31I and M31I-B have $U=1.5$, while the other three positions have U in the range 1.8-2.7. This corresponds to an average dust temperature of 20\,K, according to \citet{2014ApJ...780..172D}.

\subsection{Different velocity components}
\label{ssect:kine}
The kinematics of this area is complex with up to three different velocity components. The probability that the warp component, coming from radii larger than 15-20\,kpc, is seen in CO emission towards the centre is negligible. This is the case because of the weak metallicity of this gas, which makes it undetectable in CO. As discussed in \citet{2011A&A...536A..52M}, if we consider that the probability of chance alignment of gas is low and that the gas is inside the main galaxy, this means that we are in the presence of several structures. Indeed, an explanation in terms of non-circular motions in a possible barred potential has already been ruled out given the velocity disposition and the fact that the gas distribution is not aligned with any elongated structure, but in an off-centred ring \citep{2011A&A...536A..52M}.
The component at the systemic velocity is compatible with gas along the minor axis of the main disc, while the two counter-rotating components could be orbiting in two planes with different inclinations and orientations. They could correspond to an inclined disc and to the inner 1-kpc ring. However, in presence of observations in two positions, it is difficult to fully constrain such a configuration. Apart from the inner ring detected in infrared (see Fig. \ref{fig:ext} and \citet{2006Natur.443..832B}) and the H$_\alpha$($+[NII]$) structure detected inside \citep{1988AJ.....95..438C}, we do not have kinematic observations covering these positions. \citet{1987A&A...178...91B} got the $[NII]$ velocity field within a $3\arcmin \times 3\arcmin$ field and \citet{2006A&A...453..459N} did not integrate deep enough in the central part to detect CO. \citet{2015IAUS..309..334O} has detected two rotating components in $[OIII]$, but their two components rotate in the same direction, which adds complexity to the problem.
Interestingly, the polarisation map of \citet{2014A&A...571A..61G} corresponds to the parameters proposed by \citet{2011A&A...536A..52M} for the three-component modelling.

Finally, the amplitudes between our different velocity components are different and larger than the simulated modulation proposed by \citet{2014ApJ...788L..38D}. 

The alternative interpretation is that we are observing gas
superimposed on the central part of M31, i.e. gas in the outskirts of
the disc
\citep{2015ApJ...807..153K,2015ApJ...804...79L,2013Natur.497..224W,2012AJ....144...52L}
is ruled out. No high velocity clouds have been detected in CO
\citep{2000A&A...357...75C,2005ApJ...631L..57M}, while Galactic gas
should be around $0$\,km\,s$^{-1}$.

\subsection{$^{13}$ C depletion and evidence for a post-starburst scenario}
\label{ssect:pstarb}
 In Sect. \ref{ssec:desc}, we discussed that it is remarkable that the
 C$^{18}$O detection is at a level comparable to the $^{13}$CO
 detection, in contrast to what is observed in the Galaxy
 \citep[e.g.][]{2015ApJS..216...18N,2014A&A...564A..68S}, where the
 C$^{18}$O is usually 10 to 150 times weaker than $^{13}$ CO. In Table
 \ref{tab:etabf}, we have shown that $^{13}$CO and C$^{18}$O are
 optically thin, so we can claim that the $^{13}$CO is under-abundant
 with respect to C$^{18}$O.  In Sect. \ref{ssec:radex}, we have shown
 that these two CO components are close to thermal equilibrium. Hence,
 we can rely on the abundance ratios computed in Table \ref{table:3b};
 the deficit is five to ten times weaker than with the Galactic
 abundances. This deficit could be accounted for by a post-starburst
 scenario. This kind of $^{13}$CO deficit was first observed by
 \citet{1991A&A...251....1C} in post-merger.  \citet{2014MNRAS.445.2378D}
 also find an anti-correlation of $^{13}$CO and star formation and gas
 surface densities in local galaxies.

This could be the signature of a 200\,Myr old starburst triggered by a
collision with a galaxy like M32, as proposed by \citet{2006Natur.443..832B}. This scenario accounts for the two off-centred ring structures of M31, with most star formation in the 10\,kpc ring \citep{2013ApJ...769...55F} and a smaller ring, where we have detected CO and dense gas.

\section{Conclusion}
We have detected $^{12}$CO(2-1) and { $^{13}$CO(2-1)} lines with multiple
 velocity components along the minor axis of M31 on both sides of the
inner 1-kpc ring. We detected dense gas traced by C$^{18}$O(2-1), HCN(1-0), and HCO+(1-0)
 on the north-west side  at the systemic velocity for  C$^{18}$O(2-1) and in the blue-shifted component for HCN(1-0) and HCO+(1-0). We also detected a redshifted component for the three molecules by stacking at the CO velocity on both sides. 

\citet{2014ApJ...780..172D} and
\citet{2014A&A...567A..71V} found lower measurements for the stellar heating in this area on the north-west side, which could explain why dense gas has
only been detected directly on this side.  A careful analysis has shown that
$^{12}$CO is optically thick while the other CO lines are optically
thin and close to LTE at an excitation temperature of 17.5-20\,K. The
gas is very clumpy and the averaged beam filling factor is
0.8$\%$. The average column density of molecular hydrogen is
16\,10$^{22}$\,cm$^{-2}$. The derived abundances are all smaller than
the standard abundances, but for C$^{18}$O. The comparison of the
derived abundances with RADEX simulations shows that the HCN and HCO+
gas correspond to dense clumps with molecular hydrogen density of the
range { $1 - 6\,\times\,10^5$\,cm$^{-3}$} and an excitation temperature of
{ 9\,K.} The HCO+/HCN line ratio is comparable with that measured for
clouds detected in M31's disc \citep{2005A&A...429..153B}, while the
derived abundances are smaller than the Galactic abundances
\citep{1995ApJ...441..222B}, with a larger discrepancy for HCN(1-0).  Only one transient ULX object has been identified as a possible secondary source of heating, while the bulge stellar population is sufficient to heat
the gas and reproduce the infrared maps.  

 Our isotopic CO lines and HCN/HCO+ lines detections are all weak and close to the detection limits. We detected C$^{18}$O in a single point at the systemic velocity. Stacking of all $^{13}$CO
detections has revealed the underlying presence of C$^{18}$O gas in
the redshifted component. Both the single point and the stacking
detections have an overall intensity comparable to $^{13}$CO. As we
expect C$^{18}$O to be six times less abundant than $^{13}$CO and the
measured abundance of C$^{18}$O is compatible with the standard
abundance, $^{13}$CO should be depleted.  A coherent explanation is
that $^{12}$C has been enriched during the recent staburst and $^{13}$CO had no time to catch up, so it appears depleted. This is compatible
with the \citet{2006Natur.443..832B} scenario suggesting a frontal
collision with a nearby companion galaxy such as M\,32. This collision
explains the rings and has triggered a starburst at the epoch of the
collision. This is also in agreement with the central starburst, which
occured next to the black hole 200\,Myr ago
\citep{2012ApJ...745..121L}.

 We confirm the presence of  three kinematical components detected in molecular gas, as first discussed in \citet{2011A&A...536A..52M}. Separated by 400\,km\,s$^{-1}$ and on both sides of the systemic velocity, this configuration on the minor axis is not expected for a disc in rotation and cannot be accounted for by a bar. We have also checked that the parameters we first proposed for a simple modelling of the three kinematical components are compatible with other observations. In addition to the 77$\deg$ main disc with a position angle (PA$=35\deg$) detected in HI, there is an less inclined nuclear disc (i$=43\deg$) with PA$=53\deg$ (rather than $70\deg$, as initially thought). The position angle of this nuclear disc is well defined with different stellar sources catalogues \citep{2002ApJ...578..114K,2012ApJ...755..131R}. The 1-kpc ring with i$=40\deg$ and PA$=-35\deg$ accounts for the polarised map of \citet{2014A&A...571A..61G} very well.

\begin{acknowledgements}
We thank the IRAM staff in Granada for their help during the
observations. We are most grateful to S. Viaene for providing us with
the maps produced in the Viaene et al. 2014 paper. We used GALEX
and ROSAT archive data in Sect. \ref{ssect:heat}. We thank the anonymous referee for his useful comments, which improved the paper.
\end{acknowledgements}

≈

\appendix
\section{Additional Figures}
\begin{figure*}
  \centering
  \includegraphics[width=\textwidth]{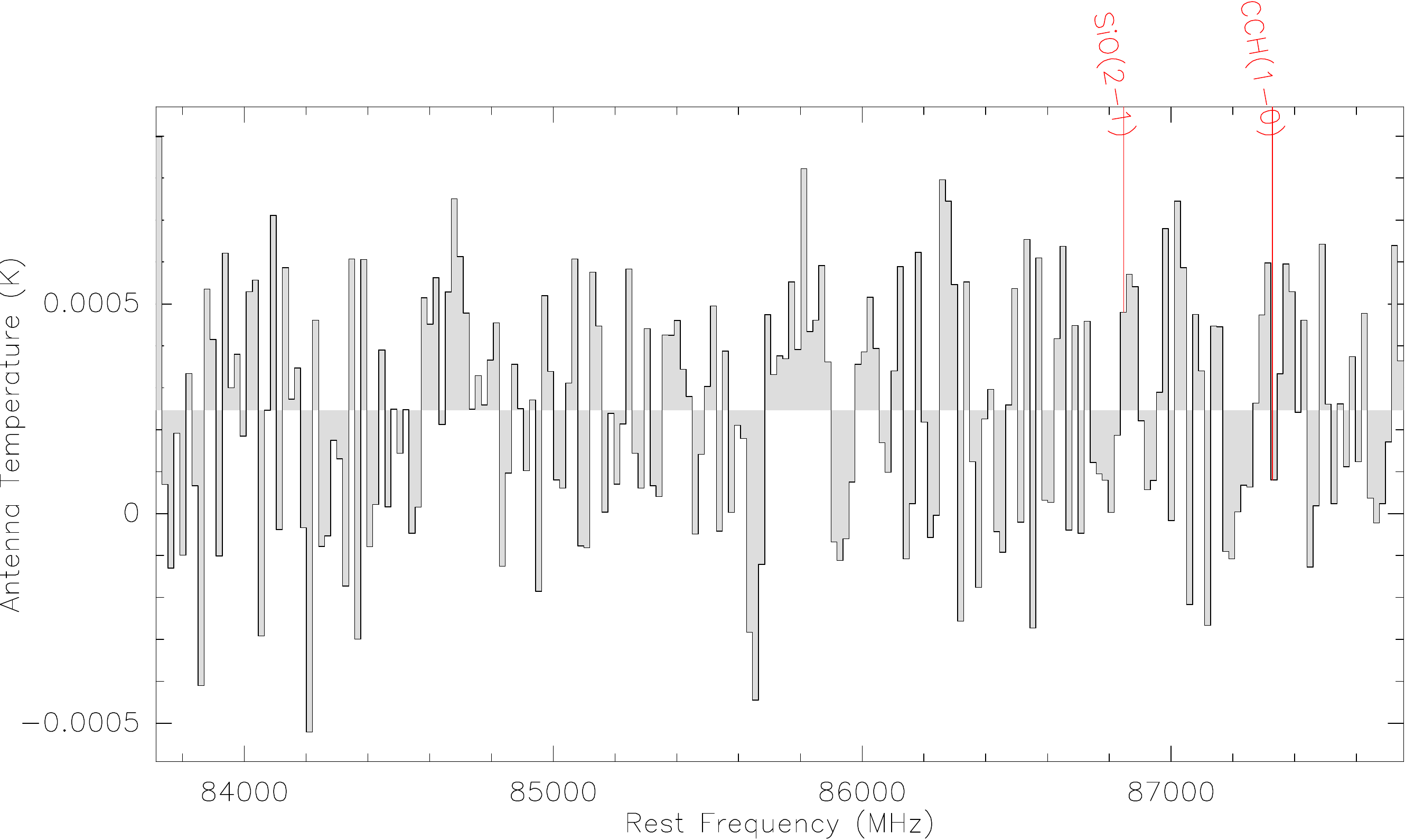}
  \includegraphics[width=\textwidth]{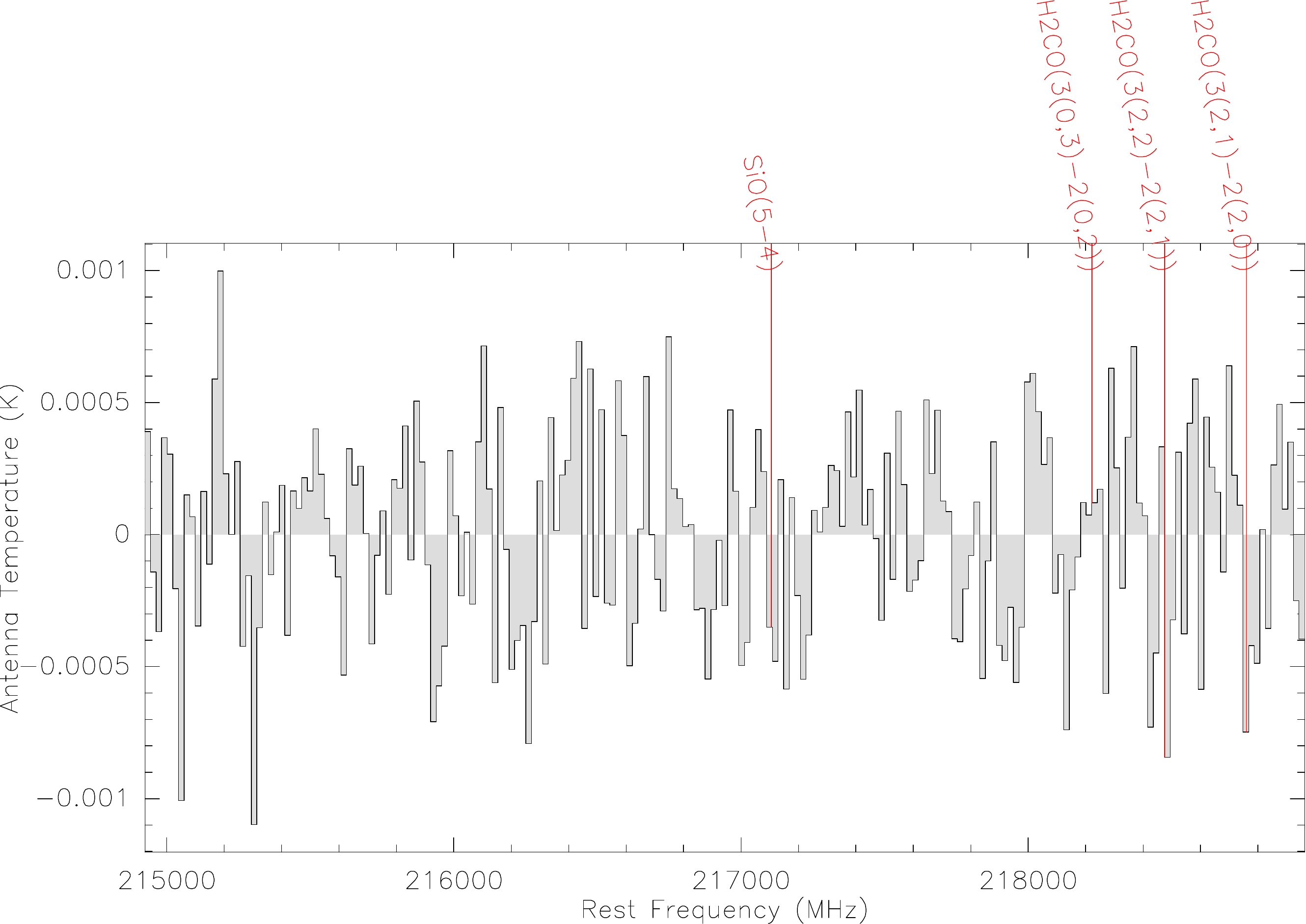}
  \caption{Averaging of the observations performed in the EMIR lower
    outer (LO) bands. The different lines present in the bandwidth
    are plotted in the rest-frame frequency. No signal is
    detected. This has been used to derived detection limit in Table
    \ref{table:3}.}
  \label{fig:stackpllo}%
\end{figure*}

\begin{figure}
  \centering \includegraphics[width=7.95cm]{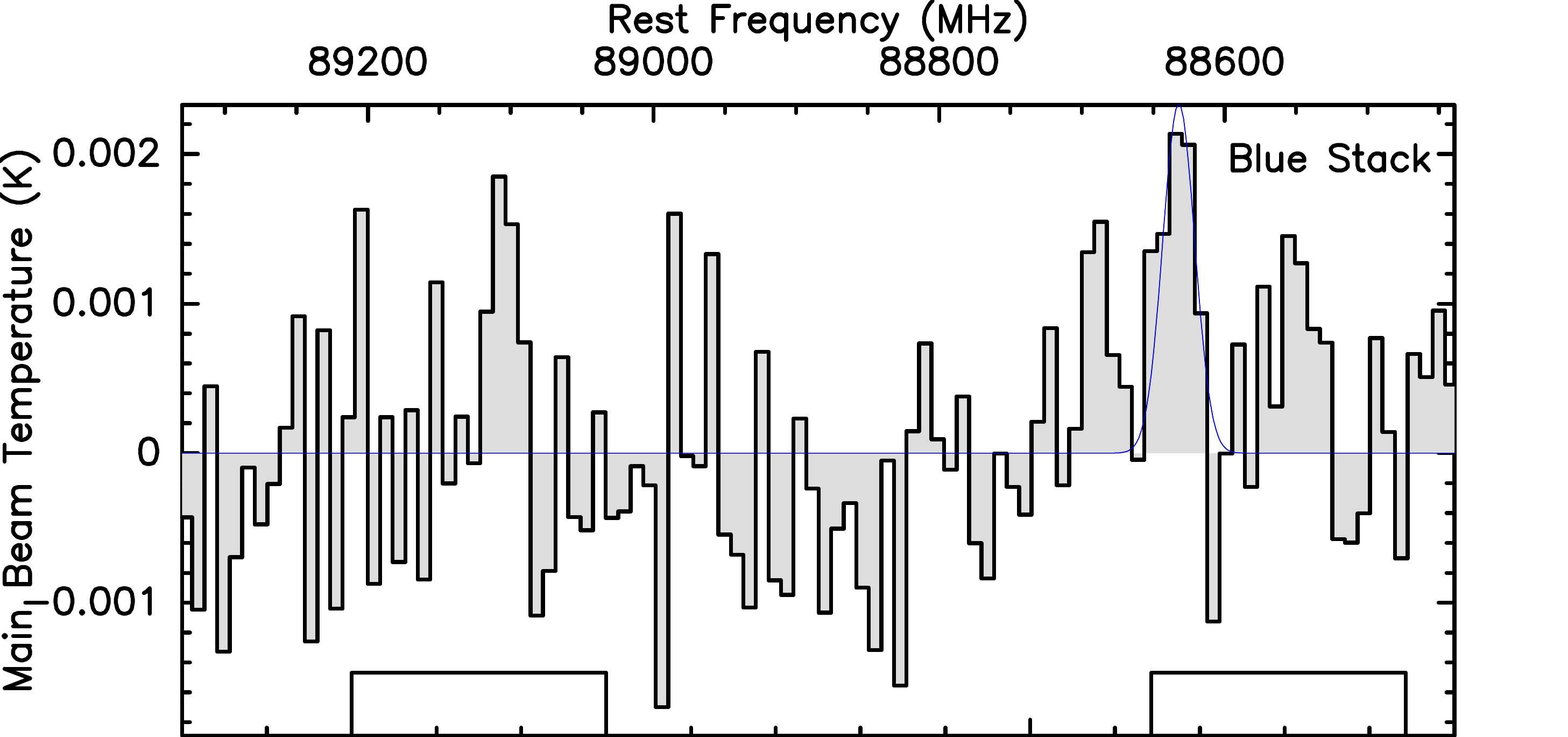} \includegraphics[width=7.95cm]{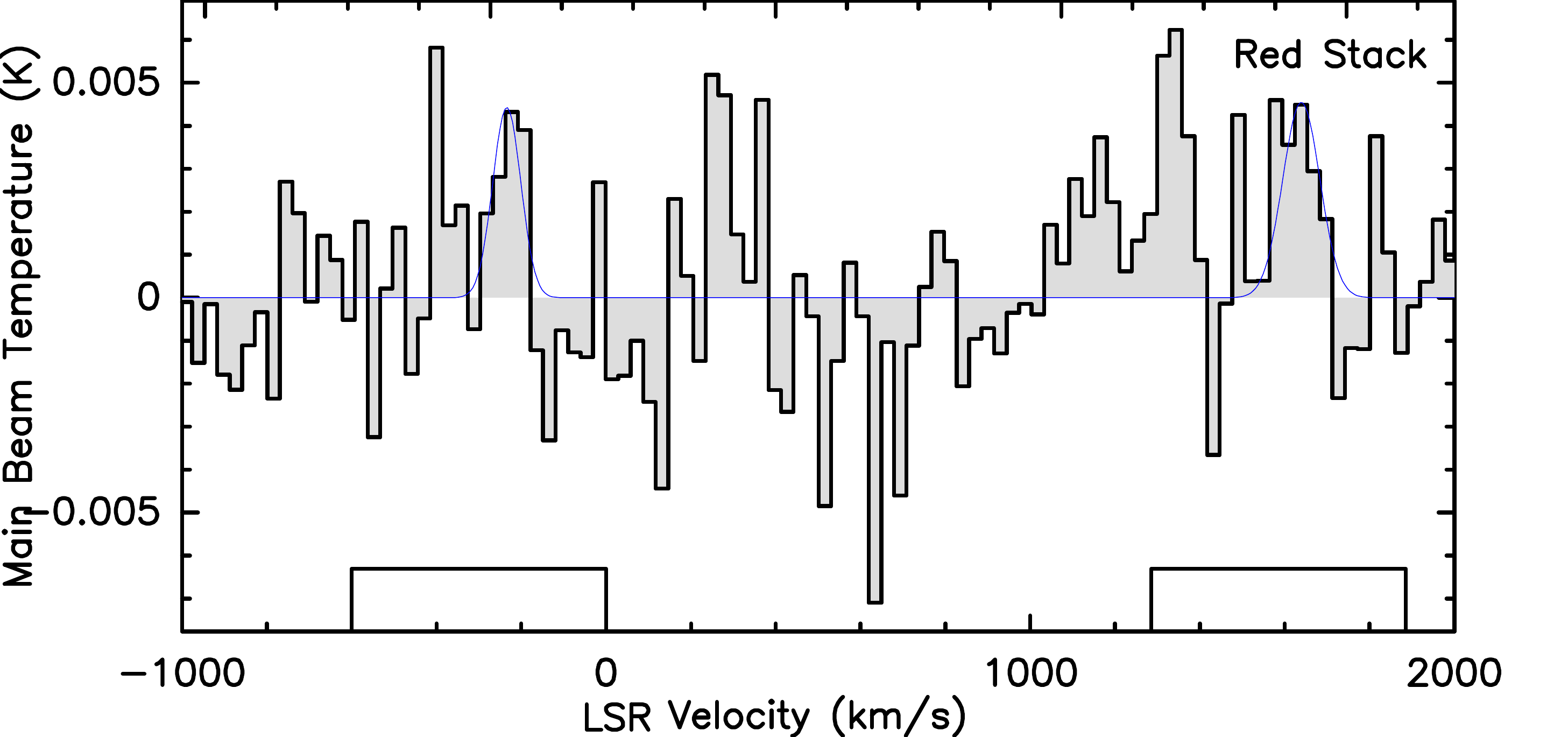} \caption{{
  HCN(1-0) and HCO+(1-0) stacking of the south-east positions. The
  stacking is based on the $^{13}$CO} velocities. The top (bottom)
  panel shows the stacking corresponding to $> -300$\,km\,s$^{-1}$
  ($<-300$\,km\,s$^{-1}$) velocities. The velocities are shown in the
  rest-frame frequency of the HCO+(1-0) line.}  \label{fig:appstack}%
\end{figure}

\begin{figure*}
  \centering
  \includegraphics[width=\textwidth]{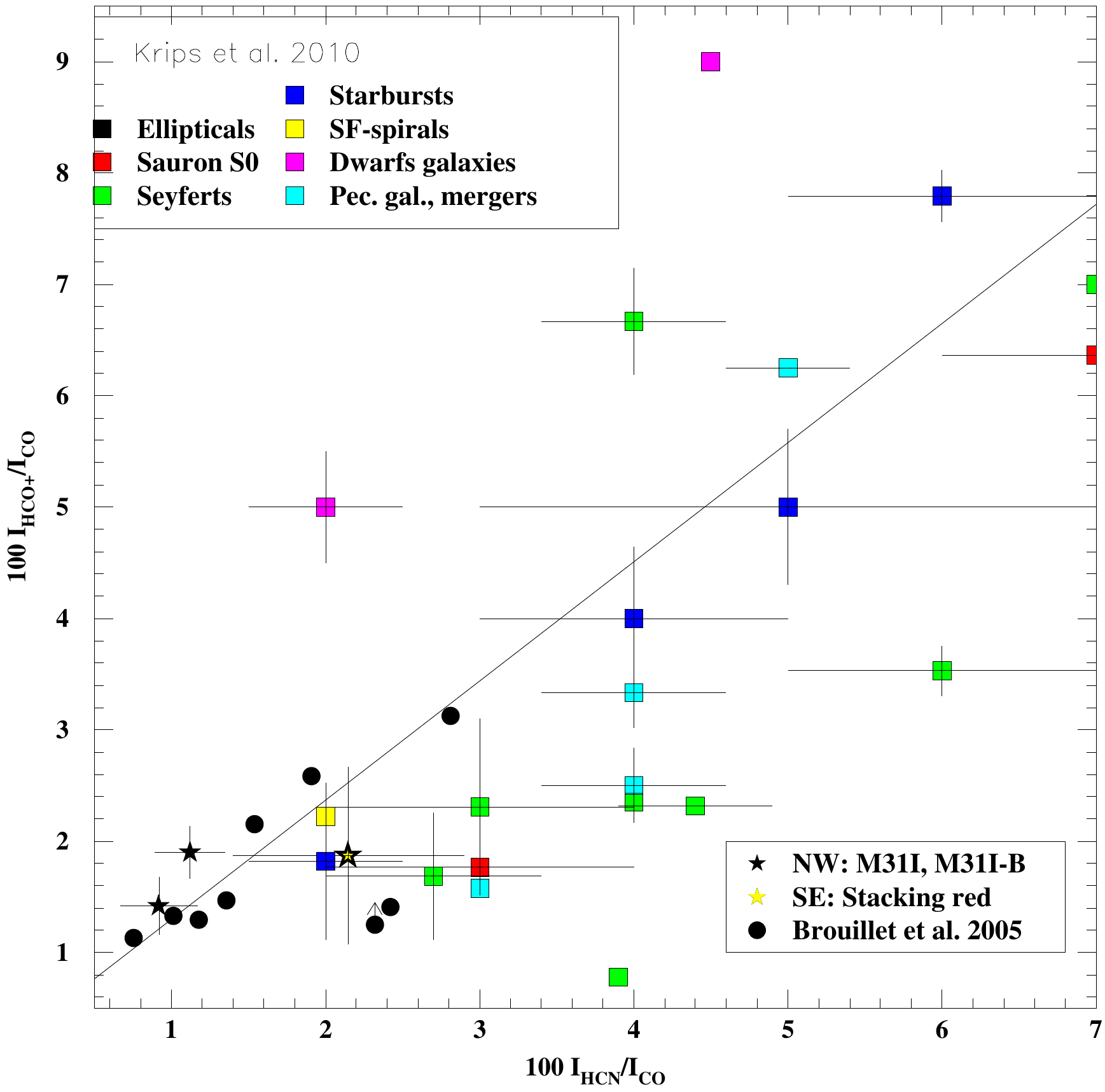}
  \caption{Comparison with other catalogues (2). HCO+/CO(1-0) line
    ratio versus HCN/CO line ratio. The two detections presented here (M31I and M31I-B) are superimposed on previous
    measurements, namely: detections across M31's main disc by
    \citet{2005A&A...429..153B} and a compilation of various types of
    galaxies from \citet{2010MNRAS.407.2261K}. The $^{12}$CO(2-1)
    measurements presented here have been corrected by a CO
    2-1/1-0 line ratio of 0.8, as measured in
    \citet{2011A&A...536A..52M} in a complex M31G within 10 and
    30\,$\arcsec$ from the observed positions. The line indicates the
    best fit provided by \citet{2005A&A...429..153B} for M31's main
    disc. Our points are systematically above the best fit they
    provide, suggesting an excess of HCO+. The overall distribution
    of \citet{2010MNRAS.407.2261K} does not exhibit any clear
    pattern. }
  \label{fig:lrat_dense2}
\end{figure*}

\begin{figure*}
  \centering
  \includegraphics[width=\textwidth]{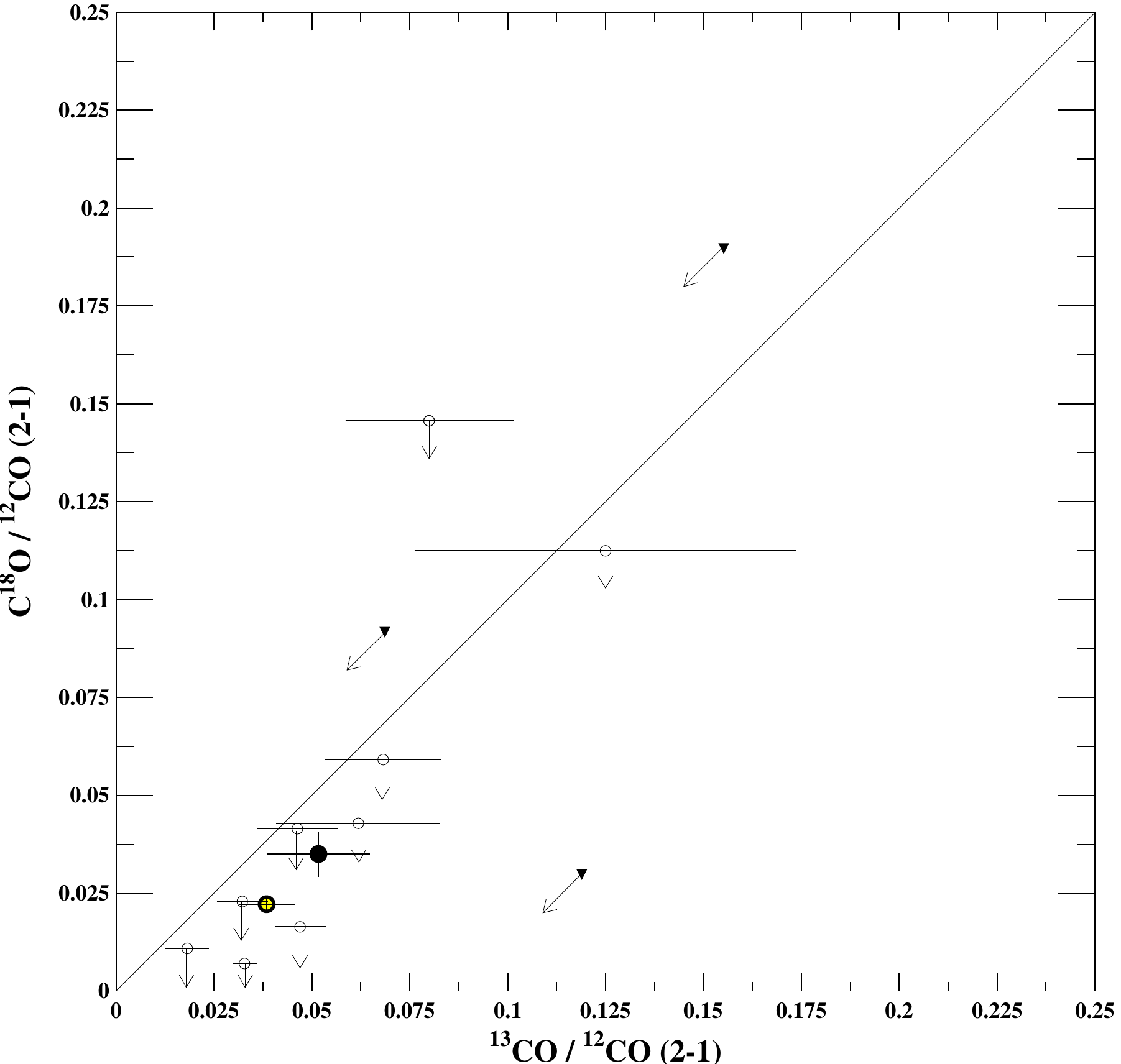}
  \caption{C$^{18}$O/$^{12}$CO (2-1) line ratios versus the
    $^{13}$CO/$^{12}$CO(2-1) line ratios presented in this paper. Arrows
    indicated 3$\sigma$ upper limits.  The full black (empty yellow) point  with error bars corresponds to our direct detection (our red stack detection). The line indicates a slope 1.}
  \label{fig:coli}%
\end{figure*}
   
\begin{figure*}
 \centering
  \includegraphics[width=\textwidth]{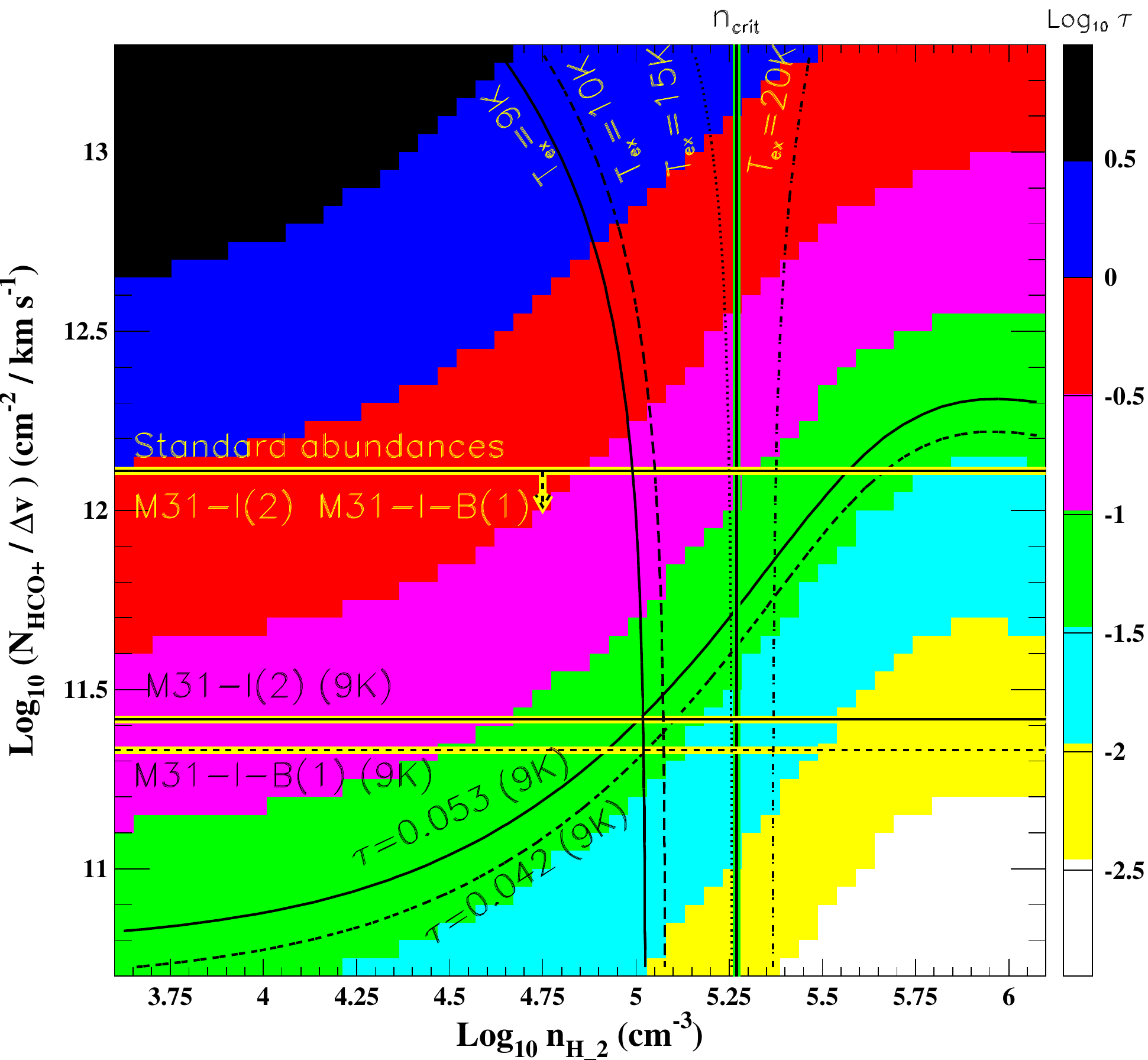}
  \caption{Main physical parameters corresponding to the HCO+(1-0) gas
    observed in M31-I(2) and M31-I-B(1). The optical depths are
    shown as function of the HCO+(1-0) column density (per
    km\,s$^{-1}$) and the molecular hydrogen density $n_{H_2}$. The
    vertical (green) line indicates the critical density computed for a
    collisional temperature of 20\,K (see Table \ref{table:2}). The
    lower horizontal full (dashed) line shows column densities
    measured for M31-I(2) (M31-I-B(1)) at 8\,K. The upper
    horizontal line corresponds to the column densities computed with
    standard abundances \citep{1995ApJ...441..222B} and the
    corresponding molecular hydrogen column densities. For M31-I-B(1),
    we derive an upper limit as the molecular hydrogen column density is
    only available for a smaller beam (see Table \ref{table:3}) and we
    expect some dilution of the signal, as is the case for M31-I(2).  The contour levels correspond
    to excitation temperatures of { 9\,K} (full line), 10\,K (dashed
    line), 15\,K (dotted line), and { 20\,K} (dash-dotted line).}
   \label{fig:rdxhc}%
\end{figure*}
\end{document}